%% file: dynamics_nearPotentialGames.tex
\documentclass[preprint,authoryear,12pt]{elsarticle}
\usepackage[margin=1in]{geometry}
\usepackage{mdwlist}

\usepackage{subfig}
\usepackage{tikz}
\usetikzlibrary{matrix,arrows,automata}
\usepackage{amsmath,amssymb,amsthm,url}
\usepackage{graphics}
\usepackage{float}
\usepackage{enumerate}
\usepackage{verbatim}
\usepackage{pgf}
\usepackage{comment}

\usepackage{array}

\newtheorem{theorem}{Theorem}[section]
\newtheorem{lemma}{Lemma}[section]
\newtheorem{proposition}{Proposition}[section]

\newtheorem{definition}{Definition}[section]
\newtheorem{corollary}{Corollary}[section]

\newtheorem{example}{Example}[section]

\newcommand{\be}{\begin{equation}}
\newcommand{\ee}{\end{equation}}

\newcommand{\bydef}{\stackrel{\bigtriangleup}{=}}
\newcommand{\eps}{\epsilon}

\bibpunct{(}{)}{,}{a}{}{;}
\biboptions{sort}

 \makeatletter
\def\ps@pprintTitle{%
  \let\@oddhead\@empty
  \let\@evenhead\@empty
  \let\@oddfoot\@empty
  \let\@evenfoot\@oddfoot
}
\makeatother
\let\today\relax

\begin{document}
\begin{frontmatter}
\title{\LARGE \bf
Dynamics in  Near-Potential Games
}

\author[lids]{Ozan Candogan\corref{cor1}}
\ead{candogan@mit.edu}

\author[lids]{Asuman Ozdaglar}
\ead{asuman@mit.edu}

\author[lids]{Pablo Parrilo}
\ead{parrilo@mit.edu}

\cortext[cor1]{Corresponding author. Phone: (+1)(617) 253-9842, Fax: 617-253-3578.}
\address[lids]{Laboratory of Information and Decision Systems,
       Massachusetts Institute of Technology,      77 Massachusetts Avenue, Room 32-D608     Cambridge, MA 02139.}
\begin{abstract}
Except for special classes of games, there is no systematic
framework for analyzing the dynamical properties of multi-agent
strategic interactions. Potential games are  one such special but
restrictive class of games that allow for tractable dynamic
analysis. Intuitively, games that are ``close'' to a potential game
should share similar properties. In this paper, we formalize and
develop this idea by quantifying  to what extent the dynamic
features of potential games extend to ``near-potential'' games.

We study convergence of three commonly studied classes of adaptive
dynamics: discrete-time better/best response, logit response, and
discrete-time fictitious play dynamics. For better/best response
dynamics, we focus on the evolution of the sequence of pure strategy
profiles and show that this sequence converges to a (pure)
approximate equilibrium set, whose size is a function of the
``distance" from a close potential game. We then study logit
response dynamics parametrized by a smoothing parameter that
determines the frequency with which the best response strategy is
played. Our analysis uses a Markov chain representation for the
evolution of pure strategy profiles. We provide a characterization
of the stationary distribution of this Markov chain in terms of the
distance of the game from a close potential game and the
corresponding potential function. We further show that the
stochastically stable strategy profiles (defined as those that have
positive probability under the stationary distribution in the limit
as the smoothing parameter goes to 0) are pure approximate
equilibria. Finally, we turn attention to fictitious play, and
establish that in near-potential games, the sequence of empirical
frequencies of player actions converges to a neighborhood of (mixed)
equilibria of the game, where the size of the neighborhood increases
with distance of the game to a potential game. Thus, our results suggest
that games that are close to a potential game inherit the dynamical
properties of potential games. Since  a close potential game to a
given game can be found by solving a convex optimization problem,
our approach also provides a systematic framework for studying
convergence behavior of adaptive learning dynamics in arbitrary
finite strategic form games.

 \end{abstract}

\end{frontmatter}

\section{Introduction}

The study of multi-agent strategic interactions both in economics
and engineering mainly relies on the concept of Nash equilibrium.
This raises the question  whether Nash equilibrium makes
approximately accurate predictions of the user behavior. One possible
justification for  Nash equilibrium is that it arises as the long
run outcome of dynamical processes, in which less than fully
rational players search for optimality over time. However, unless
the game belongs to special (but restrictive) classes of games, such
dynamics do not converge to a Nash equilibrium, and there is no
systematic analysis of their limiting behavior
\citep{jordan1993three,fudenberg1998tlg,shapley1964some}.

Potential games is a class of games for which  many of the  simple
user dynamics, such as best response dynamics and fictitious play,
converge to a Nash equilibrium  \citep{monderer1996pg,monderer1996fpp,fudenberg1998tlg,sandholm2010population,young2004strategic}. Intuitively,
dynamics in potential games and dynamics in games that are ``close''
(in terms of the payoffs of the players) to potential games should
be related. Our goal in this paper is to make this intuition precise
and provide a systematic framework for studying dynamics in finite
strategic form games by exploiting their relation to close potential
games.

We start by illustrating via examples that this ``continuity"
property of limiting dynamics need not hold for arbitrary games,
i.e., games that are close in terms of payoffs may have
significantly different limiting behavior under simple user
dynamics. Our first example focuses on better response dynamics in
which at each step or strategy profile, a player (chosen
consecutively or at random) updates its strategy unilaterally to one
that yields a better payoff.
\footnote{
Consider a game where players are not indifferent between their strategies at any strategy profile. Arbitrarily small payoff perturbations of this game lead to games which have the same better response structure as the original game. Hence, for a given game there may exist a close enough game such that  the outcome of the better response dynamics in two games are identical. However, for payoff differences of given size it is always possible to find games with different better response properties as illustrated in Example \ref{ex:nonpotGameConv_2x2}.}
\begin{example} \label{ex:nonpotGameConv_2x2}

Consider two games with two players and payoffs given in Figure
\ref{fig:noConv_2x2}.
The entries of these tables indexed by row X and column Y show
payoffs of the players when the first player uses strategy X and the
second player uses strategy Y. Let $0< \theta \ll 1$. Both games
have a unique Nash equilibrium: $(B,B)$ for ${\cal G}_1$,
and the mixed strategy profile $\left( \frac{2}{3}A+ \frac{1}{3}
B,\frac{\theta}{1+\theta} A + \frac{1}{1+\theta}B\right)$ for ${\cal G}_2$.

We consider convergence of the sequence of pure strategy profiles
generated by the better response dynamics. In ${\cal G}_1$,
the sequence converges to strategy profile $(B,B)$. In ${\cal G}_2$, the sequence does not converge (it can be shown that the
sequence follows the better response cycle  $(A,A)$, $(B,A)$,
$(B,B)$ and $(A,B)$). Thus, trajectories are not contained in any
$\epsilon$-equilibrium set for $\epsilon<2$.
  \begin{figure}[h]
   \begin{center}
   \begin{tabular}{ | c | c | c  |}
    \hline
      & A     & B      \\ \hline
    A  &  0, 1 & 0, 0   \\ \hline
    B  &  1, 0 & $\theta$, 2 \\ \hline
\multicolumn{3}{c}{ ${\cal G}_1$ }
    \end{tabular}
    \qquad
     \begin{tabular}{ | c | c |  c |}
      \hline
          & A     & B    \\ \hline
      A  &  0, 1 & 0, 0 \\ \hline
      B  & 1, 0 & $-\theta$, 2 \\ \hline
\multicolumn{3}{c}{ ${\cal G}_2$ }
      \end{tabular}
   \end{center}
   \caption{A small change in payoffs
      results in significantly different behavior for the pure strategy profiles generated
      by the better response dynamics.}
         \label{fig:noConv_2x2}
   \end{figure}
  \end{example}

The second example considers fictitious play dynamics, where at each
step, each player maintains an (independent) empirical frequency
distribution of other player's strategies and plays a best response
against it.

 \begin{example} \label{ex:nonpotGameConv_shapley}
Consider  two games with two players and payoffs given in Figure
\ref{fig:noConv_shapley}. Let $0< \theta \ll 1$.
It can be seen that ${\cal G}_1$ has multiple equilibria (including pure equilibria $(A,A)$,
$(B,B)$ and $(C,C)$), whereas
${\cal G}_2$
 has a unique
equilibrium given by the mixed strategy profile where both players
assign $1/3$ probability to each of its strategies.

  \begin{figure}[h]
   \begin{center}
   \begin{tabular}{ | c | c | c | c |}
    \hline
      & A     & B  & C    \\ \hline
    A  &  1, 1 & 1, 0      & 0, 1   \\ \hline
    B  &  0, 1 & 1, 1       & 1, 0  \\ \hline
    C  &  1, 0 & 0, 1       & 1, 1 \\ \hline
\multicolumn{4}{c}{ ${\cal G}_1$ }
    \end{tabular}
 \qquad
     \begin{tabular}{ | c | c | c | c |}
       \hline
         & A     & B  & C  \\ \hline
       A  &  $1-\theta, 1-\theta$  & 1, 0      & 0, 1   \\ \hline
       B  &  0, 1 & $1-\theta, 1-\theta$        & 1, 0  \\ \hline
       C  &  1, 0 & 0, 1        & $1-\theta, 1-\theta$ \\ \hline
\multicolumn{4}{c}{ ${\cal G}_2$ }
       \end{tabular}
   \end{center}
      \caption{A small change in payoffs
      results in significantly different behavior for the empirical frequencies generated
      by the fictitious play dynamics.}
            \label{fig:noConv_shapley}
   \end{figure}

We focus on the convergence of the sequence of empirical frequencies
generated by the fictitious play dynamics (under the assumption that
initial empirical frequency distribution assigns probability $1$ to
a pure strategy profile, and whenever players are indifferent
between different strategies, they choose the lexicographically
smaller one). In ${\cal G}_1$, this sequence converges to a
pure equilibrium starting from any pure strategy profile. In ${\cal G}_2$, the sequence displays oscillations similar to
those seen in the Shapley game
(see \citet{shapley1964some,fudenberg1998tlg}). To see this,
assume that the initial empirical frequency distribution assigns
probability 1 to the strategy profile $(A,A)$. Observe that since
the underlying game is a symmetric game,
 empirical frequency distribution of each player will be identical at all steps.
Starting from $(A,A)$, both players update their strategy to $C$.
After sufficiently many updates, the empirical frequency of  $A$
falls below $\theta/(1+\theta)$, and that of $C$ exceeds $1/(1+\theta)$. Thus, the
payoff specifications suggest that both players start using strategy
$B$. Similarly, after empirical frequency of $B$ exceeds $1/(1+\theta)$,
and that of $C$ falls below $\theta/(1+\theta)$, then both players start
playing $A$. Observe that update to a new strategy takes place only
when one of the strategies is being used with very high probability
(recall that $\theta\ll 1$) and this feature of empirical
frequencies is preserved throughout. For this reason the sequence of
empirical frequencies does not converge to $(1/3,1/3,1/3)$, the unique Nash
equilibrium of ${\cal G}_2$.
 \end{example}

In this paper, in contrast with the preceding examples, we will show
that games that are close (in terms of payoffs of players) to
potential games have similar limiting dynamics to those in potential
games. In particular, many reasonable adaptive dynamics ``converge"
to an approximate equilibrium set, whose size is a function of the
distance of the game to a close potential game. Our approach relies
on using the potential function of a close potential game for the
analysis of commonly studied update rules.\footnote{Throughout the
paper, we use the terms {\em learning dynamics} and {\em update
rules} interchangeably.} We note that our results hold for arbitrary
strategic form games, however
our characterization of limiting behavior of dynamics is more
informative for games that are close to potential games. We
therefore focus our investigation to such games in this paper and
refer to them as {\em near-potential games}.

We start our analysis by introducing  \emph{maximum pairwise difference}, a measure of ``closeness"  of games.
Let  $\textbf{p}$ and $\textbf{q}$ be two strategy profiles, which differ in the strategy of a single player, say player $m$.
We refer to the change in the payoff of player $m$ between these two strategy profiles, as the pairwise comparison of $\textbf{p}$ and $\textbf{q}$.
Intuitively,  this quantity captures how much player $m$ can improve its utility by unilaterally deviating from strategy profile $\textbf{p}$  to  strategy profile $\textbf{q}$.
For given games,  the maximum pairwise difference is defined as the maximum difference between the pairwise comparisons of these games.
Thus, the maximum pairwise difference captures how different two games are in terms of the utility  improvements due to unilateral deviations.
Since equilibria of games, and strategy updates in various update rules (such as better/best response dynamics) can be expressed in terms of unilateral deviations, maximum pairwise difference provides a measure of strategic similarities of games.
We  show that the closest potential  game to a given game, in the sense of maximum pairwise difference, can be obtained by solving a convex optimization problem.
This provides a systematic way of approximating a given game with a
potential game that has a similar equilibrium set and dynamic properties,
as illustrated in Example \ref{ex:graph3game}.

\begin{example}
\label{ex:graph3game}
Consider a two-player game $\cal G$, which is not a potential game, and the  closest potential game to this game (in terms of maximum pairwise difference), $\hat{\cal G}$, given in Figure \ref{fig:MPD_EX}.
The maximum pairwise difference of these games is $2$, since the utility improvements in these games due to unilateral deviations differ by at most $2$
(For instance consider the deviation of the column player from $(A,A)$ to $(A,B)$. In $\cal G$ this leads to a utility improvement of $6$, whereas, in $\hat{G}$ the improvement  amount is $4$).
It can be seen that for both games $(B,B)$ is the unique equilibrium. Moreover,   trajectories of better response dynamics and empirical frequencies of fictitious play dynamics converge to this equilibrium in both games.
 \begin{figure}[h]
   \begin{center}
   \begin{tabular}{ | c | c | c  |}
    \hline
      & A     & B      \\ \hline
    A  &  8, 2 & 8, 8   \\ \hline
    B  &  2, 2 & 12, 10 \\ \hline
\multicolumn{3}{c}{ ${\cal G}$ }
    \end{tabular}
    \qquad
     \begin{tabular}{ | c | c |  c |}
      \hline
          & A     & B    \\ \hline
      A  &  7, 3 & 9, 7 \\ \hline
      B  & 3, 1 & 11, 11 \\ \hline
\multicolumn{3}{c}{ $\hat{\cal G}$ }
      \end{tabular}
   \end{center}
   \caption{A game ($\cal G$) and a nearby potential game ($\hat{\cal G}$) share similar equilibrium set and dynamic properties.}
            \label{fig:MPD_EX}
   \end{figure}
\end{example}

We focus on three commonly studied user dynamics: discrete-time
better/best response, logit response, and discrete-time fictitious
play dynamics, and establish different notions of convergence for
each. We first study {\em better/best response dynamics}.  It is
known that the sequence of pure strategy profiles, which we refer to
as {\it trajectories}, generated by these update rules converge to
pure Nash equilibria in potential games
\citep{monderer1996pg,young2004strategic}. In near-potential games, a
pure Nash equilibrium need not even exist. For this reason we focus
on the notion of \emph{pure approximate equilibria} or
\emph{$\eps$-equilibria}, and show that in near-potential games
trajectories of these update rules converge to a pure approximate
equilibrium set.
The size of this set only depends on the distance of the original game from a potential game,
and is independent of the payoffs in the original game.

We then focus on \emph{logit response} dynamics. With this update
rule, agents,  when updating their strategies, choose their
best responses with high probability, but  also explore other
strategies with a nonzero probability. Logit response induces a
Markov chain on the set of pure strategy profiles. The stationary
distribution of this Markov chain is used to explain the limiting
behavior of this update rule \citep{young1993evolution,blume1997population,Blume1993387,alos2010logit,marden2008rll}.
In potential games, the
stationary distribution can be  expressed in closed form in terms of
the potential function of the game. Additionally,  the
\emph{stochastically stable strategy profiles}, i.e., the strategy
profiles which have nonzero stationary distribution as the
exploration probability goes to zero, are those that maximize the
potential function \citep{alos2010logit,blume1997population,marden2008rll}. Exploiting
their relation to close potential games, we obtain similar results
for near-potential games: (i) we obtain an explicit characterization
of the stationary distribution in terms of the distance of the game
from a close potential game and the corresponding potential
function, and (ii) we show that the stochastically stable strategy
profiles are the strategy profiles that approximately maximize the
potential of a close potential game, implying that they are pure
approximate equilibria of the game. Our analysis relies on a novel
perturbation result
 for Markov chains
(see Theorem~\ref{theo:statDist})
which provides bounds on deviations from a stationary distribution
when transition probabilities of a Markov chain are {\em
multiplicatively} perturbed, and therefore may be of independent
interest.

A summary of our convergence results on better/best response and
logit response dynamics   can be found in Table
\ref{table:summary1}.
\begin{table}[H]
\centering
\begin{tabular}{ | m{3.6cm} |  m{12cm} | }
\hline Update Rule                       &  Convergence Result  \\
\hline Better/Best Response Dynamics                & (Theorem
\ref{theo:BRconv}) Trajectories of dynamics  converge to ${\cal
X}_{\delta h}$, i.e.,
 the $\delta h$-equilibrium set of $\cal G$. \\ \hline
Logit Response Dynamics (with parameter $\tau$)              & (Corollary \ref{cor:logitNearPot}) Stationary distribution  $\mu_{\tau}$ of logit response dynamics is such that
$\left|\mu_{\tau}({\mathbf{p}})-
 \frac{e^{\frac{1}{\tau}\phi(\mathbf{p})}}{\sum_{\mathbf{q}\in E}e^{\frac{1}{\tau}\phi(\mathbf{q})}}
  \right|
  \leq
  \frac{e^{\frac{2\delta (h-1)}{\tau}}-1}{e^{\frac{2\delta (h-1)}{\tau}}+1}$, for all $\mathbf{p}$.
   \\ \hline
Logit Response Dynamics                  & (Corollary \ref{cor:logitSS})
 Stochastically stable strategy profiles of $\cal G$ are (i)~contained in $S=\{\mathbf{p} | \phi(\mathbf{p}) \geq \max_{\mathbf{q} } \phi(\mathbf{q}) - 4\delta(h-1)\}$, (ii)  $4\delta h$-equilibria of $\cal G$.   \\ \hline
\end{tabular}
 \caption{Convergence properties of better/best response and logit response dynamics in near-potential games.
 Given a game $\cal G$, we use $\hat{\cal G}$ to denote a
 nearby potential game with potential function $\phi$ such that the distance
 (in terms of the maximum pairwise difference, defined in Section \ref{sec:pre}) between the two games is $\delta$.
   We use the notation ${\cal X}_\epsilon$ to denote the $\epsilon$-equilibrium set of the original game,
   $h$ to denote the number of  strategy profiles,
    $\mu_{\tau}$ and $\hat{\mu}_\tau$ to denote the stationary distributions of logit response dynamics in $\cal G$ and $\hat{\cal G}$, respectively.}
 \label{table:summary1}
\end{table}

We finally analyze {\em fictitious play dynamics} in near-potential
games. In potential games trajectories of fictitious play need not
converge to a Nash equilibrium, but  the empirical frequencies of
the played strategies converge to a (mixed) Nash equilibrium
\citep{monderer1996fpp,Shamma:2004p1450}.
In our analysis of fictitious play dynamics, we first show that in
near-potential games  if the empirical frequencies are outside some
$\epsilon$-equilibrium set, then the potential of the close
potential game (evaluated at the empirical frequency distribution)
increases with each strategy update. Using this result we establish
convergence of fictitious play dynamics to   a set which can be
characterized in terms of the $\epsilon$-equilibrium set of the game
and the level sets of the potential function of a close potential
game. This result suggests that in near-potential games, the
empirical frequencies of fictitious play converge to a set of mixed
strategies
that (in the close potential game) have potential almost as large as the potential of
Nash equilibria. 
Moreover, exploring the
property that for small $\epsilon$, $\epsilon$-equilibria are
contained in disjoint neighborhoods of equilibria, we strengthen our
result and establish that if a game is sufficiently close to a
potential game, then empirical frequencies of fictitious play
dynamics converge to a small neighborhood of equilibria. This result
recovers as a special case convergence of empirical frequencies to
Nash equilibria in potential games.

A summary of our results on convergence of fictitious play dynamics
is given in Table \ref{table:summary2}.
\begin{table}[H]
\centering
\begin{tabular}{ | m{3cm} |  m{12.7cm} | }
\hline Update Rule                      &  Convergence Result  \\
\hline Fictitious Play                 & (Corollary
\ref{cor:discreteFP1})
 Empirical frequencies of dynamics converge to
the set of mixed strategies with large enough potential: $\{\mathbf{x}\in \prod_m \Delta E^m| \phi(\mathbf{x}) \geq \min_{\mathbf{y}\in {\cal X}_{ M \delta}} \phi(\mathbf{y})  \}$
 \\ \hline
Fictitious Play     & (Theorem \ref{theo:fpNEW})
Assume that  $\cal G$ has finitely many equilibria.
There exists some $\bar{\delta}>0$, and $\bar{\epsilon}>0$ (which are functions of utilities of $\cal G$ but not $\delta$) such that if $\delta<\bar{\delta}$, then
 the empirical frequencies of fictitious play converge to
$$   \left\{\mathbf{x} \left| ~ || \mathbf{x} -\mathbf{x}_k|| \leq \frac{ 4f(M\delta) ML}{\epsilon} +f(M\delta+\epsilon), \mbox{ for some equilibrium $\mathbf{x}_k$}  \right. \right\}, $$
for any $\epsilon$ such that $\bar{\epsilon}\geq \epsilon>0$, where
$f: \mathbb{R}_+ \rightarrow \mathbb{R}_+ $ is an upper semicontinuous function such that $f(x) \rightarrow 0$ as $x \rightarrow 0$.
 \\ \hline
\end{tabular}
 \caption{Convergence properties of fictitious play dynamics in near-potential games.
 We denote the number of players in the game by  $M$,
 set of mixed strategies of player $m$ by $\Delta E^m$,
 and the Lipschitz constant of the mixed extension of $\phi$ by $L$. Rest of the notation is the same as  in Table \ref{table:summary1}.
 }
 \label{table:summary2}
\end{table}

The framework provided in this paper enables us to study the
limiting behavior of adaptive user dynamics in arbitrary finite
strategic form games. In particular, for a given game we can use the proposed convex
optimization formulation to find a nearby potential game and use the
distance between these games to obtain a quantitative
characterization of the
limiting approximate equilibrium
set. The characterization this approach provides will be tighter if the original
game is closer to a potential game.

\emph{Related Literature:}
Potential games play an important role in game-theoretic analysis
because of  existence of  pure strategy Nash equilibrium, and  the
stability (under various learning dynamics such as better/best response dynamics) of pure Nash equilibria
in these games \citep{monderer1996pg,
fudenberg1998tlg,young2004strategic}.
Because of these properties, potential games found applications in
various control and resource allocation problems
\citep{monderer1996pg,marden2009cooperative, Candogan2009Pricing,
arslan2007autonomous}.

There is no systematic framework for analyzing the limiting behavior
of many of the adaptive update rules in general games
\citep{jordan1993three,fudenberg1998tlg,shapley1964some}. However,
for potential games there is  a long line of literature establishing
convergence of natural adaptive dynamics such as better/best response dynamics \citep{monderer1996pg,young2004strategic},
fictitious play
\citep{monderer1996fpp,Shamma:2004p1450,marden2005jsf,hofbauer2002gcs} and logit
response dynamics
\citep{Blume1993387,blume1997population,alos2010logit,marden2008rll}.

It was shown in recent work that a close potential game to a given
game can be obtained by solving a convex optimization problem
(see \citet{candogan2010flows,candogan2010_WeightedProjection}).  It
was also proved  that   equilibria of a given game  can be
characterized by first approximating this game with a potential
game, and then using the equilibrium properties of close potential
games
\citep{candogan2010flows,candogan2010_WeightedProjection}. This
paper builds on this line of work to study dynamics in games by
exploiting their relation to a close potential game.

 \emph{Paper Organization:}
The rest of the paper is organized as follows: We present the game theoretic preliminaries for our work in Section \ref{sec:pre}.
In Section \ref{sec:projection}, we explain how a close potential game to a given game can be found, and discuss possible extensions of this approach.
We present an analysis of better  and best response dynamics in near-potential games in Section \ref{sec:betterBest}.
In Section \ref{sec:logit}, we extend our analysis to logit response, and focus on the stationary distribution and stochastically stable stable states of logit response.
We present the results on fictitious play, and its extensions in Section \ref{sec:fictitiousP}.
 We close in Section \ref{sec:conclusions} with concluding remarks and future work.

\section{Preliminaries} \label{sec:pre}

In this section, we present the game-theoretic background that is
relevant to our work. Additionally, we introduce the closeness
measure for games, which is used in the rest of the paper.
\subsection{Finite Strategic Form Games}
Our focus in this paper is on finite strategic form games.
A (noncooperative) finite game in strategic form  consists of:
\begin{itemize}
\item A finite set of players, denoted by ${\cal M}=\{1, \dots, M\}$.
\item {Strategy spaces:} A finite set of strategies (or actions) $E^m$, for every $m\in {\cal M}$.
\item {Utility functions:} $u^m:\prod_{k\in{\cal M}} E^k\rightarrow {\mathbb R}$,  for every  $m\in {\cal M}$.
\end{itemize}
We denote a (strategic form) game instance by the tuple
$\langle {\cal M},\{E^m\}_{m\in{\cal M}},\{u^m \}_{m\in{\cal M}} \rangle$, and the joint strategy space of this game instance by $E=\prod_{m\in{\cal M}} E^m$.
We refer to a collection of strategies of all players as a \emph{strategy profile} and denote it by $\textbf{p}=({ p}^1,\dots,{  p }^M)\in E$. The collection of strategies of all players but the $m$th one is denoted by $\textbf{p}^{-m}$.

The basic solution concept in a noncooperative game is that of a Nash Equilibrium (NE).
A (pure) Nash equilibrium is a strategy profile
from which no player can unilaterally deviate and improve its payoff. Formally,
${ \bf p}$ is a Nash equilibrium if
\begin{equation*} \label{eq:nash_basic}
u^m({ q }^m,{ \bf p}^{-m}) - u^m({ p }^m,{\bf  p}^{-m}) \leq 0, 
\end{equation*}
{for  every   ${{  q }^m \in E^m} $ and } $m \in \mathcal{M}$.

To address  strategy profiles that are approximately a Nash equilibrium, we use the concept of  $\epsilon$-equilibrium. A  strategy profile ${ \bf p} \triangleq ({ {  p}}^1,\dots,{{  p} }^M)$ is an
$\epsilon$-equilibrium ($\eps \geq 0$)  if
\begin{equation*} \label{eq:epsNash_basic}
u^m({ q }^m,{ \bf p}^{-m}) -  u^m({  p}^m,{ { \bf p}}^{-m}) \leq \epsilon 
\end{equation*}
{for  every   ${{  q }^m \in E^m} $ and } $m \in \mathcal{M}$.
We denote the set of $\epsilon$-equilibria in a game $\cal G$ by ${\cal X}_{\epsilon}$.
Note that a Nash equilibrium is an $\epsilon$-equilibrium with $\epsilon=0$.

\subsection{Potential Games}
We next describe a particular class of games that is central in
this paper, the class of potential games \citep{monderer1996pg}.
\begin{definition}[Potential Game]
\label{def:ExactPot}
A potential game is a noncooperative game for which there exists a
function $\phi:E \rightarrow \mathbb{R}$ satisfying
\begin{equation} \label{eq:condForExact}
u^m({p}^m,{ \bf p}^{-m})-u^m({q}^m,{ \bf p}^{-m})=\phi({p}^m,{ \bf p}^{-m})-  \phi({q}^m,{ \bf p}^{-m}),
\end{equation}
for every $m\in{\cal M}$, ${p}^m, {q}^m \in E^m $, ${ \bf
  p}^{-m} \in E^{-m}$. The function $\phi$ is referred to as a
\emph{potential} function of the game.
\end{definition}
This definition ensures that the change in the utility of a player
who unilaterally deviates to a new strategy, coincides exactly with
the corresponding change in the potential function. Extensions of this definition
in which equation \eqref{eq:condForExact} holds when each  utility
function is multiplied with a (possibly different) positive weight,
 or changes in utility and potential only agree in sign,
give rise to weighted and ordinal potential games that share similar properties to potential games.
We briefly discuss some of these extensions in Section \ref{sec:projection}. However, our main focus in this paper is on potential games in Definition \ref{def:ExactPot}.

Some properties that are specific to potential games are evident from the definition. For instance, it can be seen that unilateral deviations from  a strategy profile that  maximizes the potential function
(weakly) decrease the utility of the deviating player. Hence, this strategy profile corresponds to a Nash equilibrium, and it follows that every potential game has a pure Nash equilibrium.

Another important property of potential games, which will be used
for characterizing the limiting behavior of dynamics in near-potential games,
is that the total unilateral utility improvement  around a ``closed
path" is equal to zero. Before we formally state this result, we
first provide some necessary definitions, which are also used in Section \ref{sec:betterBest}  when we analyze better/best response dynamics in near-potential games.
\begin{definition}[Path  -- Closed Path -- Improvement Path] \label{def:path}
A \emph{path} is a  collection of strategy profiles $\gamma=({{ \bf p}}_0, \dots {{ \bf p}}_N)$ such that ${{ \bf p}}_i $ and ${{ \bf p}}_{i+1}$ differ in the strategy of exactly one player.  A path is a \emph{closed path} (or a \emph{cycle}) if ${{ \bf p}}_0={{ \bf p}}_N$. A path is an \emph{improvement path} if $u^{m_i}({ \bf p}_{i}) \geq u^{m_i}({ \bf p}_{i-1})$ where $m_i$ is the player who modifies its strategy when the strategy profile is updated from  ${ \bf p}_{i-1}$ to ${ \bf p}_i$.
\end{definition}

The transition from strategy profile ${{ \bf p}}_{i-1} $ to ${{ \bf p}}_{i}$ is referred to as \emph{step $i$ of the path}. The length of a path is
 equal to its  number of steps, i.e., the length of the path  $\gamma=({{ \bf p}}_0, \dots, {{ \bf p}}_N)$ is $N$.
We say that a closed path is \emph{simple} if no strategy profile
other than the first and the last strategy profiles is repeated
along the path. For any path $\gamma=({{ \bf p}}_0, \dots, {{ \bf
p}}_N)$ let $I(\gamma)$ represent the total utility improvement
along the path, i.e.,
\begin{equation*}
I(\gamma)=\sum_{i=1}^N u^{m_i}({{ \bf p}}_i)-u^{m_{i}}({{ \bf p}}_{i-1}),
\end{equation*}
where $m_i$ is the index of the player that modifies its strategy in the $i$th step of the path.
The following proposition provides a necessary and sufficient condition under which a given game is a potential game.
\begin{proposition}[\citet{monderer1996pg}]  \label{theo:mondererShapley}
A game  is a potential game if and only if $I(\gamma)=0$ for all simple closed paths $\gamma$. 
\end{proposition}

 We conclude this section by formally defining the  measure of ``closeness'' of  games, used in the subsequent sections.
\begin{definition}[Maximum Pairwise Difference] \label{def:MPD}
Let $\cal G$ and $\hat{\cal G}$ be two games with set of players $\cal M$,   set of strategy profiles $E$, and collections of utility functions   $\{u^m\}_{m\in {\cal M}}$ and $\{\hat{u}^m \}_{m\in {\cal M}}$ respectively. The maximum pairwise difference (MPD) between these games is defined as
\begin{equation*}
d({\cal G}, \hat{\cal G})\bydef
 \max_{ \mathbf{p}\in E, m\in {\cal M},  q^m\in E^m}
\left| \left( u^m(q^m,\textbf{p}^{-m})-u^m(p^m,\textbf{p}^{-m}) \right) - \left( \hat{u}^m(q^m,\textbf{p}^{-m})-\hat{u}^m(p^m,\textbf{p}^{-m}) \right) \right|.
\end{equation*}
\end{definition}
Note that the pairwise difference $ u^m(q^m,\textbf{p}^{-m})-u^m(p^m,\textbf{p}^{-m})$ quantifies how much player $m$ can improve its utility by unilaterally deviating from strategy profile $(p^m,\textbf{p}^{-m})$ to  strategy profile $(q^m,\textbf{p}^{-m})$.
Thus, the MPD captures how different two games are in terms of the utility  improvements due to unilateral deviations.\footnote{
An alternative distance measure can be given by
\begin{equation*}
{d}_2({\cal G}, \hat{\cal G})\bydef
\left(  \sum_{\mathbf{p}\in E} \sum_{m\in {\cal M}, q^m\in E^m}
 \left( \left( u^m(q^m,\textbf{p}^{-m})-u^m(p^m,\textbf{p}^{-m}) \right) - \left( \hat{u}^m(q^m,\textbf{p}^{-m})-\hat{u}^m(p^m,\textbf{p}^{-m}) \right) \right)^2 \right)^\frac{1}{2},
\end{equation*}
and this quantity corresponds to the 2-norm of the difference of $\cal G$ and $\hat{\cal G}$ in terms of the
utility improvements due to unilateral deviations.
Our analysis of the limiting behavior of dynamics  relies on the maximum of
such utility improvement differences
between a game and a near-potential game.
Thus, the  measure in Definition~\ref{def:MPD} provides tighter bounds for our dynamics results, and hence is  preferred in this paper.
}
We refer to pairs of games with small MPD as \emph{close games}, and games that have a small MPD to a potential game as
\emph{near-potential games}.

The MPD measures the closeness of  games in terms of the difference of  unilateral deviations, rather than the difference of their utility functions, i.e.,  quantities of the form
$$\left| \left( u^m(q^m,\textbf{p}^{-m})-u^m(p^m,\textbf{p}^{-m}) \right) - \left( \hat{u}^m(q^m,\textbf{p}^{-m})-\hat{u}^m(p^m,\textbf{p}^{-m}) \right) \right|$$
are used to identify close games, rather than  quantities of the form
$
\left| u^m(p^m,\textbf{p}^{-m}) -\hat{u}^m(p^m,\textbf{p}^{-m})  \right|
$.
This is because the
difference in unilateral deviations provides a better characterization of the strategic similarities  (equilibrium and dynamic properties)  between two games than the difference in utility functions. This can be seen from  the following example: Consider two games with utility functions $\{u^m\}$ and $\{u^m+1\}$, i.e., in the second game players receive an additional payoff of $1$ at all strategy profiles. It can be seen from the definition of Nash equilibrium that despite the difference of their utility functions,  these two games share the same equilibrium set.
Intuitively,  since the additional payoff is obtained at all strategy profiles, it does not affect any of the strategic considerations in the game.
While the utility differences between these games is nonzero, it can be seen that the MPD is equal to zero. Hence MPD identifies a strategic equivalence between these games.
The recent work
\citet{candogan2010flows}
contains a formal treatment of strategic equivalence and its implications for strategic form games.

\section{Finding Near-Potential Games} \label{sec:projection}
In this section, we present a framework for finding the  closest potential game to a given game, where the distance between the games is measured in terms of MPD.
We formulate the problem of identifying such a game as a convex optimization problem, and discuss the extensions of this approach.
We note that a  procedure for finding  near-potential games
can be found   in \citet{candogan2010flows} and \citet{candogan2010_WeightedProjection}.
In these works the distance between games is measured in terms of a 2-norm. In this section we illustrate how similar ideas can be used when the distance is measured in terms of MPD.

It can be seen from Proposition \ref{theo:mondererShapley} that a game is a potential game if and only if it satisfies linear equalities. This suggests that the set of  potential games is convex\footnote{
A game is a weighted potential game if \eqref{eq:condForExact} in Definition \ref{def:ExactPot} is replaced by
\begin{equation*} \label{eq:condForWeighted}
\phi({p}^m,{ \bf p}^{-m})-  \phi({q}^m,{ \bf p}^{-m})=
w^m \left( u^m({p}^m,{ \bf p}^{-m})-u^m({q}^m,{ \bf p}^{-m}) \right),
\end{equation*}
for some positive player-specific weights $\{w^m\}$. If instead of holding with equality, the left and right hand sides of \eqref{eq:condForExact} only agree in sign, then the game is referred to as an ordinal potential game \citep{monderer1996pg}.
Despite the fact that weighted and ordinal potential games have similar desirable properties to potential games,
their sets are nonconvex, and finding the closest weighted/ordinal potential game to a given game requires solving a nonconvex optimization problem  \citep{candogan2010_WeightedProjection}.}, i.e.,
if ${\cal G}=\langle {\cal M}, E, \{u^m\}_m \rangle$ and $\hat{\cal G}=\langle {\cal M}, E, \{\hat{u}^m\}_m \rangle$  are potential games, then ${\cal G}_\alpha=\langle {\cal M}, E, \{ \alpha{u}^m +(1-\alpha) \hat{u}^m \}_m \rangle$, is also a potential game provided that $\alpha \in [0,1]$.

Assume that a game with utility functions $\{u^m\}_m$ is given. The closest potential game (in terms of MPD) to this game, with payoff functions $\{\hat{u}^m\}_m$, and potential function $\phi$ can be obtained by solving the following optimization problem:
\begin{equation*}
(\mbox{P}:) \quad
\begin{aligned}
     \min_{\phi, \{\hat{u}^m\}_m} \quad &
     \max_{\mathbf{p}\in E, m\in {\cal M},   q^m \in E^m} \left| \left(u^m(q^m,\mathbf{p}^{-m})-u^m(p^m,\mathbf{p}^{-m})\right) \right. \\
          & \qquad \qquad\qquad\qquad \qquad \left. -\left(\hat{u}^m(q^m,\mathbf{p}^{-m})-\hat{u}^m(p^m,\mathbf{p}^{-m})\right) \right| \\
s.t. \quad &  \phi(\bar{q}^m, \bar{\mathbf{p}}^{-m})- \phi(\bar{p}^m, \bar{\mathbf{p}}^{-m}) = \hat{u}^m(\bar{q}^m, \bar{\mathbf{p}}^{-m})- \hat{u}^m(\bar{p}^m, \bar{\mathbf{p}}^{-m}), \\
&\mbox{for all $m\in {\cal M}, ~\bar{\mathbf{p}}  \in E,~ \bar{q}^m \in E^m$.}
\end{aligned}
\end{equation*}
Note that the difference $\left(u^m(q^m,\mathbf{p}^{-m})-u^m(p^m,\mathbf{p}^{-m})\right)-\left(\hat{u}^m(q^m,\mathbf{p}^{-m})-\hat{u}^m(p^m,\mathbf{p}^{-m})\right)$ is linear in  $\{\hat{u}^m\}_m$. Thus, the  objective function is the maximum of such linear functions, and hence is convex  in $\{\hat{u}^m\}_m$. The constraints of this optimization problem  guarantee that the game with payoff functions $\{\hat{u}^m\}_m$ is a potential game with potential $\phi$. Note that these constrains are linear. Therefore, it follows that (P) is a convex optimization problem that gives the closest potential game to a given game.

Let ${\cal G}_1$ and ${\cal G}_2$ be games with utility functions $\{u^m\}_m$ and   $\{w^m u^m\}_m$ respectively, where for all $m\in {\cal M}$, $w^m\geq 1$ is a fixed weight.
It can be seen that preferences of players are identical in these two games, i.e., $u^m(x^m, x^{-m})-u^m(y^m, x^{-m}) > 0$ if and only if $w^m  u^m(x^m, x^{-m})-w^m u^m(y^m, x^{-m}) > 0 $
for any $m\in {\cal M}$, $y^m\in \Delta E^m$ and $\mathbf{x} \in \prod_{k\in {\cal M}} \Delta E^k$. Thus, it follows
 that
the equilibrium sets of these game are the same, and
   for many of the update rules in games\footnote{For formal definitions of these update rules see Sections \ref{sec:betterBest}, \ref{sec:logit}, \ref{sec:fictitiousP}.}, such as better/best response dynamics, and fictitious play (but not logit response), the trajectories of  dynamics in ${\cal G}_1$ and ${\cal G}_2$ are identical.

This observation suggests that it
may also be of interest  to find a close potential game to a ``scaled version'' of a given game.
The following optimization formulation obtains such a potential game:
\begin{equation*}
(\mbox{P2}:) \quad
\begin{aligned}
     \min_{\phi, \{\hat{u}^m\}_m, \{w^m\}_m} \quad &
     \max_{\mathbf{p}\in E, m\in {\cal M},   q^m \in E^m} \left| w^m\left(u^m(q^m,\mathbf{p}^{-m})-u^m(p^m,\mathbf{p}^{-m})\right) \right. \\
     & \qquad \qquad\qquad\qquad \qquad \left. -\left(\hat{u}^m(q^m,\mathbf{p}^{-m})-\hat{u}^m(p^m,\mathbf{p}^{-m})\right) \right| \\
s.t. \quad & w^m \geq 1 \quad \mbox{for all $m\in {\cal M}$,}\\
\quad &  \phi(\bar{q}^m, \bar{\mathbf{p}}^{-m})- \phi(\bar{p}^m, \bar{\mathbf{p}}^{-m}) = \hat{u}^m(\bar{q}^m, \bar{\mathbf{p}}^{-m})- \hat{u}^m(\bar{p}^m, \bar{\mathbf{p}}^{-m}), \\
&\mbox{for all $m\in {\cal M}, ~\bar{\mathbf{p}}  \in E,~ \bar{q}^m \in E^m$.}
\end{aligned}
\end{equation*}
The solution of (P2)  is a potential game with utility functions $\{\hat{u}^m\}_m$. Comparing (P) and (P2) it can be seen that  (P2) obtains the closest potential game\footnote{Note that the solution of (P2) is not the closest weighted potential game to the original game. Such a game can be obtained by replacing the objective function by
$$ \max_{\mathbf{p}\in E, m\in {\cal M},   q^m \in E^m} \left| \left(u^m(q^m,\mathbf{p}^{-m})- u^m(p^m,\mathbf{p}^{-m})\right)  -w^m\left(\hat{u}^m(q^m,\mathbf{p}^{-m})-\hat{u}^m(p^m,\mathbf{p}^{-m})\right) \right|.$$
However, this objective function leads to a nonconvex optimization
formulation due to the multiplication of the terms $w^m$ and
$\hat{u}^m$, and the solution of this  problem is different than
that of (P2). See  \citet{candogan2010_WeightedProjection} for
details. } (in terms of  MPD) to the game with utility functions
$\{w^m {u}^m\}_m$. Since (P2) also minimizes the objective function
over $\{w^m\}$, the solution also reveals the ``scaling'' of the
original game, which makes it as close as possible to a potential
game.

In the rest of the paper, we do not discuss how a close potential
game to a given game is obtained, but we just assume that a close
potential game  with potential $\phi$ is known and the MPD between
this game and the original game is $\delta$. We provide
characterization results on limiting dynamics for a given game in
terms of $\phi$ and $\delta$.

\section{Better Response and Best Response Dynamics} \label{sec:betterBest}
In this section, we consider better  and best response dynamics,  and
study convergence properties of these update rules in near-potential
games. All of the update rules considered in this section are
discrete-time update rules, i.e., players are allowed to update
their strategies at time instants $t\in \mathbb{Z}_+=\{1,2,\dots
\}$.

Best response dynamics is an update rule where at each time instant a player
chooses its best response to other players' current strategy profile.
In  better response dynamics, on the other hand, players choose strategies that improve their payoffs, but these strategies need not be their best responses.
Formal descriptions of these update rules are given below.
\begin{definition}[Better  and Best Response Dynamics] \label{algBR}
At each time instant $t\in \{1,2,\dots \}$, a single player is chosen at random   for updating its strategy,  using a probability distribution with full support over the set of players. Let $m$ be the player chosen at some time $t$, and let $\textbf{r}\in E$ denote the strategy profile that is used at time $t-1$.
\begin{enumerate}
  \item Better response dynamics is the update process where player $m$ does not modify its strategy if
   $u^m(\mathbf{r})= \max_{q^m} u^m(q^m,\mathbf{r}^{-m})$, and otherwise it updates its strategy to a strategy in  $\{q^m| u^m(q^m, \mathbf{r}^{-m})>u^m(\mathbf{r})\}$, chosen uniformly at random.

  \item Best response dynamics is the update process where player $m$ does not modify its strategy if
   $u^m(\mathbf{r})= \max_{q^m} u^m(q^m,\mathbf{r}^{-m})$, and otherwise it updates its strategy to a strategy in  $\arg\max_{q^m} u^m(q^m,\mathbf{r}^{-m})$,  chosen uniformly at random.

\end{enumerate}
\end{definition}
For simplicity of the analysis, we
 assume here that   users are chosen randomly to update their strategy. However, this assumption is not crucial for our results, and can be relaxed.

 We refer  to  strategies in $\arg\max_{q^m} u^m(q^m,\textbf{r}^{-m})$ as   \emph{best responses of player $m$ to $\textbf{r}^{-m}$}.
We denote the strategy profile used at time $t$ by $\mathbf{p}_t$, and we define the \emph{trajectory of the dynamics} as the sequence of strategy profiles $\{\mathbf{p}_t\}_{t=0}^\infty$. In our analysis, we assume that the trajectory is initialized at a strategy profile ${\mathbf{p}_0}\in E$ at time $0$ and it evolves according to one of the update rules described above.

The following theorem establishes that in finite games, better and best response dynamics converge to a set of $\eps$-equilibria, where the size of this set is characterized by the MPD to a close potential game.
\begin{theorem} \label{theo:BRconv}
Consider a game $\cal G$ and let $\hat{\cal G}$ be a nearby potential game such that  $d({\cal G}, \hat{\cal G})\leq \delta$.
  Assume that best response or better response dynamics are used in $\cal G$, and denote
the  number of strategy profiles in these games by $|E|=h$.

For both update processes,   the trajectories are  contained in the $\delta h$-equilibrium set of $\cal G$ after finite time with probability $1$, i.e.,
let $T$ be a random variable such that $\mathbf{p}_t\in {\cal X}_{\delta h}$,  for all $t>T$,
then $P(T<\infty)=1$.
\end{theorem}
\begin{proof}
We prove the claim by modeling the update process using a Markov chain, and   employing the improvement path condition for potential games (cf.~Proposition \ref{theo:mondererShapley}).

Using Definition \ref{algBR}, we can represent the strategy updates
in best response dynamics as  the state transitions in the following
Markov chain: (i) Each state corresponds to a strategy profile and,
(ii) there is a nonzero transition probability from state
$\mathbf{r}$ to state $\mathbf{q}\neq \mathbf{r}$, if $\mathbf{r}$
and $\mathbf{q}$ differ in the strategy of a single player, say $m$,
and  $q^m$ is a (strict) best response of player $m$ to
$\mathbf{r}^{-m}$. The probability of transition from state
$\mathbf{r}$ to state $\mathbf{q}$ is equal to the probability that
at strategy profile $\mathbf{r}$, player $m$ is chosen for update
and it chooses $q^m$ as its new strategy. In the case of
better response dynamics we allow $q^m$ to be any   strategy
strictly improving payoff of player $m$, and a similar Markov chain
representation still holds.

Since there are finitely many states, one of the recurrent classes of the Markov chain is reached in finite time (with probability 1).
Thus, to prove the claim, it is sufficient to show that any state which  belongs to some recurrent class of this Markov chain is contained in the $\epsilon$-equilibrium set of~$\cal G$.

It follows from Definition \ref{algBR} that
a recurrence class is a  singleton, only if none of the players  can strictly improve its payoff by unilaterally deviating from the corresponding strategy profile.
Thus, such a
strategy profile is a  Nash equilibrium of $\cal G$ and  is contained in the $\epsilon$-equilibrium set.

Consider  a recurrence class that is not a singleton. Let $\mathbf{r}$ be a strategy profile in this recurrence class. Since the recurrence class is not a singleton, there exists some player $m$, who can unilaterally deviate from $\mathbf{r}$ by following its best response to another strategy profile $\mathbf{q}$, and increase its payoff by some $\alpha>0$.
Since such a transition occurs with nonzero probability, $\mathbf{r}$ and $\mathbf{q}$ are in the same recurrence class, and the process when started from $\mathbf{r}$ visits $\mathbf{q}$ and returns to $\mathbf{r}$ in finitely many updates.
Since each  transition corresponds to a unilateral deviation that strictly improves the payoff of the deviating player, this constitutes a simple closed improvement path  containing $\mathbf{r}$ and $\mathbf{q}$.
Let $\gamma=({{  \mathbf{p}}}_0, \dots, {{ \mathbf{p}}}_N)$ be such an improvement path and  $\mathbf{p}_0=\mathbf{p}_N =\mathbf{r}$, $\mathbf{p}_1=\mathbf{q}$ and  $N \leq |E|=h$.
Since $u^m(\mathbf{q})-u^m(\mathbf{r}) = \alpha $, and $u^{m_i}({{ \mathbf p}}_i)-u^{m_{i}}({{ \bf p}}_{i-1})\geq 0$ at every step $i$ of the path,
this closed improvement path satisfies
\begin{equation} \label{eq:IwNewgamma}
\sum_{i=1}^N \left(u^{m_i}({{ \bf p}}_i)-u^{m_{i}}({{ \bf p}}_{i-1})\right) \geq  \alpha.
\end{equation}
On the other hand it follows by Proposition \ref{theo:mondererShapley} that  the close potential game satisfies
\begin{equation} \label{eq:IwNewgamma2}
\sum_{i=1}^N \left(\hat{u}^{m_i}({{ \bf p}}_i)-\hat{u}^{m_{i}}({{ \bf p}}_{i-1})\right) =0.
\end{equation}
Combining  \eqref{eq:IwNewgamma} and \eqref{eq:IwNewgamma2} we conclude that
\begin{equation*}
\begin{aligned}
\alpha &\leq \sum_{i=1}^N    \left(u^{m_i}({{ \bf p}}_i)-u^{m_{i}}({{ \bf p}}_{i-1})\right)-\left(\hat{u}^{m_i}({{ \bf p}}_i)-\hat{u}^{m_{i}}({{ \bf p}}_{i-1})\right) \\
& \leq  N \delta.
\end{aligned}
\end{equation*}
Since $N\leq |E|=h$, it follows that $\alpha \leq   \delta h$. The
claim then immediately follows since $\mathbf{r}$ and  the
recurrence class were chosen arbitrarily, and
our analysis shows that the payoff improvement of  player $m$
(chosen  for strategy update using
a probability distribution with full support as described in
Definition \ref{algBR}), due to its best response is bounded by $ \delta h$.
\end{proof}

As can be seen from the proof of this theorem, extending dynamical
properties of potential games to nearby games relies on special
structural properties of potential games. As a corollary of the
above theorem, we obtain that trajectories generated by better and
best response dynamics converge to a Nash equilibrium in potential
games, since if $\cal G$ is a potential game, the close potential
game $\hat{\cal G}$ can be chosen such that $d({\cal G},\hat{\cal
G})=0$.

\section{Logit Response Dynamics} \label{sec:logit}

In this section we focus on logit response dynamics. Logit response
dynamics can be viewed as a smoothened version of the best response
dynamics, in which a smoothing parameter determines the frequency
with which the best response strategy is picked. The evolution of
the pure strategy profiles can be represented in terms of a Markov
chain (with state space given by the set of pure strategy profiles).
We characterize the stationary distribution and stochastically
stable states of this Markov chain (or of the update rule)  in
near-potential games. Our approach involves identifying a close
potential game to a given game, and exploiting features of the
corresponding potential function to characterize the limiting
behavior of logit response dynamics in the original game.

In Section \ref{subsec:logProperties}, we provide a formal
definition of logit response dynamics and review some of its
properties. We also present some of the mathematical tools used in
the literature to study this update rule. In Section
\ref{subsec:logStationaryDist}, we show that  the  stationary
distribution of logit response dynamics in  a near-potential game
can be approximately characterized using the potential function of a
nearby potential game.
We also use this result to
show that the stochastically
stable strategy profiles are contained in approximate equilibrium sets in
near-potential games.

\subsection{Properties of Logit Response} \label{subsec:logProperties}

We start by providing a formal definition of logit response dynamics:
\begin{definition} \label{def:logit}
At each time instant $t\in \{1,2,\dots \}$, a single player is chosen at random   for updating its strategy,  using a probability distribution with full support over the set of players. Let $m$ be the player chosen at some time $t$, and let $\mathbf{r}\in E$ denote the strategy profile that is used at time $t-1$.

\emph{Logit response dynamics with parameter $\tau$} is the update process, where   player $m$ chooses a strategy $q^m\in E^m$ with probability
\begin{equation*}
P_\tau^m(q^m | \mathbf{r})=\frac{e^{\frac{1}{\tau} u^m(q^m, \mathbf{r}^{-m})}}{\sum_{p^m\in E^m} e^{\frac{1}{\tau} u^m(p^m, \mathbf{r}^{-m})}}.
\end{equation*}
\end{definition}
In this definition, $\tau>0$ is a fixed parameter that determines how often players choose their best responses.
The probability of not choosing a best response decreases as $\tau$ decreases, and as
$\tau \rightarrow 0$, players choose their best responses with probability $1$.
This feature suggests that  logit response dynamics  can be  viewed as a generalization of best response dynamics, where with small but nonzero probability players use a strategy that is not a best response.

For a given  $\tau>0$, this update process can be represented by a
finite aperiodic  and irreducible Markov chain
\citep{marden2008rll,alos2010logit}.
 The states of the Markov chain correspond to the strategy profiles in the game.
 Denoting the probability that player $m$ is chosen for a strategy update by $\alpha_m$,
 transition probability from strategy profile $\mathbf{p}$ to $\mathbf{q}$ can be given by (assuming $\mathbf{p}\neq \mathbf{q}$, and denoting the transition from $\mathbf{p}$ to $\mathbf{q}$ by $\mathbf{p}\rightarrow \mathbf{q}$):
\begin{equation} \label{eq:transitionProb}
 P_\tau(\mathbf{p}\rightarrow \mathbf{q}) = \left\{
 \begin{aligned}
\alpha_m P_\tau^m(q^m| \mathbf{p})  \qquad \qquad & \mbox{if $\mathbf{q}^{-m}=\mathbf{p}^{-m}$ for some $m\in {\cal M}$} \\
0\qquad \qquad & \mbox{otherwise.}
 \end{aligned}
 \right.
 \end{equation}
The chain is aperiodic and irreducible since a player updating its strategy can choose any strategy (including the current one) with positive probability.
Consequently, it has a unique stationary distribution.

We denote the stationary distribution of this Markov chain by
$\mu_\tau$ and refer to it as the stationary distribution of the
logit response dynamics. A strategy profile $\mathbf{q}$ such that
$\lim_{\tau\rightarrow 0} \mu_\tau(\mathbf{q}) >0 $ is referred to
as a \emph{stochastically stable strategy profile} of the logit
response dynamics. Intuitively, these strategy profiles are the ones
that are used with nonzero probability, as players adopt their
best responses more and more frequently in their strategy updates.

In potential games, the stationary distribution of the logit response dynamics can be written as an explicit function of the potential.
If $\cal G$ is a potential game with potential function $\phi$, the stationary distribution of the logit response dynamics is given by the distribution
\citep{marden2008rll,alos2010logit,blume1997population}:\footnote{Note that this expression is independent of  $\{\alpha_m\}$, i.e., the probability distribution that is used to choose which player updates its strategy has no effect on the stationary distribution of  logit response.}
\begin{equation} \label{eq:statOfLogit}
\mu_\tau({\mathbf{q}})= \frac{e^{\frac{1}{\tau}\phi(\mathbf{q})}}{\sum_{\mathbf{p}\in E}e^{\frac{1}{\tau}\phi(\mathbf{p})}}.
\end{equation}

It can be seen from \eqref{eq:statOfLogit} that
$\lim_{\tau \rightarrow 0} \mu_\tau({\mathbf{q}})>0$ if and only if $\mathbf{q}\in \arg\max_{\mathbf{p}\in E} \phi(\mathbf{p})$. Thus, in potential games the stochastically stable strategy profiles are those that maximize the potential function.

We next describe a method for obtaining the stationary distribution
of Markov chains. This method will be used in the next subsection in
characterizing the stationary distribution of logit response.
Assume that an irreducible Markov chain over a finite set of states
$S$, with transition probability matrix $P$ is given. Consider a
directed tree, $T$, with nodes given by the states of the Markov
chain, and assume that  an edge from node $\mathbf{q}$ to node
$\mathbf{p}$ can exist only if there is a nonzero transition
probability from $\mathbf{q}$ to $\mathbf{p}$ in the Markov chain.
We say that the tree is rooted at state $\mathbf{p}$, if  from every
state $\mathbf{q}\neq \mathbf{p}$ there  exists a unique directed
path along the tree to $\mathbf{p}$. For each state $\mathbf{p}\in
S$, denote by ${\cal T}(\mathbf{p})$ the set of all trees rooted at
$\mathbf{p}$, and define a weight $w_\mathbf{p} \geq 0$ such that
\begin{equation} \label{eq:treeWeight}
w_\mathbf{p}=\sum_{T\in {\cal T}(\mathbf{p})} \prod_{(\mathbf{q} \rightarrow \mathbf{r})\in T} P(\mathbf{q} \rightarrow \mathbf{r}).
\end{equation}
The following proposition from the Markov Chain literature
(\citet{leighton1983markov,anantharam1989proof,freidlin1998random}), known as the  Markov
chain tree theorem, expresses the stationary distribution of  Markov
chains in terms of these weights.
\begin{proposition}\label{prop:stationaryDistTree}
The stationary distribution of the Markov chain defined over set $S$ is given by
$\mu(\mathbf{p})= \frac{w_\mathbf{p}}{\sum_{\mathbf{q} \in S} w_\mathbf{q}}.$
\end{proposition}
For any $T \in {\cal T}(\mathbf{p})$, intuitively, the quantity  $\prod_{(\mathbf{q} \rightarrow \mathbf{r})\in T} P(\mathbf{q} \rightarrow \mathbf{r})$ gives a measure of likelihood of the event that   node $\mathbf{p}$ is  reached when the chain is initiated from the leaves (i.e., nodes with indegree equal to $0$) of $T$. Thus, $w_{\mathbf{p}}$ captures how likely it is that node $\mathbf{p}$ is visited in this chain, and the normalization in Proposition \ref{prop:stationaryDistTree} gives the stationary distribution.
Since for finite games logit response dynamics can be modeled as an irreducible
Markov chain,  this result can be used to
characterize its stationary distribution.

\subsection{Stationary Distribution of Logit Response Dynamics} \label{subsec:logStationaryDist}

In this section we show that the  stationary distribution of logit response dynamics in near-potential games can be approximated by exploiting the potential function of a close potential game.
We start by showing that in games with small MPD logit response dynamics have similar transition probabilities.
\begin{lemma} \label{lem:transProp}
Consider a game $\cal G$ and let $\hat{\cal G}$ be a nearby potential game such that  $d({\cal G}, \hat{\cal G})\leq \delta$.
Denote the transition probability matrices of logit response dynamics in $\cal G$ and $\hat{\cal G}$ by $P_\tau$ and $\hat{P}_\tau$ respectively.
For all strategy profiles $\mathbf{p}$ and $\mathbf{q}$ that differ in the strategy of at most one player, we have
\begin{equation*}
e^{-\frac{2\delta}{\tau}} \leq \hat{P}_\tau(\mathbf{p}\rightarrow \mathbf{q}) / {P}_\tau(\mathbf{p}\rightarrow \mathbf{q}) \leq e^{\frac{2\delta}{\tau}}.
\end{equation*}
\end{lemma}
\begin{proof}
Assume that $\mathbf{p}^{-m}=\mathbf{q}^{-m}$. In $\cal G$ the transition probability $P_\tau(\mathbf{p}\rightarrow \mathbf{q})$ can be expressed by (see
\eqref{eq:transitionProb}):
 \begin{equation*} \label{eq:transitionProb2}
  P_\tau(\mathbf{p}\rightarrow \mathbf{q}) = \left\{
  \begin{aligned}
 \alpha_m P_\tau^m(q^m| \mathbf{p})  \qquad \qquad & \mbox{if ${q}^{m}\neq{p}^{m}$ } \\
  \sum_{k\in {\cal M}}\alpha_k P_\tau^k(p^k| \mathbf{p}) \qquad \qquad & \mbox{otherwise.}
  \end{aligned}
  \right.
  \end{equation*}
A  similar expression holds for the transition probability $\hat{P}_\tau(\mathbf{p}\rightarrow \mathbf{q})$ in $\hat{\cal G}$, replacing $P_\tau^m$ by $\hat{P}^m_\tau$.
  Thus,  it is sufficient prove $ e^{-\frac{2\delta}{\tau}} \leq \hat{P}_\tau^m(q^m| \mathbf{p})/{P}_\tau^m(q^m| \mathbf{p}) \leq e^{\frac{2\delta}{\tau}} $  for all  $\mathbf{p}$, $m$, $q^m$ to prove the claim.

 Observe that by the definition of MPD
  \begin{equation} \label{eq:MODBounds}
  \begin{aligned}
   {u}^m(r^m, \mathbf{p}^{-m})-{u}^m(p^m, \mathbf{p}^{-m})-\delta  &\leq       \hat{u}^m(r^m, \mathbf{p}^{-m})-\hat{u}^m(p^m, \mathbf{p}^{-m}) \\
   &\leq {u}^m(r^m, \mathbf{p}^{-m})-{u}^m(p^m, \mathbf{p}^{-m})+\delta.
  \end{aligned}
  \end{equation}
Definition \ref{def:logit} suggests that $\hat{P}_\tau^m(q^m | \mathbf{p})$ can be written as (by dividing the numerator and the denominator by $e^{\frac{1}{\tau}  \hat{u}^m(p^m, \mathbf{p}^{-m})}$):
\begin{equation*}
\hat{P}_\tau^m(q^m | \mathbf{p})=\frac{e^{\frac{1}{\tau} (\hat{u}^m(q^m, \mathbf{p}^{-m})-\hat{u}^m(p^m, \mathbf{p}^{-m}))}}{\sum_{r^m\in E^m} e^{\frac{1}{\tau} (\hat{u}^m(r^m, \mathbf{p}^{-m})-\hat{u}^m(p^m, \mathbf{p}^{-m}))}}.
\end{equation*}
Therefore, using the bounds in \eqref{eq:MODBounds} it follows that
\begin{equation*} \label{eq:lowerBound}
\hat{P}_\tau^m(q^m | \mathbf{p}) \leq \frac{
\kappa (q^m) e^{\frac{\delta}{\tau}}}{\kappa (q^m) e^{\frac{\delta}{\tau} } + \sum_{r^m \neq q^m} \kappa (r^m)e^{\frac{-\delta }{\tau} } } .
\end{equation*}
where, $\kappa(r^m)=e^{\frac{1}{\tau} ({u}^m(r^m, \mathbf{p}^{-m})-{u}^m(p^m, \mathbf{p}^{-m}))}$ for all $r^m\in E^m$.
Dividing both the numerator and the denominator of the right hand side by
$\sum_{r^m\in E^m}\kappa(r^m)$ and observing that
${P}_\tau^m(q^m | \mathbf{p})= \frac{\kappa(q^m)}{\sum_{r^m\in E^m} \kappa(r^m)}$,
we obtain
\begin{equation*}
\hat{P}_\tau^m(q^m | \mathbf{p})
 \leq \frac{e^\frac{\delta}{\tau} {P}_\tau^m(q^m | \mathbf{p}) }{e^\frac{\delta}{\tau} {P}_\tau^m(q^m | \mathbf{p}) + e^{-\frac{\delta}{\tau}} \left({1}-{P}_\tau^m(q^m | \mathbf{p})  \right) },
\end{equation*}
or equivalently
\begin{equation*}
 \frac{ \hat{P}_\tau^m(q^m | \mathbf{p})}{{P}_\tau^m(q^m | \mathbf{p})} \leq \frac{e^\frac{\delta}{\tau}  }{e^\frac{\delta}{\tau} {P}_\tau^m(q^m | \mathbf{p}) + e^{-\frac{\delta}{\tau}} \left({1}-{P}_\tau^m(q^m | \mathbf{p})  \right) }.
\end{equation*}
It can be seen that the right hand side is decreasing in ${P}_\tau^m(q^m | \mathbf{p})$. Thus replacing ${P}_\tau^m(q^m | \mathbf{p})$ by $0$, the right hand side can be upper  bounded by $e^{\frac{2\delta}{\tau}}$. Then we obtain
$
 \hat{P}_\tau^m(q^m | \mathbf{p})/{{P}_\tau^m(q^m | \mathbf{p})}
\leq e^{\frac{2\delta}{\tau}}$. By symmetry we also conclude that $ {P}_\tau^m(q^m | \mathbf{p})/{\hat{P}_\tau^m(q^m | \mathbf{p})}
\leq e^{\frac{2\delta}{\tau}}$, and combining these bounds the claim follows.
\end{proof}

Definition \ref{def:logit} suggests that perturbation of utility functions changes the transition probabilities multiplicatively in logit response. The above lemma supports this intuition: if utility gains due to unilateral deviations are modified by  $\delta$, the ratio of the transition probabilities can change at most by $e^{\frac{2\delta}{\tau}}$.
Thus, if two games are close, then the transition probabilities  of logit response in these games should   be closely related.

This suggests using results from perturbation theory of Markov
chains to characterize the stationary distribution of logit response
in a near-potential game
\citep{haviv1984perturbation,cho2001comparison}. However, standard
perturbation results characterize changes in the stationary distribution
of a Markov chain when the transition probabilities are {\it
additively perturbed}. These results, when applied to multiplicative
perturbations, yield bounds which are uninformative. We therefore
first present a result which characterizes deviations from the
stationary distribution of a Markov chain when its transition
probabilities are multiplicatively perturbed, and therefore may be
of independent interest.\footnote{A multiplicative perturbation
bound similar to ours, can be found in \citet{freidlin1998random}.
However, this bound is looser than the one we obtain and   it does not provide a good
characterization of the stationary distribution in our setting. We
provide a tighter bound, and obtain stronger predictions on the
stationary distribution of logit response.}

\begin{theorem} \label{theo:statDist}
Let $P$ and $\hat{P}$ denote the probability transition matrices of two finite irreducible Markov chains with the same state space.
Denote the stationary distributions of these Markov chains by $\mu$ and $\hat{\mu}$ respectively, and let the cardinality of the state space be $h$. Assume that
$\alpha\geq 1$ is a given constant
 and
for any two states $\mathbf{p}$ and $\mathbf{q}$,  the following inequalities hold
\begin{equation*}
\begin{aligned}
\alpha^{-1} {P}(\mathbf{p} \rightarrow \mathbf{q}) &\leq
\hat{P}(\mathbf{p} \rightarrow \mathbf{q}) \leq \alpha {P}(\mathbf{p} \rightarrow \mathbf{q}).
\end{aligned}
\end{equation*}
Then,  for any state $\mathbf{p}$, we have
\begin{equation*}
\begin{aligned}
(i) \qquad &\frac{\alpha^{-(h-1)} \mu(\mathbf{p}) }{\alpha^{-(h-1)} \mu(\mathbf{p}) + \alpha^{h-1} (1-\mu(\mathbf{p}))}
\leq \hat{\mu}(\mathbf{p})  \leq \frac{\alpha^{h-1} \mu(\mathbf{p})}{\alpha^{h-1} \mu(\mathbf{p}) + \alpha^{-(h-1)} (1-\mu(\mathbf{p}))}, \\
(ii)  \qquad &|\mu({\mathbf{p}})-\hat{\mu}({\mathbf{p}})| \leq \frac{\alpha^{h-1}-1}{\alpha^{h-1}+1}.
\end{aligned}
\end{equation*}

\end{theorem}
\begin{proof}
As before, let ${\cal T}(\mathbf{p})$ denote the set of directed trees that are rooted at state $\mathbf{p}$. Using the characterization of the stationary distribution  in Proposition \ref{prop:stationaryDistTree}, for the Markov chain with probability transition  matrix $P$, we have
$\mu(\mathbf{p})= \frac{w_\mathbf{p}}{\sum_\mathbf{q} w_\mathbf{q}},$
where
for each state $\mathbf{p}$,
\begin{equation*}
w_\mathbf{p}=\sum_{T\in {\cal T} (\mathbf{p})} \prod_{(\mathbf{x}\rightarrow \mathbf{y})\in T} P(\mathbf{x}\rightarrow \mathbf{y}).
\end{equation*}
For the Markov chain with probability transition matrix $\hat{P}$, we define  $\hat{w}_\mathbf{p}$, by replacing $P$ in the above equation with $\hat{P}$ and $\hat{\mu}(\mathbf{p})$  similarly satisfies $\hat{\mu}(\mathbf{p})= \frac{\hat{w}_\mathbf{p}}{\sum_\mathbf{q} \hat{w}_\mathbf{q}}$.

Since the Markov chain has $h$ states, $|T| = h-1$ for all $T\in {\cal T}(\mathbf{p})$. Hence, it follows from the assumption of the theorem  and the above definitions of $w_\mathbf{p}$ and $\hat{w}_\mathbf{p}$ that
\begin{equation*}
\begin{aligned}
\alpha^{-(h-1)} w_\mathbf{p} &=\alpha^{-(h-1)} \sum_{T\in {\cal T}(\mathbf{p})} \prod_{(\mathbf{x} \rightarrow \mathbf{y})\in T} {P}(\mathbf{x}\rightarrow \mathbf{y}) \\
&\leq \hat{w}_\mathbf{p}=\sum_{T\in {\cal T}(\mathbf{p})} \prod_{(\mathbf{x}\rightarrow \mathbf{y})\in T} \hat{P}(\mathbf{x}\rightarrow \mathbf{y}) \\
 &\leq
\alpha^{h-1} \sum_{T\in {\cal T}(\mathbf{p})} \prod_{(\mathbf{x}\rightarrow \mathbf{y})\in T} {P}(\mathbf{x}\rightarrow \mathbf{y}) = \alpha^{h-1} {w}_\mathbf{p}.
\end{aligned}
\end{equation*}
This inequality implies that for all $\mathbf{q}$, $\hat{w}_\mathbf{q}$ is upper bounded by $\alpha^{h-1} w_\mathbf{q}$ and lower bounded by $\alpha^{-(h-1)}{w}_\mathbf{q}$.
Using this observation together with the identity  $\hat{\mu}(\mathbf{p})=\frac{\hat{w}_\mathbf{p}}{\sum_\mathbf{q} \hat{w}_\mathbf{q}}$, we obtain
\begin{equation*}
\begin{aligned}
\frac{\alpha^{-(h-1)} {w}_\mathbf{p}}{\alpha^{-(h-1)} {w}_\mathbf{p} + \alpha^{h-1} \sum_{\mathbf{q}\neq \mathbf{p}} {w}_\mathbf{q}}
\leq \hat{\mu}(\mathbf{p})= \frac{\hat{w}_\mathbf{p}}{\sum_\mathbf{q} \hat{w}_\mathbf{q}}
\leq \frac{\alpha^{h-1} {w}_\mathbf{p}}{\alpha^{h-1} {w}_\mathbf{p} + \alpha^{-(h-1)} \sum_{\mathbf{q}\neq \mathbf{p}} {w}_\mathbf{q}}.
\end{aligned}
\end{equation*}
Dividing the numerators and denominators of the left and right hand sides of the inequality by $\sum_\mathbf{q} w_\mathbf{q}$, using Proposition \ref{prop:stationaryDistTree}, and observing that $\sum_{\mathbf{q}\neq \mathbf{p}} \mu(\mathbf{q})=1-\mu(\mathbf{p})$ the first part of the theorem follows.

Consider functions $f$ and $g$ defined on $[0,1]$ such that $f(x)=\frac{\alpha^{h-1} x}{\alpha^{h-1} x + \alpha^{-(h-1)} (1-x)} -x$ and $g(x)=\frac{\alpha^{-(h-1)} x}{\alpha^{-(h-1)} x + \alpha^{h-1} (1-x)} -x$ for $x\in [0,1]$.
Checking the first order optimality conditions, it can be seen that $f(x)$ is maximized at $x=\frac{\alpha^{-(h-1)}}{1+\alpha^{-(h-1)}}$,
and the maximum equals to $\frac{\alpha^{h-1}-1}{\alpha^{h-1}+1}$.
Similarly,  the minimum of $g(x)$ is achieved at $x=\frac{\alpha^{h-1}}{1+\alpha^{h-1}}$
and  is equal to
$\frac{1-\alpha^{h-1}}{1+\alpha^{h-1}}$. Combining these observations with part (i), we obtain
\begin{equation*} \label{eq:boundStDiff}
\begin{aligned}
\frac{1-\alpha^{h-1}}{1+\alpha^{h-1}}
&\leq g(\mu (\mathbf{p}))
=
\frac{\alpha^{-(h-1)} \mu(\mathbf{p}) }{\alpha^{-(h-1)} \mu(\mathbf{p}) + \alpha^{h-1} (1-\mu(\mathbf{p}))} -\mu(\mathbf{p})
\leq \hat{\mu}(\mathbf{p})-\mu(\mathbf{p}) \\
&\leq \frac{\alpha^{h-1} \mu(\mathbf{p})}{\alpha^{h-1} \mu(\mathbf{p}) + \alpha^{-(h-1)} (1-\mu(\mathbf{p}))} - \mu(\mathbf{p}) = f(\mu(\mathbf{p})) \leq \frac{\alpha^{h-1}-1}{\alpha^{h-1}+1},
\end{aligned}
\end{equation*}
hence the second part of the claim  follows.
\end{proof}

Next we use the above theorem to relate the stationary distributions of logit response dynamics in nearby games.
\begin{corollary} \label{cor:closeGameDist}
Let $\cal G$ and $\hat{\cal G}$  be finite games with number of strategy profiles $|E|=h$, such that $d({\cal G}, \hat{\cal G})\leq \delta$. Denote the stationary distributions of logit response dynamics in these games by
$\mu_\tau$, and $\hat{\mu}_\tau$ respectively.
 Then, for any  strategy profile $\mathbf{p}$ we have
\begin{equation*}
\begin{aligned}
(i) \qquad &
\frac{e^{-\frac{2\delta (h-1)}{\tau}} \mu_\tau(\mathbf{p}) }{e^{-\frac{2\delta (h-1)}{\tau}} \mu_\tau(\mathbf{p}) + e^{\frac{2\delta (h-1)}{\tau}} (1-\mu_\tau(\mathbf{p}))}
\leq \hat{\mu}_\tau(\mathbf{p}) \leq \frac{e^{\frac{2\delta (h-1)}{\tau}} \mu_\tau(\mathbf{p})}{e^{\frac{2\delta (h-1)}{\tau}} \mu_\tau(\mathbf{p}) + e^{-\frac{2\delta (h-1)}{\tau}} (1-\mu_\tau(\mathbf{p}))},
\\
(ii)  \qquad & |\mu_\tau({\mathbf{p}})-\hat{\mu}_\tau({\mathbf{p}})| \leq \frac{e^{\frac{2\delta (h-1)}{\tau}}-1}{e^{\frac{2\delta (h-1)}{\tau}}+1}.
\end{aligned}
\end{equation*}
\end{corollary}
\begin{proof}
Proof follows  from
Lemma \ref{lem:transProp} and Theorem \ref{theo:statDist} by setting $\alpha=e^{\frac{2\delta}{\tau}}$.
\end{proof}
The above corollary can be adapted to near-potential games, by exploiting the relation of stationary distribution of logit response and potential function in potential games (see \eqref{eq:statOfLogit}).
We conclude this section by providing such a characterization of the stationary distribution of logit response dynamics in near-potential games.
\begin{corollary} \label{cor:logitNearPot}
Consider a game $\cal G$ and let $\hat{\cal G}$ be a nearby potential game such that  $d({\cal G}, \hat{\cal G})\leq \delta$.
Denote the potential function of   $\hat{\cal G}$ by $\phi$, and the number of strategy profiles in these games by $|E|=h$. Then, the stationary distribution $\mu_\tau$ of logit response dynamics in $\cal G$ is such that
\begin{equation*}
\begin{aligned}
(i) \qquad
&\frac{e^{\frac{1}{\tau}(\phi(\mathbf{p})-2\delta (h-1))}}{e^{\frac{1}{\tau}(\phi(\mathbf{p})-2\delta (h-1))} +
\sum_{\mathbf{q}\neq\mathbf{p}\in E}e^{\frac{1}{\tau}(\phi(\mathbf{q})+2\delta (h-1))}}
\leq \mu_\tau({\mathbf{p}})  \qquad \qquad\\
& \qquad\qquad\qquad\qquad\qquad\qquad\qquad \leq \frac{e^{\frac{1}{\tau}(\phi(\mathbf{p})+2\delta (h-1))}}{e^{\frac{1}{\tau}(\phi(\mathbf{p})+2\delta (h-1))} +
\sum_{\mathbf{q}\neq\mathbf{p}\in E}e^{\frac{1}{\tau}(\phi(\mathbf{q})-2\delta (h-1))}}, \\
(ii) \qquad
&\left|\mu_\tau({\mathbf{p}})-
 \frac{e^{\frac{1}{\tau}\phi(\mathbf{p})}}{\sum_{\mathbf{q}\in E}e^{\frac{1}{\tau}\phi(\mathbf{q})}}
  \right|
  \leq
  \frac{e^{\frac{2\delta (h-1)}{\tau}}-1}{e^{\frac{2\delta (h-1)}{\tau}}+1}.
\end{aligned}
\end{equation*}
\end{corollary}
\begin{proof}
Proof follows from Corollary \ref{cor:closeGameDist} and  \eqref{eq:statOfLogit}.
\end{proof}
With simple manipulations, it can be shown that $(e^x-1)/(e^x+1) \leq x/2$ for $x\geq 0$. Thus,  (ii) in the above corollary implies that
$\left|\mu_\tau({\mathbf{p}})-
 \frac{e^{\frac{1}{\tau}\phi(\mathbf{p})}}{\sum_{\mathbf{q}\in E}e^{\frac{1}{\tau}\phi(\mathbf{q})}}
  \right|
  \leq \frac{\delta (h-1)}{\tau}.$
Therefore, the stationary distribution of logit response dynamics in
a near-potential  game can be characterized in terms of  the
stationary distribution of this update rule in a close potential
game. When $\tau$ is fixed and $\delta \rightarrow 0$, i.e., when
the original game is arbitrarily close to a potential game, the
stationary distribution of logit response is arbitrarily close to
the stationary distribution in the potential game. On the other
hand, for a fixed $\delta$, as $\tau \rightarrow 0$, the upper bound
in (ii) becomes uninformative. This is the case since $\tau
\rightarrow 0$ implies that players adopt their best responses with
probability $1$, and thus the stationary distribution of the update
rule becomes very sensitive to the difference of the game from a
potential game.
In this case we can still characterize the stochastically stable states of logit response using the results of Corollary \ref{cor:logitNearPot},
as we show in Corollary \ref{cor:logitSS}.

\begin{corollary} \label{cor:logitSS}
Consider a game $\cal G$ and let $\hat{\cal G}$ be a nearby
potential game with potential function $\phi$ and  $d({\cal G},
\hat{\cal G})\leq\delta$. Denote the potential function of   $\hat{\cal
G}$ by $\phi$, and the number of strategy profiles in these games by
$|E|=h$.
  The stochastically stable strategy profiles of $\cal G$ are (i)~contained in $S=\{\mathbf{p} | \phi(\mathbf{p}) \geq \max_{\mathbf{q} } \phi(\mathbf{q}) - 4\delta(h-1)\}$, (ii)  $4\delta h$-equilibria of $\cal G$.
\end{corollary}
\begin{proof}
(i) The upper bound in the first part of Corollary \ref{cor:logitNearPot} implies that if $\mathbf{p}$ is a strategy profile such that $\phi(\mathbf{p}) < \max_{\mathbf{q}\in E} \phi(\mathbf{q}) - 4\delta (h-1)$,
then the stationary distribution of logit response in $\cal G$ is such that $\mu_\tau (\mathbf{p}) \rightarrow 0$ as $\tau \rightarrow 0$.  Thus, it immediately follows that the stochastically stable states in $\cal G$ are contained in
$\{\mathbf{p}\in E | \phi(\mathbf{p}) \geq \max_{\mathbf{q}\in E} \phi(\mathbf{q}) - 4\delta (h-1)\}.$

(ii)
From the definition of $S$ it follows that
in $\hat{\cal G}$, none of the players can deviate from a strategy profile in $S$ and improve its utility by more than $4\delta(h-1)$.
Since $d({\cal G},\hat{\cal G})\leq \delta$ it follows from part (i)  that in $\cal G$,
none of the players can unilaterally deviate from a
 stochastically stable strategy profile and improve its utility by more than $4\delta (h-1) + \delta \leq 4 \delta h $. Hence stochastically stable strategy profiles of $\cal G$ are  $4\delta h$-equilibria.
\end{proof}

We conclude that in near-potential games,  the stochastically stable states of logit response are the strategy profiles that approximately maximize the potential function of a close potential game.
This result enables us to characterize the  stochastically stable
states of logit response dynamics
in near-potential games,
without explicitly computing the
stationary distribution.

\section{Fictitious Play} \label{sec:fictitiousP}
In this section, we investigate the convergence behavior of
fictitious play in near-potential games. Unlike better/best response
dynamics and logit response, in fictitious play agents maintain an
empirical frequency distribution of other players' strategies and
play a best response against it. Thus, analyzing fictitious play
dynamics requires the notion of mixed strategies and some additional
definitions that are provided in Section \ref{subsec:mixed}.
 In Section \ref{subsec:dftp} we show  that in finite games  the empirical frequencies
 of fictitious play converge to a set which can be characterized in
  terms of the approximate equilibrium set of the game and the level sets of the potential function of a close potential game.
When the original game is sufficiently close to a potential game, we
strengthen this result and establish that the empirical frequencies
converge to a small neighborhood of mixed equilibria of the game,
and the size of this neighborhood is a function of the distance of
the original game from a potential game. As  a special case, our
result allows us to recover the result of
\citet{monderer1996fpp}, which states that in potential games the
empirical frequencies of fictitious play converge to the set of mixed
Nash equilibria.

\subsection{Mixed Strategies and Equilibria} \label{subsec:mixed}
In this section, we introduce some additional notation and
definitions, which will be used in Section \ref{subsec:dftp} when
studying convergence properties of fictitious play in near-potential
games.

We start by introducing the concept of mixed strategies in games.
For each player $m\in{\cal M}$, we denote by $\Delta E^m$ the set of probability distributions on $E^m$.  For $x^m \in \Delta E^m$,  $x^m({{ p}^m})$  denotes the probability player $m$ assigns to strategy ${ p}^m \in E^m$.
We refer to the distribution $x^m\in \Delta E^m$ as a {\em mixed strategy of player $m\in {\cal M}$} and to the collection $\mathbf{x}=\{x^m\}_{m\in {\cal M}}\in \prod_m \Delta E^m$ as a {\em mixed strategy profile}. The mixed strategy profile of all players but the $m$th one is denoted by $\mathbf{x}^{-m}$.
We use $|| \cdot||$ to denote the standard $2$-norm  on $\prod_m \Delta E^m$, i.e.,
for $\mathbf{x}\in \prod_m \Delta E^m$, we have $||\mathbf{x}||^2 =\sum_{m\in {\cal M}} \sum_{p^m\in E^m} \left(x^m(p^m)\right)^2$.

By  slight (but standard) abuse of  notation, we use the same notation for the mixed extension of  utility function $u^m$ of  player $m\in {\cal M}$,
i.e.,
\begin{equation} \label{eq:multiLinProb}
u^m(\mathbf{x})=\sum_{{\bf p}\in E} u^m({\bf p}) \prod_{k\in{\cal M}} x^k({{ p}^k}),
\end{equation}
 for all  $\mathbf{x}\in \prod_m \Delta E^m$.
In addition, if player $m$ uses some pure strategy ${q}^m$ and  other players use the mixed strategy profile $\mathbf{x}^{-m}$, the payoff of player $m$ is denoted by
\begin{equation*}
u^m({q}^m,\mathbf{x}^{-m})=\sum_{{\bf p}^{-m}\in E^{-m}} u^m({ q}^m, {\bf p}^{-m}) \prod_{k\in{\cal M}, k\neq m} x^k({{ p}^k}).
\end{equation*}
Similarly, we denote the mixed extension of the potential function by $\phi(\mathbf{x})$, and we use the notation $\phi(q^m, \mathbf{x}^{-m})$ to denote the potential when player $m$ uses some pure strategy ${q}^m$ and  other players use the mixed strategy profile $\mathbf{x}^{-m}$.

A mixed strategy profile $\mathbf{x}=\{x^m\}_{m\in{\cal M}} \in \prod_m \Delta E^m$ is a \emph{mixed $\epsilon$-equilibrium} if  for all $m\in{ \cal M}$ and ${p}^m \in  E^m$,
\begin{equation} \label{eq:defEpsEq}
 u^m({ p}^m, \mathbf{x}^{-m}) -u^m({x}^m, \mathbf{x}^{-m}) \leq \epsilon.
\end{equation}
Note that if the inequality holds for $\epsilon=0$, then $\mathbf{x}$ is referred to as a \emph{mixed  Nash equilibrium} of the game.
In the rest of the paper, we use the notation ${\cal X}_{\epsilon}$
to  denote the set of mixed $\epsilon$-equilibria.

Our characterization of the limiting mixed strategy set of fictitious play  depends on the number of players in the game. We use $M=|{\cal M}|$ as a short-hand notation for this number.

We conclude this section with two technical lemmas which summarize some properties of  mixed equilibria and mixed extensions of  potential and utility   functions. Proofs of these lemmas can be found in the Appendix.

The first lemma establishes the Lipschitz continuity of the mixed extensions of the payoff functions and the potential function.  It also shows a natural implication of continuity:
for any $\epsilon'>\epsilon$,  a small enough neighborhood of the $\epsilon$-equilibrium set is contained in the $\epsilon'$-equilibrium set.

\begin{lemma} \label{lem:mixedEqNew_1}
\begin{enumerate}
\item[(i)]  Let $\nu: \prod_{m\in {\cal M}} E^m \rightarrow \mathbb{R}$ be a mapping from pure strategy profiles to real numbers. Its mixed extension is  Lipschitz continuous with a Lipschitz constant of $M\sum_{p\in E} |\nu (\mathbf{p})|$ over the domain $\prod_{m\in {\cal M}} \Delta E^m $.

\item[(ii)] Let $\alpha \geq 0$ and $\gamma>0$ be given.
There exists a small enough $\theta>0$ such that for any  $||\mathbf{x}-\mathbf{y}||< \theta$
if
$\mathbf{x}\in {\cal X}_{\alpha}$, then $\mathbf{y}\in {\cal X}_{\alpha+\gamma}$.

\end{enumerate}
\end{lemma}
Lipschitz continuity follows from the fact that mixed extensions are multilinear functions \eqref{eq:multiLinProb}, with bounded domains.
The proof of the second part immediately follows from the  Lipschitz continuity of mixed extensions of payoff functions and the definition of approximate equilibria
\eqref{eq:defEpsEq}.
Note that the second part implies that
 for any $\epsilon'>0$, there exists a small enough neighborhood
 of equilibria  that is contained in the $\epsilon'$-equilibrium set of the game.

We next study the continuity properties of the approximate
equilibrium mapping. We first provide the relevant definitions (see
\citet{berge1963topological,fudenberg1991gt}).
\begin{definition}[Upper Semicontinuous Function] \label{def:uscFcn}
A function $g:X\rightarrow Y \subset \mathbb{R}$ is upper semicontinuous at $x_*$, if, for each $\epsilon>0$ there exists a neighborhood $U$ of $x_*$, such that  $g(x)< g(x_*)+\epsilon$ for all $x\in U$ .
We say $g$ is upper semicontinuous, if it is upper semicontinuous at every point in its domain.

Alternatively,   $g$ is upper semicontinuous if
$\limsup_{x_n \rightarrow x_*} g(x_n)\leq g(x_*)$
for every $x_*$  in its domain.

\end{definition}

\begin{definition}[Upper Semicontinuous Correspondence] \label{def:uscCor}
A correspondence $g:X\rightrightarrows Y$ is upper semicontinuous at $x_*$, if for any open neighborhood $V$ of $g(x_*)$ there exists a neighborhood $U$ of $x_*$ such that $g(x)\subset V$ for all $x\in U$.
We say $g$ is upper semicontinuous, if it is upper semicontinuous at every point in its domain
and
$g(x)$ is a compact set for each $x\in X$.

Alternatively, when $Y$ is compact, $g$ is upper semicontinuous  if its graph is closed, i.e.,
the set
$\{(x,y) | x\in X,  y\in g(x)  \}$ is closed.
\end{definition}

We next establish upper semicontinuity of the approximate
equilibrium mapping.\footnote{ Here we fix the game, and discuss
upper semicontinuity with respect to the $\epsilon$ parameter
characterizing the $\epsilon$-equilibrium set. We note that this is
different than the common results in the literature which discuss
upper semicontinuity of the equilibrium set with respect to changes
in the utility functions of the underlying game (see
\citet{fudenberg1991gt}).}
\begin{lemma} \label{lem:mixedEqNew_2}
\begin{enumerate}
\item[(i)]
 Let $\nu: \prod_{m\in {\cal M}} \Delta E^m \rightarrow \mathbb{R}$ be an upper semicontinuous function.
The correspondence $g:\mathbb{R} \rightrightarrows \prod_{m\in{\cal M}} \Delta E^m$
 such that $g(v)=\{\mathbf{x}| \nu(\mathbf{x}) \geq -v  \}$ is upper semicontinuous.

\item[(ii)] Let $g:\mathbb{R} \rightrightarrows \prod_{m\in{\cal M}} \Delta E^m$ be the correspondence such that $g(\alpha)={\cal X}_{\alpha}$. This correspondence is upper semicontinuous.

\end{enumerate}
\end{lemma}

Upper semicontinuity of the approximate equilibrium mapping implies that
for
any given neighborhood of the $\epsilon$-equilibrium set,  there exists an $\epsilon'>\epsilon$ such that $\epsilon'$-equilibrium set is contained in this neighborhood. In particular, this implies that every neighborhood of equilibria of the game contains an $\epsilon'$-equilibrium set for some $\epsilon'>0$.
Hence, if disjoint neighborhoods of equilibria are chosen (assuming there are finitely many equilibria), this implies that there exists some $\epsilon'>0$, such that the $\epsilon'$-equilibrium set is contained in disjoint neighborhoods of equilibria. In the next section, we use this observation to establish convergence of fictitious play to small neighborhoods of equilibria of  near-potential games.

\subsection{Discrete-Time Fictitious Play} \label{subsec:dftp}
Fictitious play is a classical update rule studied in the learning
in games literature. In this section, we   consider the  fictitious
play dynamics, proposed in \citet{brown1951iterative}, and explain
how the limiting behavior of this dynamical process can be
characterized in near-potential games. In particular, we show  that
the empirical frequencies of fictitious play converge to a set
 which can be characterized in
terms of the $\epsilon$-equilibrium set of the game, and the level sets of the potential function of a close potential game.
We also establish that for games sufficiently close to a potential game,   the empirical frequencies of fictitious play converge to a neighborhood of the (mixed) equilibrium set. Moreover, the size of this neighborhood depends on the distance of the original game from a nearby potential game.
This generalizes the result of  \citet{monderer1996fpp}, on convergence of
empirical frequencies
to mixed Nash equilibria in potential games.

In this paper, we only consider the discrete-time version of fictitious play, i.e.,
the update process
 starts at a given strategy profile at time $t=0$, and players can update their strategies
at discrete time instants $t\in \{1,2,\dots\}$. Throughout this subsection we denote the
strategy used by player $m$ at time instant $t$ by $p^m_t$, and we
denote by $\mathbf{1}(p^m_{t}=p^m)$ the indicator function which
equals to $1$ if $p^m_t=p^m$, and $0$ otherwise. A formal definition
of discrete-time fictitious play dynamics is given next.
\begin{definition}[Discrete-Time Fictitious Play]
\label{def:originalFPnew}
Let $\mu^m_T(q^m)= \frac{1}{T}\sum_{t=0}^{T-1} \mathbf{1}(p^m_t=q^m)$ denote the empirical frequency that player $m$ uses strategy $q^m$ from  time instant $0$ to time instant $T-1$, and $\mu^{-m}_T$ denote the collection of empirical frequencies of all players but $m$.
A game play, where at each time instant $t$, every player $m$, chooses a strategy $p^m_t$ such that
$$
p^m_t\in \arg\max_{q^m\in E^m} u^m(q^m, \mu^{-m}_t)
$$
is referred to as discrete-time fictitious play. That is, fictitious play dynamics is the update process, where each player chooses
its best response to the empirical frequencies of the actions of other players.
\end{definition}

We refer to $\mu^m_t$ as the distribution of  empirical frequencies of
player $m$'s strategies at time~$t$.
Note that $\mu^m_t$ can be thought of as vector with length $|E^m|$, whose entries are indexed by strategies of
player $m$, i.e., $\mu^m_t(p^m)$ denotes the entry of the vector
corresponding to the empirical frequency player $m$ uses strategy
$p^m$ with. Similarly, we define the joint empirical frequency
distribution  of all players as $\mu_t= \{\mu_t^m\}_{m\in {\cal M}}$. Note
that  $\mu^m_t \in \Delta E^m$, i.e., empirical frequency distributions
are mixed strategies, and similarly $\mu_t \in \prod_{m\in {\cal M}}
\Delta E^m$.

Observe that the evolution of this  empirical frequency distribution
can be captured by the following equation:
\begin{equation} \label{eq:empFreq}
\mu_{t+1}= \frac{t}{t+1} \mu_t + \frac{1}{t+1} I_t,
\end{equation}
where  $I_t=\{I_t^m\}_{m\in {\cal M}}$, and $I^m_t$ is a  vector   which has the same size as $\mu^m_t$ and its entry corresponding to strategy $p^m$ is given by
$I_t^m(p^m)=\mathbf{1}(p^m_t=p^m)$.
Rearranging the terms in \eqref{eq:empFreq}, and observing that
$I_t,\mu_t \in \prod_{m\in {\cal M}} \Delta E^m$ are vectors with entries in $[0,1]$
we conclude
\begin{equation}\label{eq:empFreq2}
||\mu_{t+1}-\mu_t|| = \frac{1}{t+1} || I_t-\mu_t || = O\left( \frac{1}{t}\right),
\end{equation}
where $O(\cdot)$ stands for the big-O notation, i.e.,
$f(x)=O(g(x))$, implies that there exists some $x_0$ and a constant
$c$ such that $|f(x)|\leq c |g(x)|$ for all $x\geq x_0$.

We start analyzing discrete-time fictitious play in near-potential games, by first focusing on
the change in the value of the potential function along the  fictitious play  updates in the original game.
In particular, we show that in near-potential games if the empirical frequencies are outside some $\epsilon$-equilibrium set, then the potential of the close potential game (evaluated at the empirical frequency distribution)
increases by
discrete-time fictitious play updates.\footnote{Our approach here is similar to the one used in \citet{monderer1996fpp} to analyze discrete-time fictitious play in potential games.}
\begin{lemma} \label{lem:potIncEpsEq}
Consider a game $\cal G$ and let $\hat{\cal G}$ be a close potential game such that  $d({\cal G}, \hat{\cal G}) \leq \delta$.
Denote the potential function of   $\hat{\cal G}$ by $\phi$.
Assume that in $\cal G$ players update their strategies according to discrete-time fictitious play dynamics, and at some time instant $T>0$, the  empirical frequency distribution $\mu_T$ is outside an $\epsilon$-equilibrium set of $\cal G$. Then, $$ \phi(\mu_{T+1})- \phi(\mu_T) \geq \frac{\epsilon-M\delta}{T+1} + O\left(\frac{1}{T^2}\right).$$
\end{lemma}
\begin{proof}
Consider the mixed extension  of the potential function
$\phi(\mathbf{x})=\sum_{\mathbf{p}\in E} \phi(\mathbf{p})
\prod_{m\in {\cal M}} x^m({p^m})$, where $\mathbf{x}= \{ x^m \}_m$
and $x^m({p^m})$ denotes the probability player $m$ plays strategy
$p^m$. The expression for $\phi(\mathbf{x})$ implies that Taylor
expansion of $\phi$ around $\mu_T$ satisfies
\begin{equation*}
\begin{aligned}
\phi(\mu_{T+1}) &= \phi(\mu_T) +  \sum_{m\in {\cal M}} \sum_{p^m\in E^m} (\mu_{T+1}^m(p^m)- \mu_{T}^m(p^m)) \phi(p^m, \mu_T^{-m})  + O(||\mu_{T+1}-\mu_T ||^2).
\end{aligned}
\end{equation*}
Observing from \eqref{eq:empFreq} that $\mu_{t+1}-\mu_t = \frac{1}{t+1} (I_t-\mu_t)$,
and noting from \eqref{eq:empFreq2} that $||\mu_{t+1}-\mu_t|| = O\left( \frac{1}{t} \right)$
  the above equality can be rewritten as
\begin{equation*}
\begin{aligned}
\phi(\mu_{T+1}) &= \phi(\mu_T) +  \sum_{m\in {\cal M}} \sum_{p^m\in E^m}
\frac{1}{T+1}(\mathbf{1}(p^m_{T}=p^m)-\mu_T^m(p^m)) \phi(p^m, \mu_T^{-m})  + O\left(\frac{1}{T^2}\right).
\end{aligned}
\end{equation*}
Rearranging the terms, and noting that
$\sum_{p^m\in E^m}
\mu_T^m(p^m) \phi(p^m, \mu_T^{-m})= \phi(\mu^m_T, \mu_T^{-m})$,
 it follows that
\begin{equation*}
\begin{aligned}
\phi(\mu_{T+1})&=\phi(\mu_T) +  \sum_{m\in {\cal M}}
\frac{1}{T+1}\phi(p^m_{T}, \mu_T^{-m})
-\sum_{m\in {\cal M}}
\frac{1}{T+1} \phi(\mu^m_T, \mu_T^{-m})
 + O\left(\frac{1}{T^2}\right)\\
 &=\phi(\mu_T) +   \frac{1}{T+1} \sum_{m\in {\cal M}}
 \left( \phi(p^m_{T}, \mu_T^{-m}) -\phi(\mu^m_T, \mu_T^{-m})   \right)
   +O\left(\frac{1}{T^2}\right).
\end{aligned}
\end{equation*}
Since
$d({\cal G}, \hat{\cal G}) \leq \delta$, the above equality and the definition of MPD imply
\begin{equation} \label{eq:taylorLast}
\begin{aligned}
\phi(\mu_{T+1})    & \geq \phi(\mu_T) +   \frac{1}{T+1} \sum_{m\in {\cal M}}
    \left( u^m(p^m_{T}, \mu_T^{-m}) -u^m(\mu^m_T, \mu_T^{-m}) - \delta  \right)
      + O\left(\frac{1}{T^2}\right).
\end{aligned}
\end{equation}
By definition of the fictitious play dynamics, every player $m$ plays its best response to $\mu_T^{-m}$, therefore $u^m(p^m_{T}, \mu_T^{-m}) -u^m(\mu^m_T, \mu_T^{-m})\geq 0$ for all $m$. Additionally, if $\mu_T$ is outside the $\epsilon$-equilibrium set, as in the statement of the lemma, then it follows that $u^m(p^m_{T}, \mu_T^{-m}) -u^m(\mu^m_T, \mu_T^{-m})\geq \epsilon$ for at least one player. Therefore, \eqref{eq:taylorLast} implies
\begin{equation*}
\begin{aligned}
\phi(\mu_{T+1})    & \geq \phi(\mu_T) +   \frac{\epsilon-M\delta}{T+1}
      + O\left(\frac{1}{T^2}\right),
\end{aligned}
\end{equation*}
hence, the claim follows.
\end{proof}
The above theorem implies that  if $\mu_T$ is not in the
$\epsilon$-equilibrium set
for some $\epsilon >M\delta$,
 and
 $T$  sufficiently large, then the potential evaluated at empirical frequencies increases when players update their strategies.
 Since the mixed extension of the potential is a bounded function, the potential cannot increase unboundedly, and  this observation suggests that  the $\epsilon$-equilibrium set is eventually reached by the  empirical frequency distribution.
 On the other hand, at a later  time instant $\mu_T$ can still leave this equilibrium set, and
 before it does so the potential cannot be lower than the lowest potential in this set (since $\mu_T$ itself belongs to this set).
 Moreover, after $\mu_T$ leaves the $\epsilon$-equilibrium set
 the potential keeps increasing.
Thus, the empirical frequencies are contained in the set of mixed strategy profiles, which have potential at least as large as the minimum potential in
this  approximate equilibrium set.
We next make this intuition precise, and characterize the
set of limiting mixed strategies for fictitious play in near-potential games.
We adopt the following convergence notion: we say that empirical frequencies of  fictitious play converge
to a set $S\subset \prod_{m\in {\cal M}} \Delta E^m$, if $\inf_{\mathbf{x}\in S} ||\mu_t-\mathbf{x}|| \rightarrow 0$ as $t\rightarrow \infty$.
\begin{theorem} \label{theo:FPConv1}
Consider a game $\cal G$ and let $\hat{\cal G}$ be a close potential game such that  $d({\cal G}, \hat{\cal G}) \leq \delta$.
Denote the potential function of   $\hat{\cal G}$ by $\phi$.
Assume that in $\cal G$ players update their strategies according to discrete-time fictitious play dynamics, and let ${\cal X}_\alpha$ denote the $\alpha$-equilibrium set of $\cal G$.
For any $\epsilon > 0$, there exists a time instant $T_\epsilon>0$ such that for all $t>T_\epsilon$
$$
\mu_t\in
C_\epsilon\triangleq \left\{\mathbf{x}\in  \left. \prod_{m \in {\cal M}}\Delta E^m \right|  \phi(\mathbf{x}) \geq \min_{\mathbf{y}\in {\cal X}_{{M} \delta+\epsilon}} \phi(\mathbf{y})  \right\}.$$
\end{theorem}
\begin{proof}
Let $\epsilon'$ be such that $\epsilon > \epsilon'> 0$.
It can be seen from the definition of $C_{\epsilon}$ that ${\cal X}_{M\delta+\epsilon'} \subset {\cal X}_{M\delta+\epsilon} \subset C_{\epsilon}$.
We prove the claim in two steps:
(i) We first show that in this update process ${\cal X}_{M\delta+\epsilon'}$ is visited infinitely often by $\mu_t$, i.e., for all $T'$, there exists $t>T'$  such that $\mu_t \in {\cal X}_{M\delta+\epsilon'}$, (ii) We prove that there exists a $T''$ such that if $\mu_t \in C_{\epsilon}$ for some $t>T''$, then for all $t'>t$ we have $\mu_{t'} \in C_\epsilon$. Thus, the second step guarantees that if $C_{\epsilon}$ is visited at a sufficiently later time instant, then $\mu_t$ remains in $C_\epsilon$. Since ${\cal X}_{M\delta+\epsilon'} \subset C_{\epsilon}$ the first step ensures that such a time instant exists, and the claim in the theorem immediately follows from (ii). Moreover, this time instant corresponds to $T_\epsilon$ in the theorem statement.

Proof of both steps rely on the following simple observation:
Lemma \ref{lem:potIncEpsEq} implies that  there exists a large enough $T$, such that  if the empirical frequencies do not belong to ${\cal X}_{M\delta+\epsilon'}$ at a time instant $t>T$, then $\phi$ increases:
\begin{equation} \label{eq:potIncreaseStrict}
\phi(\mu_{t+1})-\phi(\mu_t) \geq
 \frac{M\delta+\epsilon'-  M\delta}{(t+1)} + O \left(\frac{1}{t^2}\right) >
 \frac{\epsilon'}{2(t+1)}>0.
\end{equation}

We prove (i) by contradiction. Assume that there exists a $T'$ such that
$\mu_t \notin {\cal X}_{M\delta+\epsilon'}$
 for $t>T'$, and let ${T}_m=\max\{T,T'\}$.
 Then, \eqref{eq:potIncreaseStrict} holds for all $t=\{T_m+1, \dots\}$,
and  summing both sides of this inequality over this set
  we obtain
\begin{equation*}
\limsup_{t \rightarrow \infty} \phi(\mu_{t+1})-\phi(\mu_{T_m+1}) \geq
\sum_{t=T_m+1}^{\infty}
 \frac{\epsilon'}{2(t+1)}.
\end{equation*}
 Since the mixed extension of the potential is a bounded function,
 it follows that the left hand side of the above inequality is bounded, but the right hand side grows unboundedly.
Hence, we reach a contradiction, and (i) follows.

Lemma \ref{lem:mixedEqNew_1} (ii) implies that
there exists some $\theta>0$ such that
if a strategy profile $\mathbf{x}$ is an $(M\delta+\epsilon')$-equilibrium, then  any strategy profile $\mathbf{y}$ that satisfies $||\mathbf{x}-\mathbf{y}||<\theta$ is an $(M\delta+\epsilon)$-equilibrium (recall that $\epsilon> \epsilon' > 0$).
Since  $||\mu_{t+1}-\mu_{t}||=O(1/t)$ by \eqref{eq:empFreq2}, this implies that there exists some $T''>T$,  such that for all $t>T''$ if $\mu_{t}\in {\cal X}_{M\delta+\epsilon'}$, then we have
\begin{equation}\label{eq:refMuIn}
\mu_{t+1}\in {\cal X}_{M\delta+\epsilon}.
\end{equation}

Let $\mu_t \in C_\epsilon$ for some time instant $t>T''$. If  $\mu_{t}\in {\cal X}_{M\delta+\epsilon'}$, then by \eqref{eq:refMuIn}  $\mu_{t+1}\in {\cal X}_{M\delta+\epsilon} \subset C_\epsilon $. If, on the other hand, $\mu_t \in C_\epsilon- {\cal X}_{M\delta+\epsilon'}$, then by \eqref{eq:potIncreaseStrict} and the definition of $C_\epsilon$ we have
\begin{equation}
\phi(\mu_{t+1}) > \phi(\mu_{t}) \geq \min_{\mathbf{y}\in {\cal X}_{M\delta+\epsilon}} \phi(\mathbf{y}),
\end{equation}
and hence
$\mu_{t+1}\in C_\epsilon$.
Thus, we have established that there exists some $T''$ such that if $\mu_t \in C_\epsilon$ for some $t >  T''$, then $\mu_{t+1} \in C_\epsilon$, and hence (ii) follows.
\end{proof}
The above theorem establishes that after finite time $\mu_t$ is contained in the set $C_\epsilon$  for any $\epsilon>  0$.
Corollary \ref{cor:discreteFP1}, establishes that in the limit this result can be strengthened:
as $t\rightarrow \infty$, $\mu_t$ converges to a set, which is a subset of $C_\epsilon$ for every $\epsilon> 0$. The proof can be found in the Appendix.
\begin{corollary} \label{cor:discreteFP1}
The  empirical frequencies of discrete-time fictitious play converge to
\begin{equation*}
\begin{aligned}
C \triangleq \left\{\mathbf{x}\in \left. \prod_{m\in {\cal M}}\Delta E^m \right| \phi(\mathbf{x}) \geq \min_{\mathbf{y}\in {\cal X}_{ M \delta}} \phi(\mathbf{y})  \right\}.
\end{aligned}
\end{equation*}
\end{corollary}

This result suggests that  in near-potential games, the empirical frequencies of fictitious play converge to a set where the potential is at least as large as the minimum potential in an approximate equilibrium set.
For exact potential games, it is known that the empirical
frequencies converge to a Nash equilibrium \citep{monderer1996fpp}.
It can be seen from Definition \ref{def:ExactPot} that in potential
games, maximizers of the potential function are equilibria of the
game. Thus, in potential games with a unique equilibrium  the
equilibrium is the unique maximizer of the potential function.
Hence, for such games, we have $\delta=0$,  $\min_{\mathbf{y}\in
{\cal X}_{ M \delta}} \phi(\mathbf{y})= \max_{\mathbf{x}\in
\prod_{m\in{\cal M}} \Delta E^m } \phi(\mathbf{x})$, and Corollary
\ref{cor:discreteFP1} implies that  empirical frequencies of
fictitious play converge to the unique equilibrium of the game,
recovering the convergence result of \citet{monderer1996fpp}.
 However, when there are multiple equilibria
Corollary \ref{cor:discreteFP1} suggests that empirical frequencies converge to the set of mixed strategy profiles that have potential weakly larger than the minimum potential attained by the equilibria.
While this set contains equilibria, it may contain a continuum of other mixed strategy profiles.
This suggests that in games with multiple equilibria
our result may provide a loose characterization of the limiting behavior of fictitious play dynamics.

We next show that by exploiting the properties of mixed approximate
equilibrium sets, it is possible to obtain a stronger result. Before
we present our result, we discuss a feature of mixed equilibrium
sets  which will be key in our analysis: For small $\epsilon$, the
$\epsilon$-equilibrium set is contained in a small neighborhood of
equilibria (this statement follows from Lemma \ref{lem:mixedEqNew_2}
(ii) by considering the upper semicontinuity of the approximate
equilibrium correspondence $g(\alpha)$ at $\alpha=0$). This property
is illustrated in Example \ref{ex:BoSgame}.
\begin{example}[Mixed equilibrium set of Battle of the Sexes:]
\label{ex:BoSgame}
Consider the two-player battle of the sexes (BoS) game:
Each player has two possible actions $\{O, F\}$, and the payoffs of players are as given in Table
\ref{tab:BoS}.
\begin{table}[h]
 \begin{center}
 \begin{tabular}{ | c | c | c | c |}
  \hline
    & O     & F     \\ \hline
  O  &  3, 2 & 0, 0   \\ \hline
  F  &  0, 0 & 2, 3  \\ \hline
  \end{tabular}
 \caption{
Payoffs in BoS.}
 \label{tab:BoS}
 \end{center}
\end{table}
This game has three equilibria: (i) both players use $O$, (ii) Both players use $F$, (iii) Row player uses $O$ with probability $0.6$, and column player uses $O$ with probability $0.4$.
Note that since this is a game where each player has only two strategies, the probability of using strategy $O$, in the third case uniquely identifies the corresponding mixed equilibrium.
For different values of $\epsilon$, the set of $\epsilon$-equilibria of this game is shown in Figure \ref{fig:eqBoS}. It follows that the set of  $\epsilon$-equilibria is contained in disjoint neighborhoods of equilibria for small values of $\epsilon$.
\begin{figure}[h]
\begin{center}
\subfloat[$0.2$-equilibrium set.]{
\includegraphics[scale=.35]{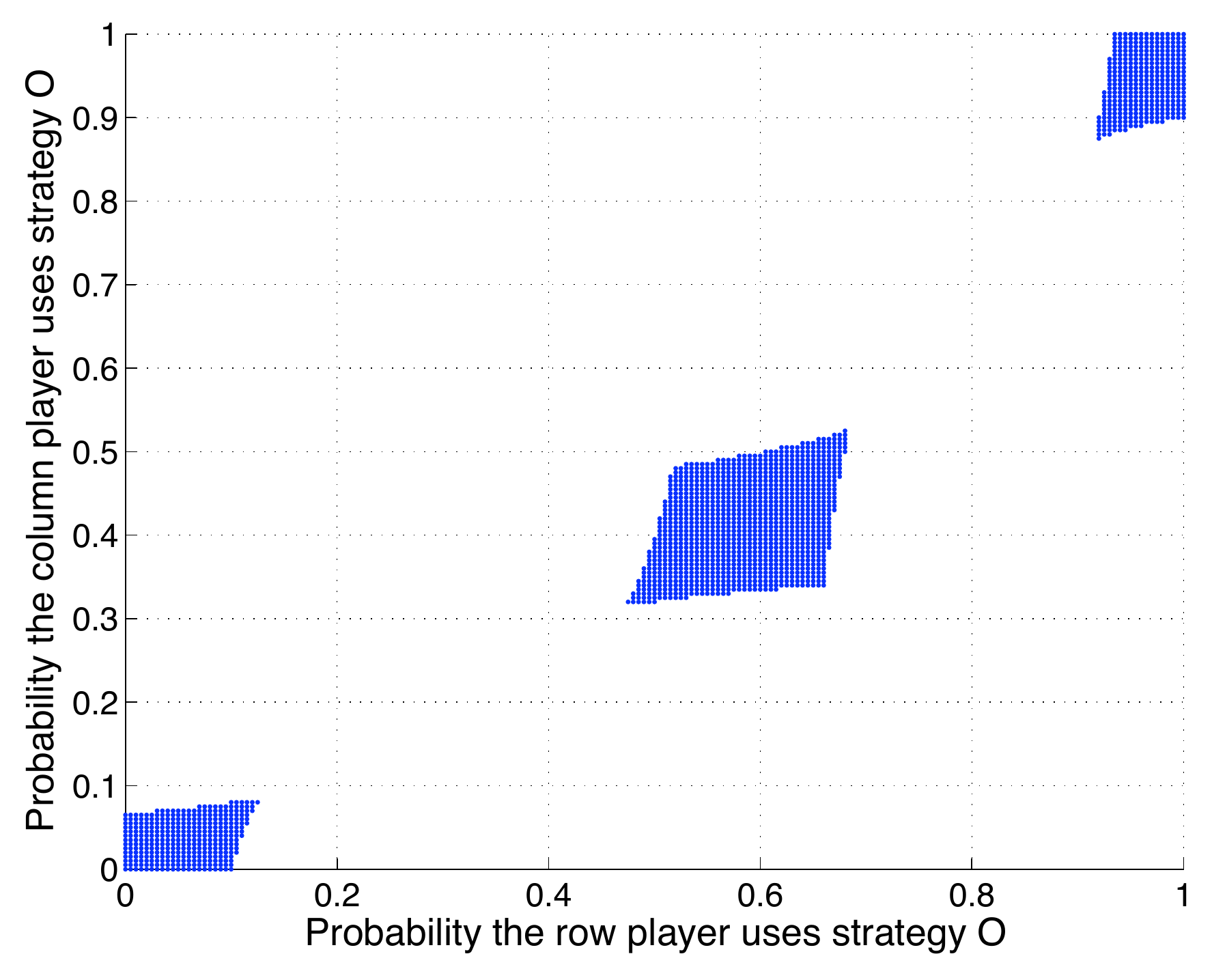}
}\qquad\qquad
\subfloat[$0.3$-equilibrium set.]{
\includegraphics[scale=.35]{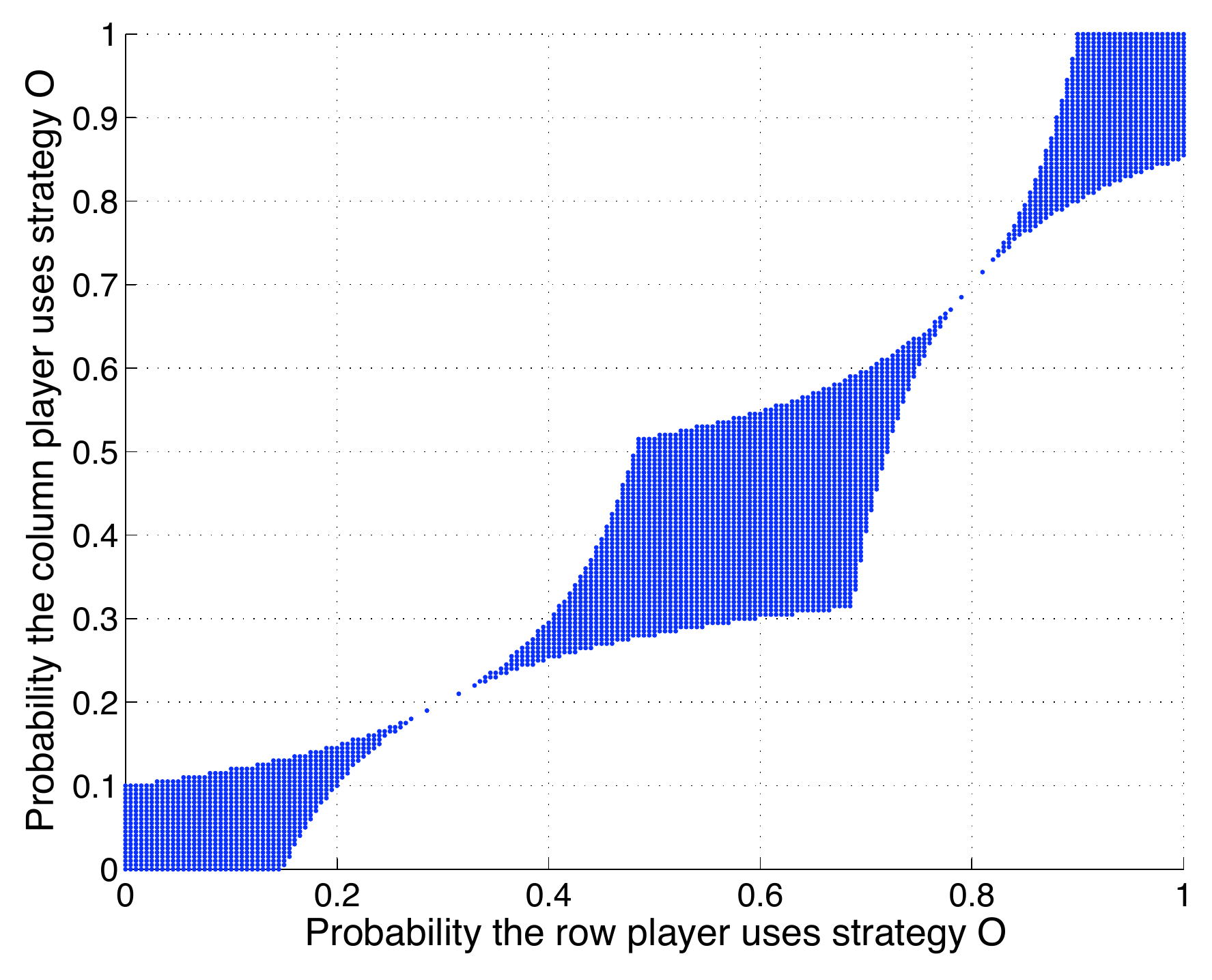}
} \\
\subfloat[$0.4$-equilibrium set.]{
\includegraphics[scale=.35]{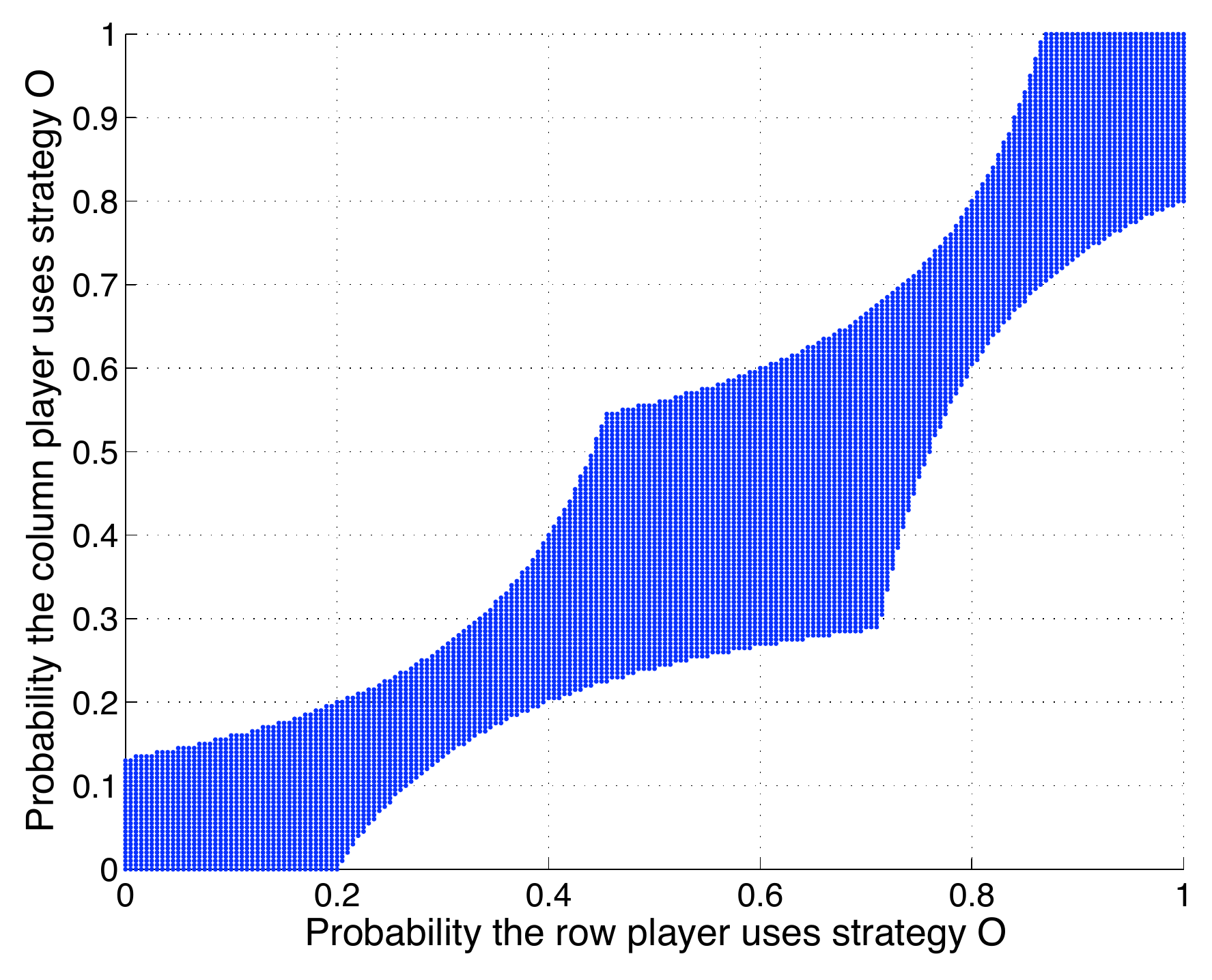}
}
\end{center}
\caption{
Approximate equilibrium sets in BoS are contained in disjoint neighborhoods of equilibria for small $\epsilon$.}
 \label{fig:eqBoS}
\end{figure}
\end{example}
 It was established
 in Lemma \ref{lem:potIncEpsEq}
 that   the potential function of a nearby  potential game (with MPD $\delta$ to the original game),
 evaluated at the  empirical frequency distribution,
  increases when   this distribution is outside  the $M \delta$-equilibrium set of the original game (where $M$ is the number of players).
  If $\delta$ is sufficiently small, then the $M\delta$-equilibria of the game will be contained in a small neighborhood of the equilibria, as illustrated above and shown in Lemma \ref{lem:mixedEqNew_2} (ii).
  Thus, for sufficiently small $\delta$, it is possible to establish that the potential of a close potential game increases outside a small neighborhood of the equilibria of the game.
  In Theorem \ref{theo:fpNEW}, we use this observation to show that
for sufficiently small $\delta$  the empirical frequencies of
  fictitious play dynamics converge to a neighborhood of an
  equilibrium.
  We state the theorem under the assumption that the original game has finitely many equilibria. This assumption generically holds, i.e., for any game a (nondegenerate) random perturbation of payoffs will lead to such a game with probability one (see \citet{fudenberg1991gt}).
 When stating our result, we make use of the Lipschitz continuity of the mixed extension of the potential function, as established in
Lemma
 \ref{lem:mixedEqNew_1}.

\begin{theorem}\label{theo:fpNEW}
Consider a game $\cal G$ and let $\hat{\cal G}$ be a close potential game such that  $d({\cal G}, \hat{\cal G}) \leq \delta$. Denote the potential function of $\hat{\cal G}$ by $\phi$,
and the Lipschitz constant of the mixed extension of $\phi$ by $L$.
Assume that $\cal G$ has finitely many equilibria, and in $\cal G$ players update their strategies according to discrete-time fictitious play dynamics.

There exists some $\bar{\delta}>0$, and $\bar{\epsilon}>0$ (which are functions of utilities of $\cal G$ but not $\delta$) such that if $\delta<\bar{\delta}$, then
 the empirical frequencies of fictitious play converge to
  \begin{equation} \label{eq:TheoStatement}
   \left\{\mathbf{x} \left| ~ || \mathbf{x} -\mathbf{x}_k|| \leq \frac{ 4f(M\delta) ML}{\epsilon} +f(M\delta+\epsilon), \mbox{ for some equilibrium $\mathbf{x}_k$}  \right. \right\},
   \end{equation}
for any $\epsilon$ such that $\bar{\epsilon}\geq \epsilon>0$, where
$f: \mathbb{R}_+ \rightarrow \mathbb{R}_+ $ is an upper semicontinuous function satisfying $f(x) \rightarrow 0$ as $x \rightarrow 0$.
\end{theorem}

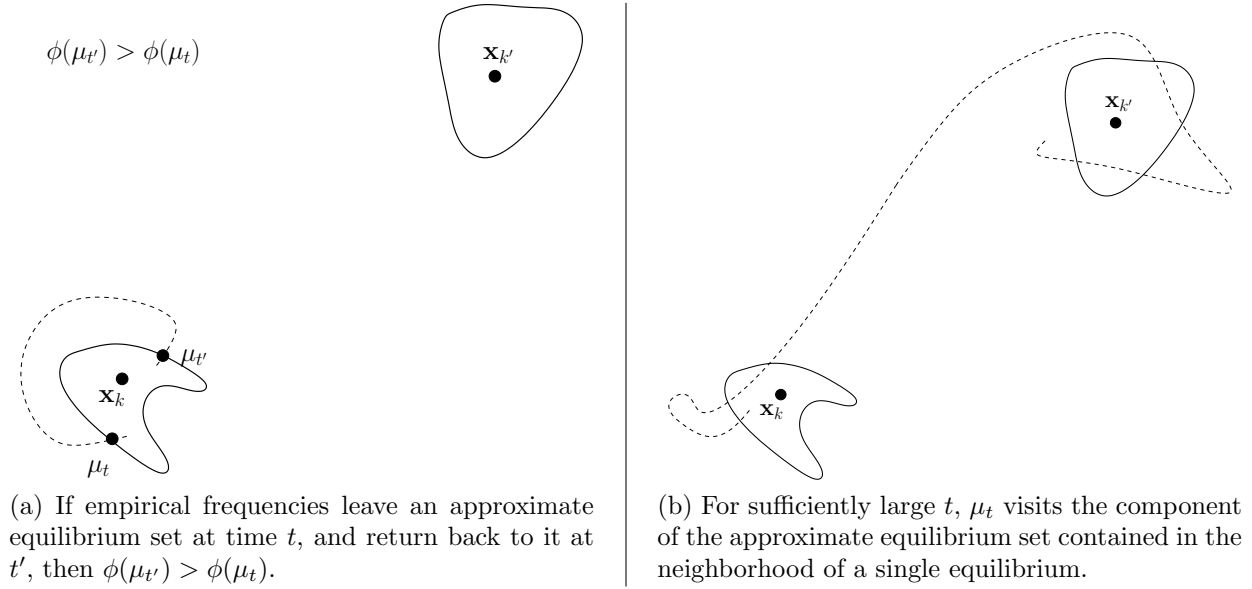
\begin{figure}[htbp]
\begin{center}
\subfloat[If empirical frequencies  leave an approximate equilibrium set at time $t$, and return back to it at $t'$, then $\phi(\mu_{t'})>\phi(\mu_{t})$.]{
\label{fig:IntuitionPart1}
\resizebox{7.5cm}{!}{\input{xFigIntuitionProofV1.tex}}
%{{ \scalebox{.43}{\input{xFigIntuitionProofV1.tex}}}}
}
\quad\vline\quad
\subfloat[For sufficiently large $t$, $\mu_t$ visits the component of the approximate equilibrium set contained in the neighborhood of a single  equilibrium.]{
\label{fig:IntuitionPart2}
\resizebox{7.5cm}{!}{\input{xFigIntuitionProofV2.tex}}
%{{ \scalebox{.43}{\input{xFigIntuitionProofV2.tex}}}}
}
\caption{For small $\delta$ and $\epsilon$, $M\delta+\epsilon$-equilibrium set (enclosed by solid lines around equilibria $\mathbf{x}_{k'}$ and $\mathbf{x}_{k}$) is contained in disjoint neighborhoods of equilibria.
If the  empirical frequency distribution, $\mu_t$, is outside this approximate equilibrium set, then the potential increases with each strategy update.
Assume that empirical frequency distribution leaves an approximate equilibrium set (at time $t$) and returns back to it at a later time instant ($t'>t$).
 We first quantify the resulting increase in the potential
 (left). If $\mu_t$ travels from the component of the approximate equilibrium set in the neighborhood of equilibrium $\mathbf{x}_{k}$ to that in the neighborhood of equilibrium $\mathbf{x}_{k'}$, then the increase in the potential is significant, and consequently $\mu_t$ cannot visit the       approximate equilibrium set in the neighborhood of equilibrium $\mathbf{x}_{k}$ at a later time instant (right).}
\label{fig:IntuitionPart0}
\end{center}
\end{figure}

\begin{figure}[htbp]
\begin{center}
 \resizebox{6cm}{!}{\input{xFigIntuitionProofv4.tex}}
%{{ \scalebox{.4}{\input{xFigIntuitionProofv4.tex}}}}
\caption{If after time $T$, $\mu_t$ only visits the approximate
equilibrium set  in the neighborhood of a single equilibrium
$\mathbf{x}_{k'}$, then we can establish that $\mu_t$ never leaves a
neighborhood of this equilibrium for $t>T$. The size of this
neighborhood is denoted by $r$ in the figure and can be expressed as
in Theorem \ref{theo:fpNEW}.} \label{fig:IntuitionPart3}
\end{center}
\end{figure}
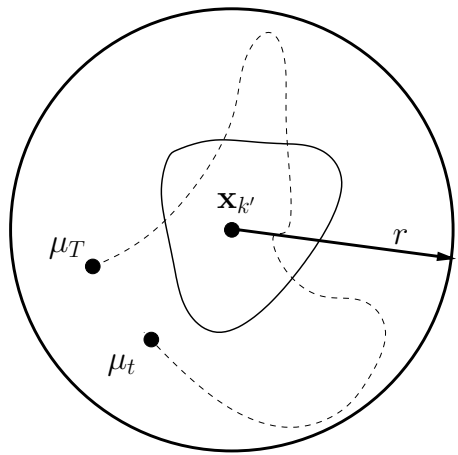

The proof of this theorem can be found in the Appendix, and it has
three main steps illustrated in Figures \ref{fig:IntuitionPart0} and
\ref{fig:IntuitionPart3}. As explained earlier, for small  $\delta$
and $\epsilon$,  the $M\delta+\epsilon$-equilibrium set of the game
is contained in disjoint neighborhoods of the equilibria of the
game.
 Lemma~\ref{lem:potIncEpsEq} implies that potential evaluated at $\mu_t$  increases outside this approximate equilibrium set with strategy updates.
In the proof, we first quantify the increase in the potential, when
$\mu_t$ leaves this approximate equilibrium set and returns back to
it at a later time instant (see Figure \ref{fig:IntuitionPart1}).
 Then, using this increase condition we show that for sufficiently large $t$,  $\mu_t$  can
visit the approximate equilibrium set infinitely often only around
one equilibrium, say $\mathbf{x}_{k'}$ (see Figure
\ref{fig:IntuitionPart2}). This holds since, the increase condition
guarantees that the potential increases significantly when $\mu_t$
leaves the neighborhood of an equilibrium $\mathbf{x}_{k}$, and
reaches to that of $\mathbf{x}_{k'}$. Finally, using the increase
condition one more time, we establish that  if  after  time $T$,
$\mu_t$ visits the approximate equilibrium set only in the
neighborhood of $\mathbf{x}_{k'}$, we can construct a neighborhood
of $\mathbf{x}_{k'}$, which contains $\mu_t$ for all $t>T$ (see
Figure \ref{fig:IntuitionPart3}). This neighborhood is expressed in
\eqref{eq:TheoStatement}.

Observe that if $\delta=0$, i.e., the original game is a potential game, then $f(M\delta)=0$, and
 Theorem  \ref{theo:fpNEW} implies that empirical frequencies of fictitious play converge to the
$f(\epsilon)$-neighborhood of equilibria  for any $\epsilon$ such that $\bar{\epsilon}\geq \epsilon>0$.
Thus, choosing $\epsilon$ arbitrarily small, and observing that $\lim_{x \rightarrow 0} f(x) =0$,
 our result implies that in potential games, empirical frequencies converge to the set of Nash equilibria. Hence, as a special  case of Theorem \ref{theo:fpNEW}, we obtain the convergence result of \citet{monderer1996fpp}.

 Assume that $\delta\neq 0$ and a small $\epsilon<\bar{\epsilon}$ is given.
If $\delta $ is sufficiently small then
  $f(M\delta)/\epsilon \approx 0$,
since $\lim_{x \rightarrow 0} f(x) =0$. Consequently,
$\frac{ 4f(M\delta) ML}{\epsilon} +f(M\delta+\epsilon)$ is small, and
Theorem  \ref{theo:fpNEW} establishes convergence of empirical frequencies to a small neighborhood of equilibria.
 Thus, we conclude that
 for games that are close to potential games, i.e.,
  for  $ \delta \ll 1$,  Theorem  \ref{theo:fpNEW} establishes convergence of empirical frequencies to a small neighborhood of equilibria.

\section{Conclusions} \label{sec:conclusions}

In this paper, we  present a framework for studying the limiting
behavior of adaptive learning dynamics in finite strategic form
games by exploiting their relation to  nearby potential games. We
restrict our attention to better/best response, logit response and
fictitious play dynamics. We show that for near-potential games trajectories of
better/best response dynamics converge to $\epsilon$-equilibrium
sets, where $\epsilon$ depends on closeness to a potential game.
 We study the
stochastically stable strategy profiles of logit response dynamics
and prove that they are contained in the set of strategy profiles
that approximately maximize the potential function of a  nearby
potential game.
In the case of fictitious play we focus on the empirical frequencies of players' actions, and establish that they converge to a small neighborhood of equilibria in near-potential games.
Our results suggest that games that are close to a
potential game inherit the dynamical properties (such as convergence
to approximate equilibrium sets) of potential games. Additionally,
since a close potential game to a given game can be found by solving
a convex optimization problem, as discussed in Section \ref{sec:projection}, this
enables us to study dynamical properties of strategic form games by
first identifying a nearby potential game to this game, and then
studying the dynamical properties of the nearby potential game.

The framework presented in this paper opens up a number of interesting research directions.
Among them, we mention the following:

\paragraph{Heterogeneous update rules:} In this paper we only analyzed the update rules in which players update their strategies using the same mechanism. For instance, we assumed that all players adopt best response, or logit response dynamics with the same parameter.
The limiting behavior of dynamic processes, where players adhere to different update rules is still an open question, even for potential games.
An interesting future research question is whether the techniques in this paper can be used to understand the limiting behavior of such update rules. For example, consider a potential game  where all players update their strategies using logit response  with different but ``close'' $\tau$ parameters. Can the outcome of this dynamic process be approximated with the outcome of logit response in a close potential game where all players use the same parameter for their updates?

\paragraph{Guaranteeing desirable limiting behavior:}
Another promising research direction is to use our understanding of  simple update rules, such as better/best response and logit response dynamics to
design mechanisms
 that guarantee desirable limiting behavior, such as low efficiency loss and ``fair'' outcomes.
It is well known that  equilibria in games can be very different in terms of such properties \citep{roughgarden2005sra}.
Hence, it is of interest to develop  update rules that  converge to a particular equilibrium, thus providing equilibrium refinement in the limit, or to find mechanisms that modify the underlying game in a way that can induce desirable limiting behavior.
It has been shown in some cases that  simple pricing mechanisms can ensure convergence to desirable equilibria in near-potential games \citep{Candogan2009Pricing}.
It is an interesting research direction to extend such mechanisms to general games.

\paragraph{Dynamics in ``near'' zero-sum and supermodular games:}
Dynamical properties of  simple update rules in zero-sum games and supermodular games are also well understood
\citep{Shamma:2004p1450,milgrom1990rationalizability}.
If  a game is close to a zero-sum game or a supermodular game, does it still inherit some of the dynamical properties of the original game?
If such ``continuity'' properties do not hold, then the results on dynamical properties of these classes of games may be fragile.
Hence, it would be interesting to
investigate whether
analogous results to the ones in this paper
can be established
for these classes of games.

\bibliographystyle{elsarticle-harv}
\bibliography{IEEEabrv,dynNearPotNew}

\appendix

\section{Proofs of Section \ref{sec:fictitiousP}}
\paragraph{Proof of Lemma \ref{lem:mixedEqNew_1}:}
(i)
The mixed extension of $\nu$ can be given as in \eqref{eq:multiLinProb}:
\begin{equation*}
\nu(\mathbf{x})=\sum_{{\bf p}\in E} \nu ({\bf p}) \prod_{k\in{\cal M}} x^k({{ p}^k}).
\end{equation*}
Hence $\nu(\mathbf{x})$ is equal to the sum of Lipschitz continuous functions $\{\nu ({\bf p}) \prod_{k\in{\cal M}} x^k({{ p}^k})\}_{\mathbf{p}\in E}$.
The claim follows since, as a function of $\mathbf{x}\in \prod_{m\in {\cal M}} \Delta E^m$,
  the Lipschitz constant of \\ $\nu ({\bf p}) \prod_{k\in{\cal M}} x^k({{ p}^k})$ is bounded by $\nu ({\bf p}) \sqrt{M} \leq \nu ({\bf p}) {M}$.

 (ii)  Let $\nu: \prod_{m\in {\cal M}} \Delta E^m \rightarrow \mathbb{R}$ be a function such that
  \begin{equation} \label{eq:nuMap_new}
  \nu(\mathbf{x})= -\max_{m\in{\cal M}, p^m \in E^m}  \left( u^m(p^m, \mathbf{x}^{-m})- u^m(x^m, \mathbf{x}^{-m}) \right).
  \end{equation}
  It follows from the definition of $\epsilon$-equilibrium that a strategy profile $\mathbf{x}$
  is an $\epsilon$-equilibrium if and only if $\nu(\mathbf{x})\geq -\epsilon$.

By (i), it follows that mixed extensions of utility functions are Lipschitz continuous.
 Thus, the difference $ u^m(p^m, \mathbf{x}^{-m})- u^m(x^m, \mathbf{x}^{-m}) $ is Lipschitz continuous in $\mathbf{x}$.
 Since $\nu$  is obtained from maximum of finitely many such functions, we conclude that it  is Lipschitz continuous with some constant $L$.
It follows that if $||\mathbf{x}-\mathbf{y}||<\theta$, then $|\nu(\mathbf{x}) - \nu(\mathbf{y})| < L \theta$. Thus, choosing $\theta< \gamma/L$, and recalling that
  $\mathbf{x}$
 is an $\epsilon$-equilibrium if and only if $\nu(\mathbf{x})\geq -\epsilon$, the claim follows.
 \qed

\paragraph{Proof of Lemma \ref{lem:mixedEqNew_2}:}
(i) Consider the graph of $g$, i.e.,
$S=\{(v,\mathbf{x}) | v\in \mathbb{R},  \mathbf{x}\in g(v)  \}$.
The definition of $g$ suggests that $S$ can alternatively be written as
\begin{equation}
S=\left\{ \left. (v,\mathbf{x})\in \mathbb{R} \times \prod_{m\in {\cal M}} \Delta E^m \right|   \nu(\mathbf{x}) \geq -v
  \right\}
\end{equation}
Since $\nu$ is upper semicontinuous,
the function $h: \mathbb{R} \times \prod_{m\in {\cal M}} \Delta E^m$, such that
$h(v,\mathbf{x})=\nu(\mathbf{x})+v$, is also an upper semicontinuous function. Since upper level sets of upper semicontinuous functions are closed (see \citet{berge1963topological}),  the set $\left\{(v,\mathbf{x}) \in \mathbb{R} \times \prod_{m\in {\cal M}} \Delta E^m | h(v,\mathbf{x}) \geq 0 \right\}$, or equivalently $S$ is a closed subset of $\mathbb{R} \times \prod_{m\in {\cal M}} \Delta E^m$. Thus, the graph of $g$ is closed, and since $\prod_{m\in {\cal M}} \Delta E^m$ is a compact set the claim follows from Definition \ref{def:uscCor}.

(ii)
Let $\nu: \prod_{m\in {\cal M}} \Delta E^m \rightarrow \mathbb{R}$ be as in \eqref{eq:nuMap_new}.  i.e., $$  \nu(\mathbf{x})= -\max_{m\in{\cal M}, p^m \in E^m}  \left( u^m(p^m, \mathbf{x}^{-m})- u^m(x^m, \mathbf{x}^{-m}) \right).$$
 As explained in the proof of Lemma \ref{lem:mixedEqNew_1}, $\mathbf{x}$ is an $\epsilon$-equilibrium if and only if $\nu(\mathbf{x}) \geq -\epsilon$.
The claim follows from part (i) by noting that $\nu$ is a continuous function, and $g(\alpha)={\cal X}_{\alpha}= \{\mathbf{x| \nu(\mathbf{x}) \geq -\alpha}\}$.
 \qed

\paragraph{Proof of Corollary \ref{cor:discreteFP1}:}
 Let $\epsilon_n= M\delta + \frac{1}{n}$ for $n\in \mathrm{Z}_+$.
Observe that since the mixed extension of the potential function is  continuous, $C$
and $C_{\epsilon_n}$ are closed sets for any $n\in \mathrm{Z}_+$.
 Since $C$ is closed $\min_{\mathbf{y}\in C } ||\mathbf{x}-\mathbf{y}||$ is well-defined for any $\mathbf{x}\in \prod_{m\in {\cal M}} \Delta E^m$.

 We claim that for any $\theta>0$ the set
\begin{equation}\label{eq:defineS}
S_\theta= \left\{ \mathbf{x}\in \prod_{m\in {\cal M}} \Delta E^m  \left|   \min_{\mathbf{y}\in C } ||\mathbf{x}-\mathbf{y}|| < \theta \right. \right\},
\end{equation}
is such that $C_{\epsilon_n} \subset S_\theta$ for some $n$.
Note that if this claim holds, then it follows from  Theorem~\ref{theo:FPConv1} that
there exists some $T_\theta$ such that for all $t>T_\theta$ we have  $\mu_t \in S_\theta$. Using the definition of $S_\theta$ given in \eqref{eq:defineS}, this implies
\begin{equation}\label{eq:supDefConv}
\limsup_{t\rightarrow \infty } ~\min_{\mathbf{x}\in C} || \mathbf{x} - \mu_t || < \theta.
\end{equation}
Moreover, since $\theta > 0$ is arbitrary, and $|| \mathbf{x} - \mu_t || \geq 0$, using \eqref{eq:supDefConv} we obtain
$$\lim_{t\rightarrow \infty }~ \min_{\mathbf{x}\in C} || \mathbf{x} - \mu_t ||=0.$$
Thus, if we prove $C_{\epsilon_n} \subset S_\theta$ for some $n$, it follows that $\mu_t$ converges to $C$.

In order to prove  $C_{\epsilon_n} \subset S_\theta$
we first obtain a certificate which can be used to guarantee that a mixed strategy profile belongs to $S_\theta$.
Then, we show that for large enough $n$ any $\mathbf{z}\in C_{\epsilon_n}$ satisfies this certificate, and hence belongs to $S_\theta$.

It follows from Lemma \ref{lem:mixedEqNew_2} (i) (by setting $\nu=\phi$ and $v= -\min_{ \mathbf{y} \in {\cal X}_{ M \delta}} \phi(\mathbf{y})$)
and definition of upper semicontinuity (Definition \ref{def:uscCor})
that
there exists $\gamma>0$
such that $\theta$ neighborhood of
$\{\mathbf{x} | \phi(\mathbf{x}) \geq \min_{ \mathbf{y} \in {\cal X}_{ M \delta}} \phi(\mathbf{y})\}$
contains
$\{\mathbf{x} | \phi(\mathbf{x}) \geq \min_{ \mathbf{y} \in {\cal X}_{ M \delta}} \phi(\mathbf{y}) -\gamma\}$. Hence,
 for any
  $\mathbf{z}$ satisfying $\phi(\mathbf{z}) \geq
\min_{ \mathbf{y} \in {\cal X}_{ M \delta}}  \phi (\mathbf{y}) -\gamma $
there exists some
$\mathbf{x}$ satisfying $\phi(\mathbf{x}) \geq \min_{ \mathbf{y} \in {\cal X}_{ M \delta}} \phi(\mathbf{y})$
and
  $|| \mathbf{x}-\mathbf{z}||<\theta$.
Note that the definition of $S_\theta$ implies that $\mathbf{z}$ for which there exists such  $\mathbf{x}$   belongs to  $S_\theta$. Thus, if $\phi(\mathbf{z}) \geq
\min_{ \mathbf{y} \in {\cal X}_{ M \delta}}  \phi (\mathbf{y}) -\gamma$ it follows that $\mathbf{z} \in S_\theta$.

We next show that for large enough $n$, any $\mathbf{z}$ which belongs to $C_{\epsilon_n}$, satisfies the above certificate and hence belongs to $S_\theta$.
Let $L$ denote the Lipschitz constant for the mixed extension of $\phi$, as given in  Lemma \ref{lem:mixedEqNew_1} (i), and define $\theta'=\gamma/L>0$.
Lemma \ref{lem:mixedEqNew_2} (ii)  and Definition \ref{def:uscCor} imply that for large enough $n$,
${\cal X}_{M\delta +\frac{1}{n}}$ is contained in $\theta'$ neighborhood of ${\cal X}_{M\delta }$, i.e.,
  if $\mathbf{y}\in{\cal X}_{M\delta +\frac{1}{n}}$ then there exists $\mathbf{x}\in {\cal X}_{M\delta}$ such that $|| \mathbf{x}-\mathbf{y}||< \theta'$.
Moreover, by Lemma \ref{lem:mixedEqNew_1} (i), it  follows that  $\phi(\mathbf{y}) \geq \phi(\mathbf{x})- L \theta'= \phi(\mathbf{x})- \gamma$. Thus, we conclude
that there exists  large enough $n$ such that
\begin{equation}\label{eq:relationOfMin}
\min_{ \mathbf{y} \in {\cal X}_{ M \delta+1/n}} \phi(\mathbf{y}) \geq
\min_{ \mathbf{y} \in {\cal X}_{ M \delta}}  \phi (\mathbf{y}) - \gamma.
\end{equation}

Let $\mathbf{z} \in C_{\epsilon_n}$ for some $n$ for which \eqref{eq:relationOfMin} holds. By definition of $C_\epsilon$ it follows that $\phi(\mathbf{z}) \geq \min_{ \mathbf{y} \in {\cal X}_{ M \delta+1/n}} \phi(\mathbf{y}) $. Thus,
\eqref{eq:relationOfMin} implies that
$\phi(\mathbf{z}) \geq \min_{ \mathbf{y} \in {\cal X}_{ M \delta}}  \phi (\mathbf{y}) - \gamma$. However, as argued before such $\mathbf{z}$ belong to $S_\theta$.
Hence, we have established that for  large enough $n$, if $\mathbf{z}\in C_{\epsilon_n}$ then $\mathbf{z}\in S_\theta$. Therefore, the claim follows.
\qed

\begin{figure}[htbp]
\begin{center}
 \resizebox{5cm}{!}{\input{definitionSizeF.tex}}
\caption{Consider a game with a unique equilibrium $\mathbf{x}_k$. The $\alpha$-equilibrium set of the game (enclosed by a solid line around $\mathbf{x}_k$) is contained in the $f(\alpha)$ neighborhood of this equilibrium.}
\label{fig:fDef}
\end{center}
\end{figure}
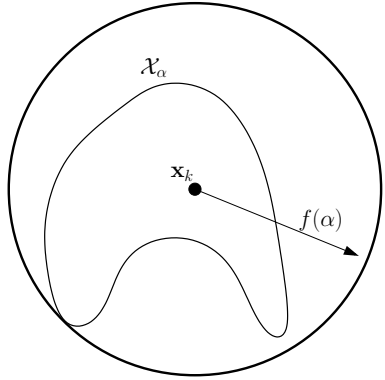

\paragraph{Proof of Theorem \ref{theo:fpNEW}:}
Assume that $\cal G$ has $l$ equilibria, denoted by $\mathbf{x}_1, \dots, \mathbf{x}_l$. Define the minimum pairwise distance between the equilibria as $d\triangleq \min_{i\neq j} ||\mathbf{x}_i - \mathbf{x}_j ||$.
Let $f: \mathbb{R}_+ \rightarrow \mathbb{R}_+ $ be a function such that
\begin{equation} \label{eq:distanceFcn}
f(\alpha)= \max_{\mathbf{x}\in {\cal X}_{\alpha}} \min_{k\in \{1,\dots, l\}} || \mathbf{x} - \mathbf{x}_k ||,
\end{equation}
for all $\alpha \in  \mathbb{R}_+$. Note that $\min_{k\in \{1,\dots, l\}} || \mathbf{x} - \mathbf{x}_k ||$ is   continuous in $\mathbf{x}$, since it is minimum of finitely many continuous functions. Moreover, $ {\cal X}_\alpha$ is a compact set, since $\epsilon$-equilibria are defined by finitely many inequality constraints of the form
\eqref{eq:defEpsEq}. Therefore,  in \eqref{eq:distanceFcn} maximum is achieved and $f$ is well-defined for all $\alpha \geq 0$.
From the definition of $f$, it follows that the union of closed balls of radius $f(\alpha)$, centered at equilibria, contain $\alpha$-equilibrium set of the game.
Thus, intuitively, $f(\alpha)$ captures the size of a closed neighborhood of equilibria, which contains $\alpha$-equilibria of the underlying game. This is illustrated in Figure \ref{fig:fDef}.

Let $a>0$ be such that $f(a)<d/4$, i.e.,  every $a$-equilibrium is at most  $d/4$ distant from an equilibrium of  a game.
Lemma \ref{lem:mixedEqNew_2} (ii) implies (using upper semicontinuity at $\alpha=0$)  that such $a$ exists.
Since $d$
is defined as the minimum pairwise distance between the equilibria, it follows that
$a$-equilibria of the game are contained in disjoint $f(a)<d/4$ neighborhoods around equilibria of the game, i.e., if $\mathbf{x}\in {\cal X}_{a}$, then $||\mathbf{x} -\mathbf{x}_k|| \leq f(a)$ for exactly one equilibrium $\mathbf{x}_k$.
Moreover, for  $a_1 \leq a$, since ${\cal X}_{a_1} \subset {\cal X}_a$, it follows that $a_1$-equilibria of the game are contained in disjoint neighborhoods of equilibria.

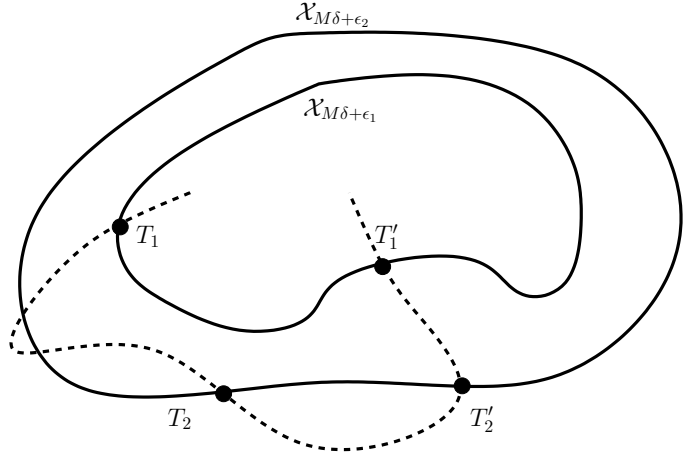
\begin{figure}[htbp]
\begin{center}
 \resizebox{9cm}{!}{\input{xfigDynTrajectory2.tex}}
\caption{Trajectory of $\mu_t$ (initialized at the left end of the dashed line) is illustrated. $T_1$ and $T_2$ correspond to the time instants $\mu_t$ leaves ${\cal X}_{M\delta+\epsilon_1}$ and ${\cal X}_{M\delta+\epsilon_2} $ respectively.
$T_1'$ and $T_2'$ correspond to the time instants $\mu_t$ enters ${\cal X}_{M\delta+\epsilon_1}$ and ${\cal X}_{M\delta+\epsilon_2} $ respectively.
}
\label{figure:TDef}
\end{center}
\end{figure}

We prove  the theorem in 5 steps summarized below. First two steps
explore the properties of function $f$, and define $\bar{\delta}$
and $\bar{\epsilon}$ presented in the theorem statement. Last three
steps are the main steps of the proof, where we establish
convergence of fictitious play to a neighborhood of equilibria.
\begin{itemize}
\item {\em Step 1:} We first show that $f$ is (i) weakly increasing, (ii)  upper semicontinuous, and it satisfies (iii) $f(0)=0$, (iv) $f(x)\rightarrow 0$ as $x\rightarrow 0$.

\item {\em Step 2:} We show that there exists some $\bar{\delta}>0$
and $\bar{\epsilon}>0$ such that
the following inequalities hold:
\begin{equation}\label{eq:deltaReq1}
 M\bar{\delta} + \bar{\epsilon}<a,
\end{equation}
and
 \begin{equation}\label{eq:deltaReq2}
f(M\bar{\delta} + \bar{\epsilon})< \frac{(a- M \bar{\delta}) d}{24LM}.
 \end{equation}
   We will prove the statement of the theorem assuming that  $0\leq \delta<\bar{\delta}$, and establish  convergence
   to the set in \eqref{eq:TheoStatement},
   for any $\epsilon$ such that $0<\epsilon \leq \bar{\epsilon}$.
 As can be seen from the definition of $a$ and $f$ (see \eqref{eq:distanceFcn}),
 the first inequality guarantees that  $ M\bar{\delta} + \bar{\epsilon}$-equilibrium set is contained in disjoint neighborhoods of equilibria, and the second one guarantees that these neighborhoods are small.
 In Step 4, we will exploit this observation, and use
 the inequalities in \eqref{eq:deltaReq1} and \eqref{eq:deltaReq2} to establish  that  the empirical frequency distribution $\mu_t$ can visit the component of ${\cal X}_{M\delta+\bar{\epsilon}}$
contained in the neighborhood of only a single equilibrium infinitely often.

\item {\em Step 3:}
  Let  $\epsilon_1, \epsilon_2$ be such that $\epsilon_2>\epsilon_1>0$.
  Assume that (i) at some time instant $T$, $\mu_t$ is contained in ${\cal X}_{M\delta+\epsilon_1}$, (ii)
  at time instants $T_1$ and $T_2$ (such that
  $T_2>T_1> T $)  $\mu_t$ leaves ${\cal X}_{M\delta+\epsilon_1}$ and  ${\cal X}_{M\delta+\epsilon_2}$ respectively and (iii) at time instants $T_2'$ and $T_1'$ (such that
    $T_1'>T_2'> T_2 $)  $\mu_t $  returns back to  ${\cal X}_{M\delta+\epsilon_2}$ and $ {\cal X}_{M\delta+\epsilon_1}$ respectively.
    In Figure~\ref{figure:TDef},
  the path $\mu_t$ follows between $T_1$ and $T_1'$ is illustrated.

  In this step,
  we provide a lower bound  on $\phi(\mu_{T_1'})-\phi(\mu_{T_1})$,  i.e.,
      the increase in the potential when $\mu_t$ follows such a path.
This  lower bound   holds for any  $\epsilon_1$ and $\epsilon_2$ provided that $\epsilon_2>\epsilon_1>0$.
   We use this result by choosing different  values for   $\epsilon_1$ and $\epsilon_2$  in Steps 4 and 5.

    Our lower bound  in Step 3 is a function of  ${\epsilon_2}$.
  In addition to this lower bound, in Steps 4 and 5, we use the ${M\delta+\epsilon_1}$ equilibrium set and Lipschitz continuity of the potential to provide an upper bound on $\phi(\mu_{T_1'})-\phi(\mu_{T_1})$ as a function of $\epsilon_1$.
  Thus, properties of ${M\delta+\epsilon_1}$ and ${M\delta+\epsilon_2}$ equilibrium sets
  are exploited for obtaining upper and lower bounds on $\phi(\mu_{T_1'})-\phi(\mu_{T_1})$ respectively.
We establish convergence of fictitious play updates to a neighborhood of an equilibrium by using these bounds together in Steps 4 and 5.
    We emphasize that allowing for two different approximate equilibrium sets
  leads to better bounds on   $\phi(\mu_{T_1'})-\phi(\mu_{T_1})$, and  a more informative characterization of the limiting behavior of fictitious play,
  as opposed to using a single approximate equilibrium set, i.e., setting $\epsilon_1=\epsilon_2$.

 \item {\em Step 4:}
 Our objective in this step is to establish that fictitious play can visit the component of an approximate equilibrium set contained in the
  neighborhood of  only one equilibrium infinitely often.

  Let $\epsilon_1=\bar{\epsilon}$ and $\epsilon_2=a-M\bar{\delta}$.
 By \eqref{eq:deltaReq1}
we have $\epsilon_1<\epsilon_2$, and using the definition of $a$
  we establish that ${\cal X}_{M\delta + \epsilon_1}$ and ${\cal X}_{M\delta+\epsilon_2}$ are contained in disjoint neighborhoods of equilibria.
  Assume that $\mu_t$ leaves the components of ${\cal X}_{M\delta + \epsilon_1}$ and ${\cal X}_{M\delta+\epsilon_2}$ in the neighborhood of equilibrium $\mathbf{x}_{k}$, and reaches to a similar neighborhood around equilibrium $\mathbf{x}_{k'}$.
  Using  Step 3
  we establish a lower bound on the increase in the potential when $\mu_t $ follows such a trajectory.
  We also provide an upper bound,
using the Lipschitz continuity of the potential and inequalities   \eqref{eq:deltaReq1} and \eqref{eq:deltaReq2}. Comparing these bounds, we establish that
  the maximum potential in the neighborhood of equilibrium $\mathbf{x}_{k}$ is lower than the minimum potential in the neighborhood of $\mathbf{x}_{k'}$.
  Since,  $\mathbf{x}_k$ and $\mathbf{x}_{k'}$ are arbitrary,
this observation implies that $\mu_t$ cannot visit  the component of
 ${\cal X}_{M\delta + \epsilon_1}$ contained in
the neighborhood of $\mathbf{x}_{k}$
at a later time instant. Hence, it follows that $\mu_t$ visits only one such component infinitely often.

 \item {\em Step 5:} In this step we show that $\mu_t$ converges to the approximate equilibrium set given in the theorem statement.

Let   $\epsilon_1, \epsilon_2$ be such that
$0<\epsilon_1<\epsilon_2\leq \bar{\epsilon}$. We consider the
equilibrium,  whose neighborhood is visited infinitely often (as
obtained in Step 4), and a trajectory of $\mu_t$  which leaves the
components of ${\cal X}_{M\delta + \epsilon_1}$ and ${\cal
X}_{M\delta + \epsilon_2}$  contained in the neighborhood of this
equilibrium and returns back to these sets at a later time instant
(as illustrated in Figure \ref{figure:TDef}). As in Step 4,
Lipschitz continuity of $\phi$ is used to obtain an upper bound on
the  increase in the potential between the end points of this
trajectory. Together with the lower bound obtained in Step 3,  this
provides a bound on how far $\mu_t$ can get from the component of
${\cal X}_{M\delta + \epsilon_2}$ contained in this neighborhood.
Choosing $\epsilon_1$ arbitrarily small  (for a fixed $\epsilon_2$)
we obtain the tightest such bound. Using this result, we quantify
how far $\mu_t$ can get from the equilibria of the game (after
sufficient time) and the theorem follows.

\end{itemize}
Next we prove each of these steps.

\paragraph{Step 1:}
By definition ${\cal X}_{\alpha_1} \subset {\cal X}_{\alpha}$ for any $\alpha_1 \leq \alpha$.
Since the feasible set of the maximization problem in \eqref{eq:distanceFcn} is given by ${\cal X}_\alpha$, this implies that $f(\alpha_1) \leq f(\alpha)$, i.e., $f$ is a weakly increasing function of its argument.
Note that the feasible set of the maximization problem in \eqref{eq:distanceFcn} can be given by the correspondence $g(\alpha)={\cal X}_\alpha$, which is upper semi continuous in $\alpha$ as shown in Lemma \ref{lem:mixedEqNew_2} (ii).
Since as a function of $\mathbf{x}$, $\min_{k\in \{1,\dots, l\}} || \mathbf{x} - \mathbf{x}_k ||$ is continuous it follows from Berge's maximum theorem (see \citet{berge1963topological}) that for $\alpha\geq 0$, $f(\alpha)$ is an upper semicontinuous function.

The set ${\cal X}_0$ corresponds to the set of equilibria of the game, hence ${\cal X}_0= \{\mathbf{x}_1,\dots,\mathbf{x}_l\}$. Thus, the definition of $f$ implies that $f(0)=0$.
Moreover, upper semicontinuity of $f$ implies that for any $\epsilon>0$, there exists  some neighborhood  $V$ of $0$, such that $f(x) \leq  \epsilon$ for all $x\in V$. Since,
 $f(x) \geq 0$ by definition, this implies that $\lim_{x\rightarrow 0} f(x)$ exists and equals to $0$.

\paragraph{Step 2:}
Let $\bar{\delta}>0$ be small enough such that $M\bar{\delta}< a/2$.
Since $\lim_{x\rightarrow 0} f(x)=0$, it follows that for sufficiently small $\bar{\delta}$ and $\bar{\epsilon}$, we obtain $f(M\bar{\delta} +\bar{\epsilon} ) < \frac{ad}{48 LM}
<\frac{(a-M\bar{\delta})d}{24 LM}$ and $ M\bar{\delta} + \bar{\epsilon}<a$.

\paragraph{Step 3:}
Let $\epsilon_1, \epsilon_2$ be such that $0<\epsilon_1< \epsilon_2  $.
Assume $T>0$ is large enough so that for $t>T$,
\begin{equation} \label{eq:incBounds}
\begin{aligned}
&\mbox{$\phi(\mu_{t+1}) - \phi(\mu_t) \geq \frac{2\epsilon_1}{3(t+1)}$ if $\mu_t \notin {\cal X}_{M\delta+\epsilon_1}$, and similarly }\\
&\mbox{$\phi(\mu_{t+1}) - \phi(\mu_t) \geq \frac{2\epsilon_2}{3(t+1)}$ if $\mu_t \notin {\cal X}_{M\delta+\epsilon_2}$. }
\end{aligned}
\end{equation}
Existence of $T$ satisfying these inequalities follows from Lemma \ref{lem:potIncEpsEq},
since for large $T$ and $t>T$, this lemma implies $\phi(\mu_{t+1}) - \phi(\mu_t) \geq
\frac{\epsilon_1}{(t+1)} + O\left(\frac{1}{t^2}\right) \geq \frac{2\epsilon_1}{3(t+1)}$  if $\mu_t \notin {\cal X}_{M\delta+\epsilon_1}$, and similarly if $\mu_t \notin {\cal X}_{M\delta+\epsilon_2}$.

Since $\phi(\mu_t)$ increases outside $M\delta+\epsilon_1$-equilibrium set for $t>T$, as \eqref{eq:incBounds} suggests, it follows that $\mu_t$ visits  ${\cal X}_{M\delta+\epsilon_1}$ (and ${\cal X}_{M\delta+\epsilon_2}$ since ${\cal X}_{M\delta+\epsilon_1} \subset {\cal X}_{M\delta+\epsilon_2}$) infinitely often.  Otherwise $\phi(\mu_t)$  increases  unboundedly, and we reach a contradiction since mixed extension of the potential is a bounded function.

Assume that at some time after $T$, $\mu_t$ leaves   ${\cal X}_{M\delta+\epsilon_1}$  and
${\cal X}_{M\delta+\epsilon_2}$
and returns back to ${\cal X}_{M\delta+\epsilon_1}$ at a later time instant.
In this step, we quantify how much the potential increases when $\mu_t$ follows such a path.
We first
define time instants
$T_1$, $T_2$, $T_1'$, and $T_2'$  satisfying
$T<T_1\leq T_2<T_2'\leq T_1'$,
as follows:
\begin{itemize}
\item $T_1$ is a time instant when $\mu_t$ leaves ${\cal X}_{M\delta+\epsilon_1}$, i.e.,
$\mu_{T_1-1}\in {\cal X}_{M\delta+\epsilon_1}$ and $\mu_t \notin {\cal X}_{M\delta+\epsilon_1}$
for $T_1\leq t<T_1'$.
\item  $T_2$ is a time instant when $\mu_t$ leaves ${\cal X}_{M\delta+\epsilon_2}$, i.e.,
$\mu_{T_2-1}\in {\cal X}_{M\delta+\epsilon_2}$ and $\mu_t \notin {\cal X}_{M\delta+\epsilon_2}$
for $T_2\leq t<T_2'$.
\item $T_2'$ is the
first time  instant
after $T_2$
when $\mu_t$ returns back to
 ${\cal X}_{M\delta+\epsilon_2}$, i.e.,
$\mu_{T_2'-1}\notin {\cal X}_{M\delta+\epsilon_2}$ and $\mu_{T_2'} \in {\cal X}_{M\delta+\epsilon_2}$.

\item $T_1'$ is the first time instant after $T_1$ when $\mu_t$ returns back to
 ${\cal X}_{M\delta+\epsilon_1}$, i.e.,
$\mu_{T_1'-1}\notin {\cal X}_{M\delta+\epsilon_1}$ and $\mu_{T_1'} \in {\cal X}_{M\delta+\epsilon_1}$.
\end{itemize}
The definitions are illustrated in Figure \ref{figure:TDef}.
We next provide a lower bound on the quantity $\phi(\mu_{T_1'})-\phi(\mu_{T_1})$.
 Note that if there are multiple time instants between $T_1$ and $T_1'$ for which  $\mu_t$ leaves ${\cal X}_{M\delta+\epsilon_2}$ (as in the figure), any of these time instants can be chosen as $T_2$ to obtain a lower bound.

By definition, for $t$ such that $T_2 \leq t < T_2'$,  we have $\mu_t \notin  {\cal X}_{M\delta+\epsilon_2}$, and for $t$ such that $T_1 \leq t < T_2$ or $T_2'\leq t< T_1'$, we have $\mu_t \notin  {\cal X}_{M\delta+\epsilon_1}$. Thus, it follows from   \eqref{eq:incBounds} that
\begin{equation}
\phi(\mu_{t+1})-\phi(\mu_{t}) \geq  \frac{2\epsilon_2}{3(t+1)} \qquad \mbox{ for $T_2 \leq t < T_2'$},
\end{equation}
and consequently,
\begin{equation} \label{eq:phiNewB1}
\phi(\mu_{T_2'})-\phi(\mu_{T_2}) = \sum_{t=T_2}^{T_2'-1} \phi(\mu_{t+1}) -\phi(\mu_t)
\geq \sum_{t=T_2}^{T_2'-1} \frac{2\epsilon_2}{3(t+1)}.
\end{equation}
Similarly,  since   $\mu_t \notin  {\cal X}_{M\delta+\epsilon_1}$ for $t$ such that $T_1 \leq t < T_2$ or $T_2'\leq t< T_1'$, using \eqref{eq:incBounds}
we establish
\begin{equation}\label{eq:phiNewB2}
\phi(\mu_{T_1'})-\phi(\mu_{T_2'})
= \sum_{t=T_2'}^{T_1'-1} \phi(\mu_{t+1}) -\phi(\mu_t)
\geq \sum_{t=T_2'}^{T_1'-1} \frac{2\epsilon_1}{3(t+1)},
\end{equation}
\begin{equation}\label{eq:phiNewB3}
\phi(\mu_{T_2})-\phi(\mu_{T_1})
= \sum_{t=T_1}^{T_2-1} \phi(\mu_{t+1}) -\phi(\mu_t)
 \geq \sum_{t=T_1}^{T_2-1} \frac{2\epsilon_1}{3(t+1)}.
\end{equation}
Since $\phi(\mu_{T_1'})-\phi(\mu_{T_1})=
\left(  \phi(\mu_{T_1'})-\phi(\mu_{T_2'}) \right)+ \left( \phi(\mu_{T_2'})-\phi(\mu_{T_2})\right)+
\left( \phi(\mu_{T_2})-\phi(\mu_{T_1})  \right)$, it follows
from \eqref{eq:phiNewB1}, \eqref{eq:phiNewB2} and \eqref{eq:phiNewB3}
that
 \begin{equation} \label{eq:increasePot}
 \phi(\mu_{T_1'})-\phi(\mu_{T_1}) \geq
  \sum_{t=T_2}^{T_2'-1} \frac{2\epsilon_2}{3(t+1)}.
 \end{equation}

 \paragraph{Step 4:}
 Let $\epsilon_2=a-M\bar{\delta}$, and $\epsilon_1=\bar{\epsilon}$. By definition of $\bar{\epsilon}$ and $\bar{\delta}$ (see Step 2), it follows that $\epsilon_2>\epsilon_1 >0$. Assume that $\delta<\bar{\delta}$.
Since $a=  M \bar{\delta} +{\epsilon_2}
 > M \delta +{\epsilon_2} > M\delta+ {\epsilon_1}$
we obtain
${\cal X}_{M\delta+\epsilon_1} \subset {\cal X}_{M\delta+\epsilon_2} \subset {\cal X}_a$.
By definition of $a$,  ${\cal X}_a$ is contained in disjoint neighborhoods of equilibria. Thus, it follows that
components of $ {\cal X}_{M\delta+\epsilon_1}$ and ${\cal X}_{M\delta+\epsilon_2}$ are also contained in disjoint neighborhoods of equilibria.
Hence, the definition of $f$ suggests that
if $\mathbf{x} \in {\cal X}_{M\delta+\epsilon_1}$
then $|| \mathbf{x}_k - \mathbf{x}|| \leq f({M\delta+\epsilon_1})$ (similarly if $\mathbf{x} \in {\cal X}_{M\delta+\epsilon_2}$ , then
 $|| \mathbf{x}_k - \mathbf{x}|| \leq f({M\delta+\epsilon_2})$) for exactly one
equilibrium $\mathbf{x}_k$.

Let $T_1, T_2, T_1'$ and $ T_2'$ be defined as in Step 3.
In this step,
by obtaining an upper bound on $ \phi(\mu_{T_1'})-\phi(\mu_{T_1})$ and refining the lower bound obtained in Step 3 for given values of $\epsilon_1$ and $\epsilon_2$,  we  prove that after sufficient time $\mu_t$ can visit the  component of
 ${\cal X}_{M\delta + \epsilon_1}$
in the neighborhood   of a single equilibrium.

 Assume that $\mu_t$ leaves the component of the ${M\delta + \epsilon_1}$-equilibrium set in the neighborhood of equilibrium $\mathbf{x}_{k}$, and it reaches to another component in the neighborhood of  equilibrium $\mathbf{x}_{k'}$.
Since, by definition $\mu_{T_1-1}, \mu_{T_1'} \in {\cal X}_{M\delta + \epsilon_1}$,
and $\mu_{T_2-1}, \mu_{T_2'} \in {\cal X}_{M\delta + \epsilon_2}$,
it follows that
   $\mu_{T_1-1}$ and $\mu_{T_2-1}$  belong to  neighborhoods of equilibrium  $\mathbf{x}_{k}$, whereas,
$\mu_{T_1'}$ and $\mu_{T_2'}$ belong to  neighborhoods of
$\mathbf{x}_{k'}$,
i.e.,
\begin{align} \label{eq:xVsMu_1}
|| \mathbf{x}_{k} - \mu_{T_1-1}|| \leq f({M\delta+\epsilon_1})
\qquad
&\mbox{and}
\qquad
|| \mathbf{x}_{k} - \mu_{T_2-1}|| \leq f({M\delta+\epsilon_2}) \mbox{, whereas,} \\
 || \mathbf{x}_{k'} - \mu_{T_1'}|| \leq f({M\delta+\epsilon_1})
 \qquad
 &\mbox{and}
 \qquad
  || \mathbf{x}_{k'} - \mu_{T_2'}|| \leq f({M\delta+\epsilon_2}).  \label{eq:xVsMu_2}
\end{align}

By definition of $d$
 we have $||\mathbf{x}_k - \mathbf{x}_{k'}|| \geq d$.
 Since $a> M\delta+\epsilon_2$, it follows that
 $f(M\delta+\epsilon_2) \leq f(a) < d/4$, and hence the second  inequalities in \eqref{eq:xVsMu_1}
 and  \eqref{eq:xVsMu_2}
 imply
\begin{equation} \label{eq:muDeltaChange}
|| \mu_{T_2'}- \mu_{T_2-1}|| >\frac{d}{2}.
\end{equation}

Using this inequality,
 we next refine the lower bound
 on $\phi(\mu_{T_1'})-\phi(\mu_{T_1})$
 obtained in Step~3.
 By \eqref{eq:empFreq}, with an update at time $t$,  the empirical frequency distribution can change by at most
 \begin{equation} \label{eq:lengthOfDelMu}
 || \mu_{t+1}-\mu_{t}|| = \frac{1}{t+1} ||\mu_t - I_t || \leq \frac{1}{t+1} (||\mu_t|| + || I_t ||) \leq   \frac{2M}{t+1},
 \end{equation}
 where the last inequality follows from the fact that $\mu_t=\{\mu_t^m\}_{m\in {\cal M}}$,  and $I_t=\{I_t^m\}_{m\in {\cal M}}$, and $||\mu_t^m||, ||I_t^m|| \leq 1$, since $I_t^m, \mu_t^m \in \Delta E^m$. Hence, if  $T_2$ is sufficiently large,
 then $|| \mu_{T_2} - \mu_{T_2-1} ||$ is small enough so that
 \eqref{eq:muDeltaChange} implies
 $|| \mu_{T_2'}- \mu_{T_2}|| >\frac{d}{2}.$
Using this together with \eqref{eq:lengthOfDelMu}, we conclude
 \begin{equation} \label{eq:distIdentity}
 \sum_{t=T_2}^{T_2'-1} \frac{2M}{t+1}\geq
 \sum_{t=T_2}^{T_2'-1}|| \mu_{t+1}-\mu_t|| \geq
 ||\sum_{t=T_2}^{T_2'-1} \mu_{t+1}-\mu_t|| =
  || \mu_{T_2'}- \mu_{T_2}|| >\frac{d}{2}.
 \end{equation}
Thus, the lower bound on  $ \phi(\mu_{T_1'}) - \phi(\mu_{T_1})$ provided in
 \eqref{eq:increasePot} takes the following form:
 \begin{equation} \label{eq:poteMuT2T1}
 \phi(\mu_{T_1'}) - \phi(\mu_{T_1}) \geq
  \sum_{t=T_2}^{T_2'-1} \frac{2\epsilon_2}{3(t+1)} \geq \frac{\epsilon_2 d}{6M}.
 \end{equation}

Next we provide an upper bound on $\phi(\mu_{T_1'})-\phi(\mu_{T_1})$, using Lipschitz continuity of the potential and the properties of the $M\delta+\epsilon_1$ equilibrium set.
 Let
$ \overline{\phi}_k = \max_{\{\mathbf{x}~|~ ||\mathbf{x}-\mathbf{x}_k|| \leq f(M\delta + \epsilon_1) \}} \phi (\mathbf{x}),$
 and define $\mathbf{y}_k$ as a strategy profile which achieves this maximum.
 Similarly, let
$
 \underline{\phi}_{k'} = \min_{\{\mathbf{x}~|~ ||\mathbf{x}-\mathbf{x}_{k'}|| \leq f(M\delta + \epsilon_1) \}} \phi (\mathbf{x})
 $
 and define $\mathbf{y}_{k'}$ as a strategy profile which achieves this minimum.
 Observe that
 \begin{equation} \label{eq:unOvPot}
 \begin{aligned}
 \underline{\phi}_{k'}-
 \overline{\phi}_{k} &=\phi(\mathbf{y}_{k'}) -\phi(\mathbf{y}_{k}) \\
 &=
 \left(\phi(\mathbf{y}_{k'}) -\phi(\mu_{T_1'}) \right)
 +\left(\phi(\mu_{T_1'}) -\phi(\mu_{T_1}) \right)
 +\left(\phi(\mu_{T_1}) -\phi(\mathbf{y}_{k}) \right).
 \end{aligned}
 \end{equation}
 Note that by \eqref{eq:xVsMu_1} and \eqref{eq:xVsMu_2},  and  the definitions of $\mathbf{y}_{k}$ and $\mathbf{y}_{k'}$, we have
 $\mu_{T_1'}, \mathbf{y}_{k'} \in \{\mathbf{x}~|~ ||\mathbf{x}-\mathbf{x}_{k'}|| \leq f(M\delta + \epsilon_1) \}$, and
 $\mu_{T_1-1}, \mathbf{y}_{k} \in \{\mathbf{x}~|~ ||\mathbf{x}-\mathbf{x}_{k}|| \leq f(M\delta + \epsilon_1) \}$.
 Hence, using  Lipschitz continuity of $\phi$ (and denoting the Lipschitz constant by $L$) it follows that $\phi(\mathbf{y}_{k'}) -\phi(\mu_{T_1'}) \geq -2Lf(M\delta + \epsilon_1)$, and $\phi(\mu_{T_1-1})-\phi(\mathbf{y}_{k})  \geq -2Lf(M\delta + \epsilon_1)$.
Moreover, \eqref{eq:lengthOfDelMu} and Lipschitz continuity of $\phi$ imply that
$\phi(\mu_{T_1})-\phi(\mu_{T_1-1})= O\left(\frac{1}{T_1} \right)$.
 Thus, using \eqref{eq:unOvPot} we obtain the following upper bound on $\phi(\mu_{T_1'}) -\phi(\mu_{T_1}) $:
\begin{equation} \label{eq:unOvPot2}
 \begin{aligned}
 \underline{\phi}_{k'}-
 \overline{\phi}_{k}
+ 4Lf(M\delta + \epsilon_1) + O\left( \frac{1}{T_1}\right)
  \geq
\phi(\mu_{T_1'}) -\phi(\mu_{T_1}).
 \end{aligned}
 \end{equation}

Using the lower and upper bounds we obtained
in \eqref{eq:poteMuT2T1} and \eqref{eq:unOvPot2}, it follows that
\begin{equation}\label{eq:unOvPot3}
\underline{\phi}_{k'}-
 \overline{\phi}_{k}
+ 4Lf(M\delta + \epsilon_1) +O \left( \frac{1}{T_1}\right)
  \geq \frac{\epsilon_2 d}{6M}.
\end{equation}

 Since
 $\epsilon_2=a-M\bar{\delta}$, and $\epsilon_1=\bar{\epsilon}$, using the fact that $f$ is an increasing function and $\delta <\bar{\delta}$, it follows from \eqref{eq:unOvPot3}  that
  \begin{equation*}
  \begin{aligned}
  \underline{\phi}_{k'}-
  \overline{\phi}_{k}
  \geq \frac{(a-M\bar{\delta}) d}{6M} -4Lf(M\delta + \bar{\epsilon})+ O\left( \frac{1}{T_1}\right)
  \geq \frac{(a-M\bar{\delta}) d}{6M} -4Lf(M\bar{\delta} + \bar{\epsilon})+ O\left( \frac{1}{T_1}\right)
  \end{aligned}
  \end{equation*}
  Note that \eqref{eq:deltaReq2} implies  $\frac{(a-M\bar{\delta}) d}{6M} -4Lf(M\bar{\delta} + \bar{\epsilon})>0$. Thus, for sufficiently large $T_1$
we obtain
 $\underline{\phi}_{k'}-
 \overline{\phi}_{k}>0$. Therefore, we conclude when $\mu_t$ leaves the component of ${\cal X}_{M\delta + \epsilon_1}$ contained in the neighborhood of some equilibrium $\mathbf{x}_k$, and enters that of another equilibrium $\mathbf{x}_{k'}$, then the minimum potential in the new neighborhood is strictly larger than the maximum potential in the older one (for sufficiently large $T_1$).
 Since this is true for arbitrary equilibria $\mathbf{x}_k$ and $\mathbf{x}_{k'}$, it follows that
 after entering the component of
 ${\cal X}_{M\delta + \epsilon_1}$ in the
  neighborhood of $\mathbf{x}_{k'}$,
 $\mu_t$ cannot return to the component in the neighborhood of $\mathbf{x}_k$, as doing so contradicts with the
relation between the minimum and maximum potentials in these neighborhoods.
 Thus, after sufficient  time, $\mu_t$ can visit
the component of
 ${\cal X}_{M\delta + \epsilon_1}$
 (or equivalently  ${\cal X}_{M\delta + \bar{\epsilon}}$)
in the neighborhood   of a single equilibrium.

 \paragraph{Step 5:}
 Let $\epsilon_1$, and $\epsilon_2$ be such that
 $0<\epsilon_1<\epsilon_2 \leq \bar{\epsilon}$.
 As established in Step 4, there exists some $T$, such that for $t>T,$ $\mu_t$ visits the
 component of ${\cal X}_{M\delta + \bar{\epsilon}}$, in the neighborhood  of a single equilibrium, say $\mathbf{x}_k$.

Assume that $T_1, T_2, T_1'$ and $ T_2'$ are defined as in Step 3,
and  let $T_1 >T+1$. Since  $\epsilon_1<\epsilon_2 \leq
\bar{\epsilon}$,  we have ${\cal X}_{M\delta + {\epsilon}_1} \subset
{\cal X}_{M\delta + {\epsilon_2}} \subset {\cal X}_{M\delta +
\bar{\epsilon}}$, and $T_1 >T+1$ implies that $\mu_t$ can only visit
the components of ${\cal X}_{M\delta + {\epsilon}_1}$ and  ${\cal
X}_{M\delta + {\epsilon}_2}$ contained in the neighborhood of
${\mathbf{x}}_k$. Following a similar approach to Step 4, we next
obtain upper and lower bounds on $\phi(\mu_{T_1'})-\phi(\mu_{T_1})$,
and use these bounds to establish convergence to the mixed
equilibrium set given in the theorem statement.

Define $d^*$ as the maximum distance of $\mu_t$ from  ${\cal X}_{M\delta +\epsilon_2}$ for $t$ such that $T+1<T_2\leq t \leq T_2'-1 $, i.e.,
 $$d^* = \max_{\{t| T_2\leq t \leq T_2'-1 \}} \min_{\mathbf{x}\in {\cal X}_{M\delta +\epsilon_2}} || \mu_t - \mathbf{x} ||.$$
 Since $\mu_{T_2-1}, \mu_{T_2'} \in {\cal X}_{M\delta + \epsilon_2}$ by definition, the total length of the trajectory between   ${T_2-1}$ and ${T_2'}$   is an upper bound on $2d^*$, i.e.,
 \begin{equation*}
 2d^* \leq \sum_{t=T_2-1}^{T_2'-1} || \mu_{t+1} - \mu_{t} ||.
 \end{equation*}
 As explained in \eqref{eq:lengthOfDelMu}, $|| \mu_{t+1} - \mu_{t} || \leq \frac{2M}{t+1}$, thus the above inequality implies
 \begin{equation} \label{eq:boundDs}
 2d^* \leq \sum_{t=T_2-1}^{T_2'-1} \frac{2M}{t+1}=
 \sum_{t=T_2}^{T_2'-1} \frac{2M}{t+1} + \frac{2M}{T_2}.
 \end{equation}
Using this inequality, the lower bound in  \eqref{eq:increasePot} implies
 \begin{equation} \label{eq:boundeDsNew}
\phi(\mu_{T_1'})- \phi(\mu_{T_1})\geq
  \sum_{t=T_2}^{T_2'-1} \frac{2\epsilon_2}{3(t+1)}
  \geq
\left( d^* -\frac{M}{T_2} \right) \frac{2\epsilon_2}{3M}
 \end{equation}

 We next obtain an upper bound on $\phi(\mu_{T_1'})- \phi(\mu_{T_1})$.
By definition of $f$,
 ${\cal X}_{M\delta+\epsilon_1}$ is contained in $f(M\delta+\epsilon_1)$ neighborhoods of equilibria.
For $T_1> T+1$, $\mu_t$ can only visit the component of
${\cal X}_{M\delta+\epsilon_1}$ in the
 neighborhood of $\mathbf{x}_k$, as can be seen from the definition of $T$.
 Thus, since $\mu_{T_1-1},\mu_{T_1'} \in {\cal X}_{M\delta+\epsilon_1}$, it follows that
  $\mu_{T_1-1},\mu_{T_1'} \in
  \{ \mathbf{x} ~|~ ||\mathbf{x} - \mathbf{x}_k|| \leq f(M\delta+\epsilon_1)\} $.
   By Lipschitz continuity of the potential function it follows that $\phi(\mu_{T_1'})- \phi(\mu_{T_1-1}) \leq
 2f(M\delta+\epsilon_1)L$. Additionally, by \eqref{eq:lengthOfDelMu} Lipschitz continuity also implies that
$ \phi(\mu_{T_1})-\phi(\mu_{T_1-1})\leq \frac{2ML}{T_1}$. Combining these we obtain the following upper bound on $\phi(\mu_{T_1'})- \phi(\mu_{T_1})$:
\begin{equation}\label{eq:upperBFinal}
\phi(\mu_{T_1'})- \phi(\mu_{T_1}) \leq  2f(M\delta+\epsilon_1)L+ \frac{2ML}{T_1}.
\end{equation}
It follows from the upper and lower bounds on $\phi(\mu_{T_1'})- \phi(\mu_{T_1})$ given in
\eqref{eq:boundeDsNew} and \eqref{eq:upperBFinal} that
 \begin{equation*}
 \left( d^* -\frac{M}{T_2} \right) \frac{2\epsilon_2}{3M}
 \leq  2f(M\delta+\epsilon_1)L+ \frac{2ML}{T_1}
 \end{equation*}
 Thus, for sufficiently large $T_1$ (and hence $T_2$), we obtain
 \begin{equation} \label{eq:preLimSup}
 d^*
  \leq  \frac{3f(M\delta+\epsilon_1)ML}{\epsilon_2} + \frac{3M^2L}{{\epsilon_2}T_1} + \frac{M}{T_2}
 \leq  \frac{4f(M\delta+\epsilon_1) ML }{\epsilon_2}.
 \end{equation}

 Note that in the above derivation $\epsilon_1$ is an arbitrary number that satisfies $0<\epsilon_1 <\epsilon_2$. Thus, \eqref{eq:preLimSup} implies that
 \begin{equation} \label{eq:upBoundD}
 d^* \leq \limsup_{\epsilon_1 \rightarrow 0}  \frac{4f(M\delta+\epsilon_1) ML }{\epsilon_2} \leq \frac{4f(M\delta) ML }{\epsilon_2},
 \end{equation}
 where the last inequality follows by upper semicontinuity of $f$.
Thus,
by definition of $d^*$, we conclude that $\mu_t$ converges  $d^*$ neighborhood of ${\cal X}_{M\delta+\epsilon_2}$.
Hence, using  \eqref{eq:upBoundD},  we can establish convergence of $\mu_t$ to
  \begin{equation} \label{eq:theoResult1}
  \left\{\mathbf{x} \left| ~ || \mathbf{x} -\mathbf{y}|| \leq \frac{ 4f(M\delta) ML}{\epsilon_2}, \mbox{ for some }  \mathbf{y}\in {\cal X}_{M\delta+\epsilon_2} \right. \right\}.
  \end{equation}
Observe that definition of $f$ implies if $\mathbf{y} \in {\cal X}_{M\delta+\epsilon_2}$,
then for some equilibrium $\mathbf{x}_k$ we have $||\mathbf{x}_k - \mathbf{y}|| \leq f(M\delta+\epsilon_2)$.
Thus, using \eqref{eq:theoResult1} and triangle inequality,  we conclude that $\mu_t$ converges to
 \begin{equation} \label{eq:theoResult2}
  \left\{\mathbf{x} \left| ~ || \mathbf{x} -\mathbf{x}_k|| \leq \frac{ 4f(M\delta) ML}{\epsilon_2} +f(M\delta+\epsilon_2), \mbox{ for some equilibrium $\mathbf{x}_k$}  \right. \right\}.
  \end{equation}
 Noting that in   \eqref{eq:theoResult2}
  $\epsilon_2$ is an arbitrary number satisfying $0<\epsilon_2 \leq \bar{\epsilon}$, the theorem follows. \qed

\end{document}

%% file: xFigIntuitionProofV1.tex
\begin{picture}(0,0)%
\includegraphics{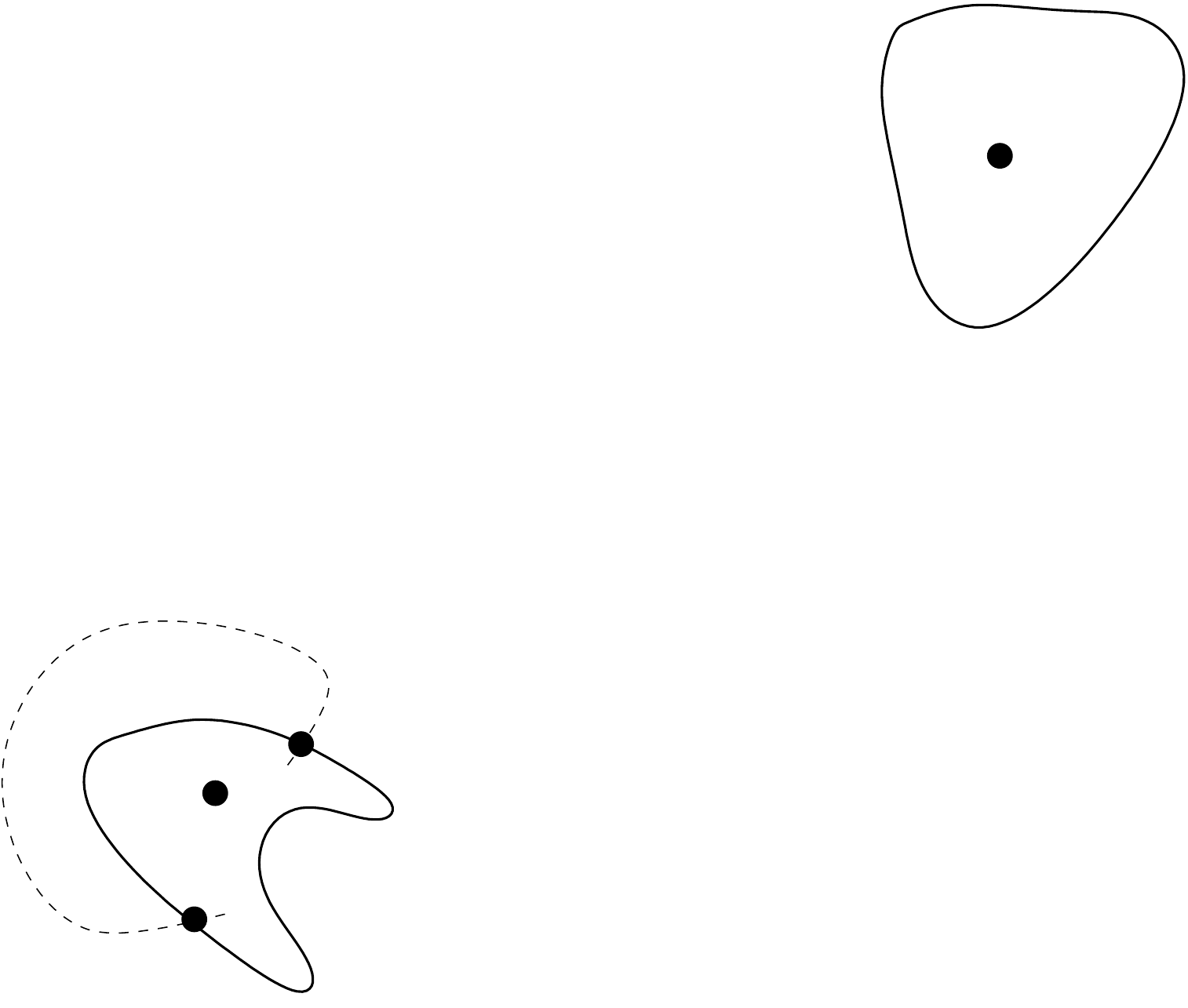}%
\end{picture}%
\setlength{\unitlength}{3947sp}%
\begingroup\makeatletter\ifx\SetFigFont\undefined%
\gdef\SetFigFont#1#2#3#4#5{%
  \reset@font\fontsize{#1}{#2pt}%
  \fontfamily{#3}\fontseries{#4}\fontshape{#5}%
  \selectfont}%
\fi\endgroup%
\begin{picture}(7264,6159)(485,49325)
\put(826,54764){\makebox(0,0)[lb]{\smash{{\SetFigFont{20}{24.0}{\familydefault}{\mddefault}{\updefault}{\color[rgb]{0,0,0}$\phi(\mu_{t'})>\phi(\mu_t)$}%
}}}}
\put(1501,50339){\makebox(0,0)[lb]{\smash{{\SetFigFont{20}{24.0}{\familydefault}{\mddefault}{\updefault}{\color[rgb]{0,0,0}$\mathbf{x}_{k}$}%
}}}}
\put(1351,49439){\makebox(0,0)[lb]{\smash{{\SetFigFont{20}{24.0}{\familydefault}{\mddefault}{\updefault}{\color[rgb]{0,0,0}$\mu_{t}$}%
}}}}
\put(6451,54764){\makebox(0,0)[lb]{\smash{{\SetFigFont{20}{24.0}{\familydefault}{\mddefault}{\updefault}{\color[rgb]{0,0,0}$\mathbf{x}_{k'}$}%
}}}}
\put(2551,50864){\makebox(0,0)[lb]{\smash{{\SetFigFont{20}{24.0}{\familydefault}{\mddefault}{\updefault}{\color[rgb]{0,0,0}$\mu_{t'}$}%
}}}}
\end{picture}%

%% file: xFigIntuitionProofV2.tex
\begin{picture}(0,0)%
\includegraphics{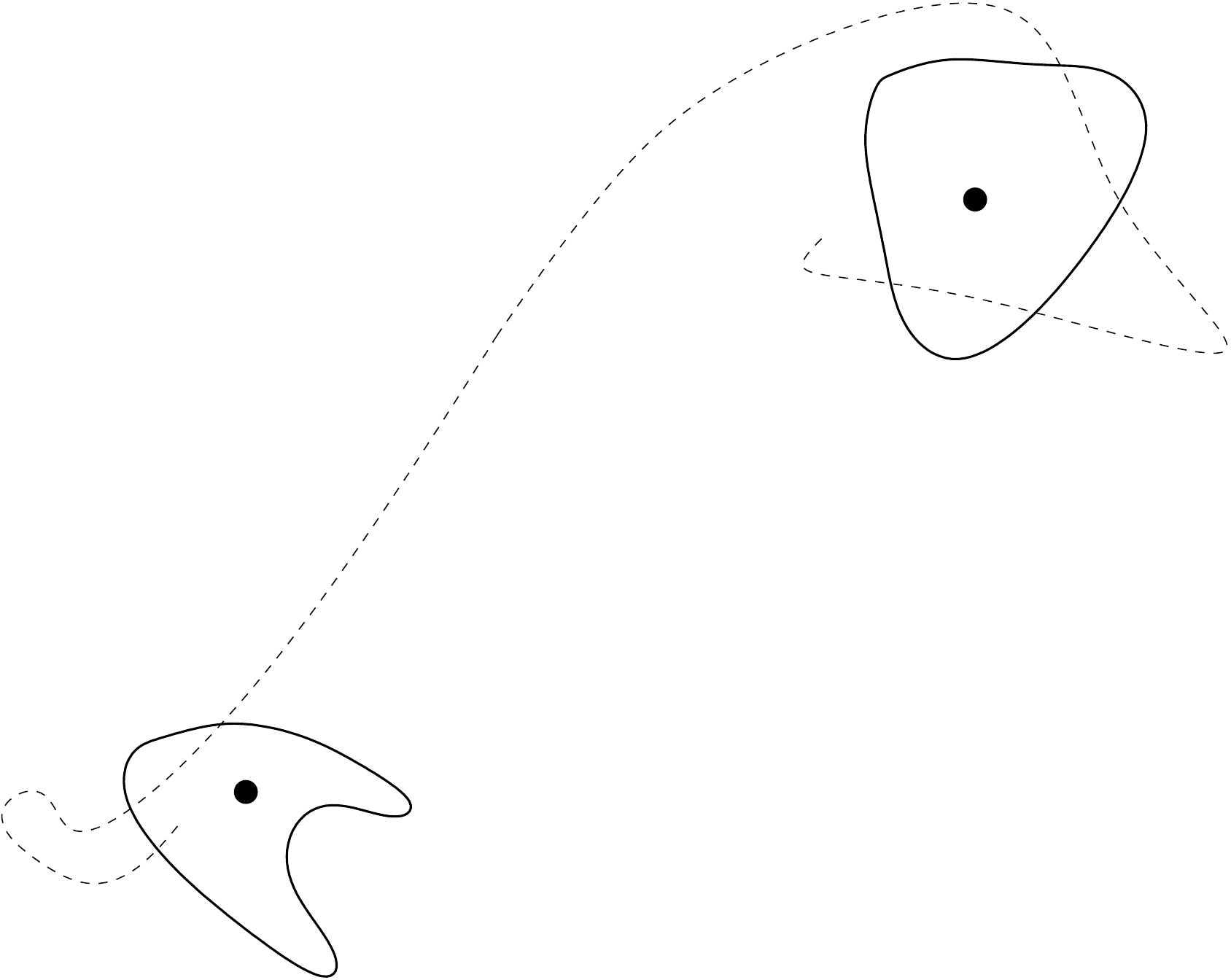}%
\end{picture}%
\setlength{\unitlength}{3947sp}%
\begingroup\makeatletter\ifx\SetFigFont\undefined%
\gdef\SetFigFont#1#2#3#4#5{%
  \reset@font\fontsize{#1}{#2pt}%
  \fontfamily{#3}\fontseries{#4}\fontshape{#5}%
  \selectfont}%
\fi\endgroup%
\begin{picture}(8088,6441)(183,49402)
\put(6451,54764){\makebox(0,0)[lb]{\smash{{\SetFigFont{20}{24.0}{\familydefault}{\mddefault}{\updefault}{\color[rgb]{0,0,0}$\mathbf{x}_{k'}$}%
}}}}
\put(1501,50339){\makebox(0,0)[lb]{\smash{{\SetFigFont{20}{24.0}{\familydefault}{\mddefault}{\updefault}{\color[rgb]{0,0,0}$\mathbf{x}_{k}$}%
}}}}
\end{picture}%

%% file: xFigIntuitionProofv4.tex
\begin{picture}(0,0)%
\includegraphics{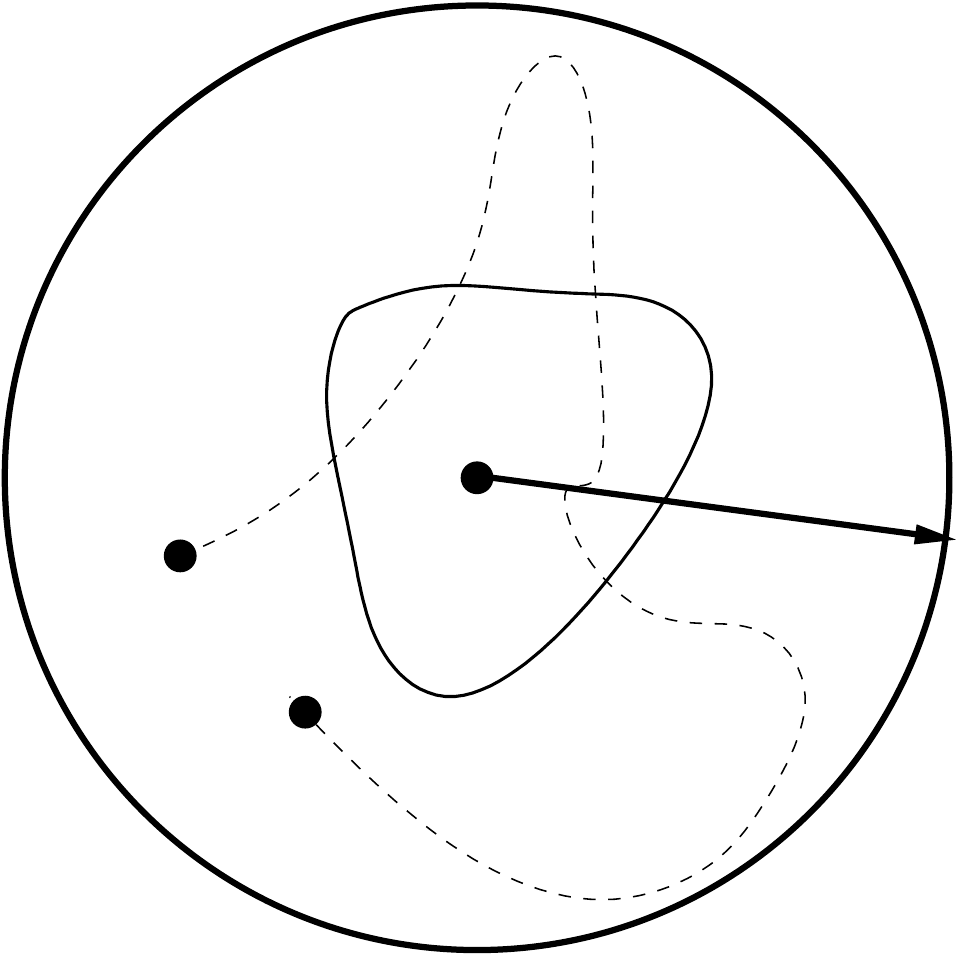}%
\end{picture}%
\setlength{\unitlength}{3947sp}%
\begingroup\makeatletter\ifx\SetFigFont\undefined%
\gdef\SetFigFont#1#2#3#4#5{%
  \reset@font\fontsize{#1}{#2pt}%
  \fontfamily{#3}\fontseries{#4}\fontshape{#5}%
  \selectfont}%
\fi\endgroup%
\begin{picture}(4624,4580)(4311,52249)
\put(8251,54389){\makebox(0,0)[lb]{\smash{{\SetFigFont{20}{24.0}{\familydefault}{\mddefault}{\updefault}{\color[rgb]{0,0,0}$r$}%
}}}}
\put(6451,54764){\makebox(0,0)[lb]{\smash{{\SetFigFont{20}{24.0}{\familydefault}{\mddefault}{\updefault}{\color[rgb]{0,0,0}$\mathbf{x}_{k'}$}%
}}}}
\put(4726,54314){\makebox(0,0)[lb]{\smash{{\SetFigFont{20}{24.0}{\familydefault}{\mddefault}{\updefault}{\color[rgb]{0,0,0}$\mu_T$}%
}}}}
\put(5326,53114){\makebox(0,0)[lb]{\smash{{\SetFigFont{20}{24.0}{\familydefault}{\mddefault}{\updefault}{\color[rgb]{0,0,0}$\mu_t $}%
}}}}
\end{picture}%

%% file: definitionSizeF.tex
\begin{picture}(0,0)%
\includegraphics{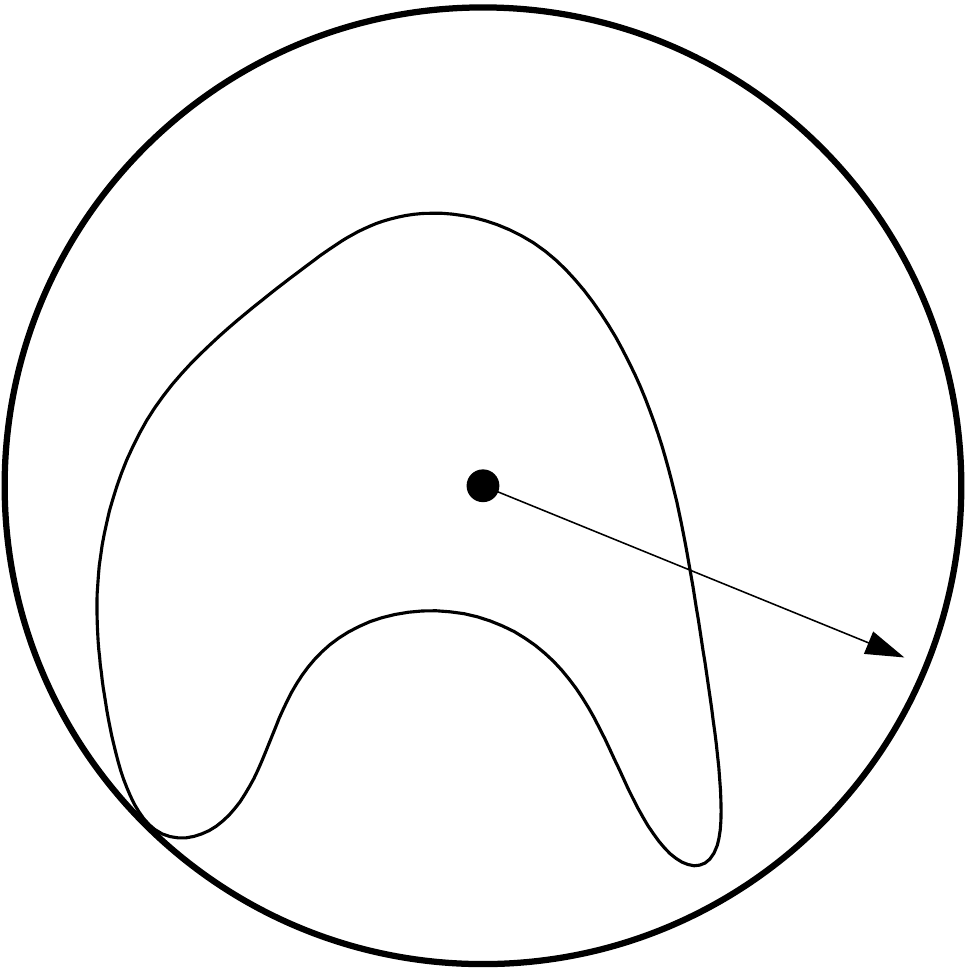}%
\end{picture}%
\setlength{\unitlength}{3947sp}%
\begingroup\makeatletter\ifx\SetFigFont\undefined%
\gdef\SetFigFont#1#2#3#4#5{%
  \reset@font\fontsize{#1}{#2pt}%
  \fontfamily{#3}\fontseries{#4}\fontshape{#5}%
  \selectfont}%
\fi\endgroup%
\begin{picture}(4638,4636)(307,-4104)
\put(1951,-361){\makebox(0,0)[lb]{\smash{{\SetFigFont{17}{20.4}{\rmdefault}{\mddefault}{\updefault}{\color[rgb]{0,0,0}${\cal X}_\alpha$}%
}}}}
\put(2326,-1636){\makebox(0,0)[lb]{\smash{{\SetFigFont{17}{20.4}{\rmdefault}{\mddefault}{\updefault}{\color[rgb]{0,0,0}$\mathbf{x}_k$}%
}}}}
\put(3901,-2236){\makebox(0,0)[lb]{\smash{{\SetFigFont{17}{20.4}{\rmdefault}{\mddefault}{\updefault}{\color[rgb]{0,0,0}$f(\alpha)$}%
}}}}
\end{picture}%

%% file: xfigDynTrajectory2.tex
\begin{picture}(0,0)%
\includegraphics{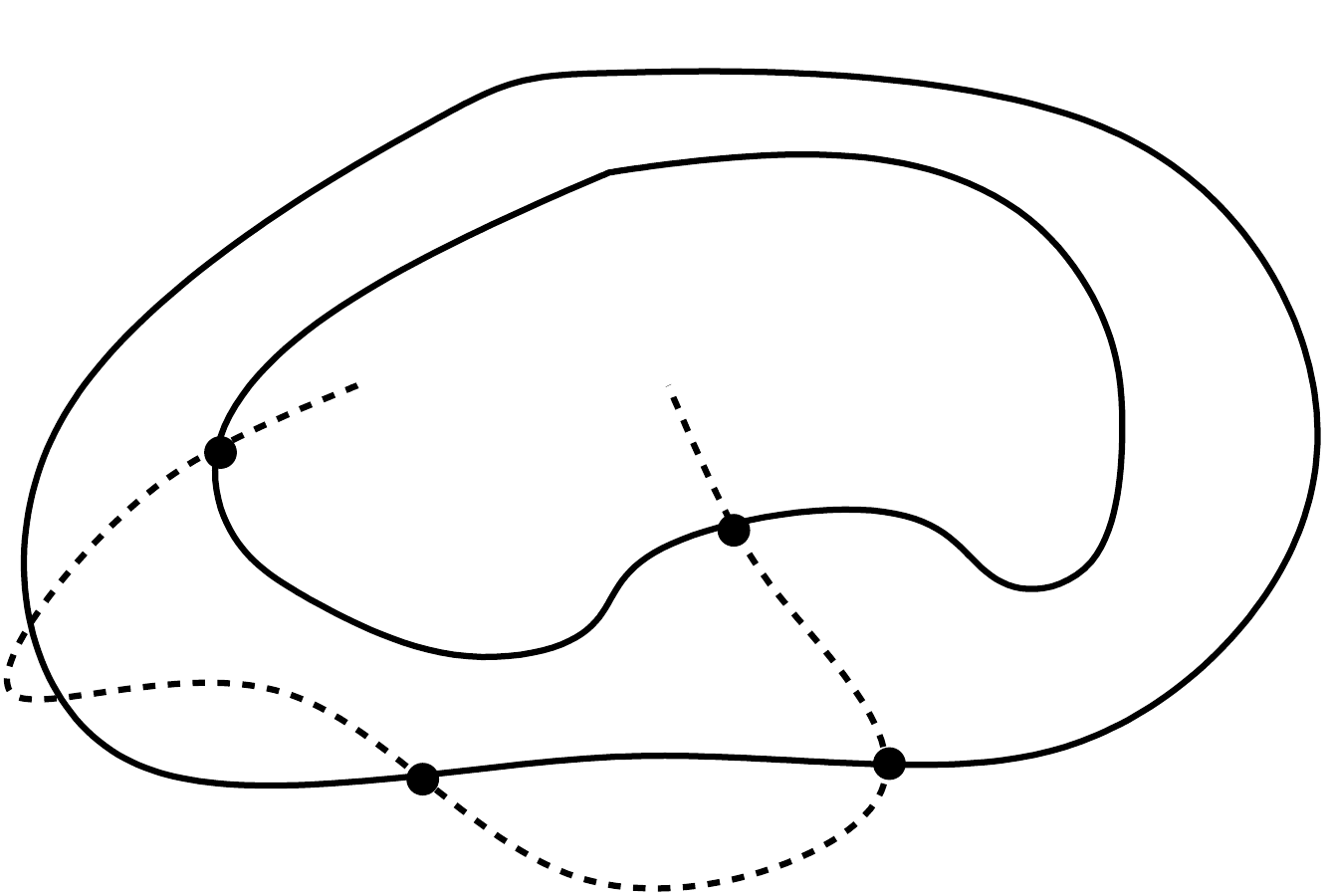}%
\end{picture}%
\setlength{\unitlength}{3947sp}%
\begingroup\makeatletter\ifx\SetFigFont\undefined%
\gdef\SetFigFont#1#2#3#4#5{%
  \reset@font\fontsize{#1}{#2pt}%
  \fontfamily{#3}\fontseries{#4}\fontshape{#5}%
  \selectfont}%
\fi\endgroup%
\begin{picture}(6385,4316)(663,44379)
\put(2176,44639){\makebox(0,0)[lb]{\smash{{\SetFigFont{14}{16.8}{\familydefault}{\mddefault}{\updefault}{\color[rgb]{0,0,0}$T_2$}%
}}}}
\put(3376,48464){\makebox(0,0)[lb]{\smash{{\SetFigFont{14}{16.8}{\familydefault}{\mddefault}{\updefault}{\color[rgb]{0,0,0}${\cal X}_{M\delta+\epsilon_2}$}%
}}}}
\put(3451,47564){\makebox(0,0)[lb]{\smash{{\SetFigFont{14}{16.8}{\familydefault}{\mddefault}{\updefault}{\color[rgb]{0,0,0}${\cal X}_{M\delta+\epsilon_1}$}%
}}}}
\put(1876,46364){\makebox(0,0)[lb]{\smash{{\SetFigFont{14}{16.8}{\familydefault}{\mddefault}{\updefault}{\color[rgb]{0,0,0}$T_1$}%
}}}}
\put(5026,44639){\makebox(0,0)[lb]{\smash{{\SetFigFont{14}{16.8}{\familydefault}{\mddefault}{\updefault}{\color[rgb]{0,0,0}$T_2'$}%
}}}}
\put(4126,46364){\makebox(0,0)[lb]{\smash{{\SetFigFont{14}{16.8}{\familydefault}{\mddefault}{\updefault}{\color[rgb]{0,0,0}$T_1'$}%
}}}}
\end{picture}%